\pdfoutput=1
\documentclass[11pt]{article}\usepackage[]{graphicx}\usepackage[]{color}
\makeatletter
\def\maxwidth{ %
  \ifdim\Gin@nat@width>\linewidth
    \linewidth
  \else
    \Gin@nat@width
  \fi
}
\makeatother

\definecolor{fgcolor}{rgb}{0.345, 0.345, 0.345}
\newcommand{\hlnum}[1]{\textcolor[rgb]{0.686,0.059,0.569}{#1}}%
\newcommand{\hlstr}[1]{\textcolor[rgb]{0.192,0.494,0.8}{#1}}%
\newcommand{\hlcom}[1]{\textcolor[rgb]{0.678,0.584,0.686}{\textit{#1}}}%
\newcommand{\hlopt}[1]{\textcolor[rgb]{0,0,0}{#1}}%
\newcommand{\hlstd}[1]{\textcolor[rgb]{0.345,0.345,0.345}{#1}}%
\newcommand{\hlkwa}[1]{\textcolor[rgb]{0.161,0.373,0.58}{\textbf{#1}}}%
\newcommand{\hlkwb}[1]{\textcolor[rgb]{0.69,0.353,0.396}{#1}}%
\newcommand{\hlkwc}[1]{\textcolor[rgb]{0.333,0.667,0.333}{#1}}%
\newcommand{\hlkwd}[1]{\textcolor[rgb]{0.737,0.353,0.396}{\textbf{#1}}}%

\usepackage{framed}
\makeatletter
\newenvironment{kframe}{%
 \def\at@end@of@kframe{}%
 \ifinner\ifhmode%
  \def\at@end@of@kframe{\end{minipage}}%
  \begin{minipage}{\columnwidth}%
 \fi\fi%
 \def\FrameCommand##1{\hskip\@totalleftmargin \hskip-\fboxsep
 \colorbox{shadecolor}{##1}\hskip-\fboxsep
     \hskip-\linewidth \hskip-\@totalleftmargin \hskip\columnwidth}%
 \MakeFramed {\advance\hsize-\width
   \@totalleftmargin\z@ \linewidth\hsize
   \@setminipage}}%
 {\par\unskip\endMakeFramed%
 \at@end@of@kframe}
\makeatother

\definecolor{shadecolor}{rgb}{.97, .97, .97}
\definecolor{messagecolor}{rgb}{0, 0, 0}
\definecolor{warningcolor}{rgb}{1, 0, 1}
\definecolor{errorcolor}{rgb}{1, 0, 0}
\newenvironment{knitrout}{}{} 

\usepackage{alltt}
\usepackage{amsfonts, amsmath, amssymb, amsthm}
\usepackage[english]{babel}
\usepackage{booktabs}
\usepackage{bm}
\usepackage[labelsep=period]{caption}
\usepackage{eucal}
\usepackage{enumerate}
\usepackage{float}
\usepackage{dsfont}
\usepackage{epsfig}
\usepackage {fontenc}
\usepackage[hmargin=1.6cm,vmargin=3cm]{geometry}
\usepackage{lscape}
\usepackage{mathrsfs}
\usepackage{mdframed}
\usepackage{natbib}
\usepackage{rotating}
\usepackage{verbatim}
\usepackage{subfigure}
\usepackage[table]{xcolor}
\usepackage{enumitem}

\renewcommand{\thefootnote}{$\dagger$}

\IfFileExists{upquote.sty}{\usepackage{upquote}}{}
\begin{document}
\title{\vspace{-2cm} {Bayesian Bootstrap Inference for the ROC Surface}}
\author{{Vanda~\textsc{In\'acio de Carvalho}, Miguel~\textsc{de Carvalho}, and Adam~\textsc{Branscum}}}
\date{}
\maketitle

\begin{abstract}
\footnotesize{Accurate diagnosis of disease is of great importance in clinical practice and medical research. The receiver operating characteristic (ROC) surface is a popular tool for evaluating the discriminatory ability of continuous diagnostic test outcomes when there exist three ordered disease classes (e.g., no disease, mild disease, advanced disease). We propose the Bayesian bootstrap, a fully nonparametric method, for conducting inference about the ROC surface and its functionals, such as the volume under the surface. The proposed method is based on a simple, yet interesting, representation of the ROC surface in terms of placement variables. Results from a simulation study demonstrate the ability of our method to successfully recover the true ROC surface and to produce valid inferences in a variety of complex scenarios. An application to data from the Trail Making Test to assess cognitive impairment in Parkinson's disease patients is provided.}
\end{abstract}
\let\thefootnote\relax\footnotetext{Vanda In\'acio de Carvalho is Lecturer in Statistics, School of Mathematics, University of Edinburgh, Scotland, UK (\textit{Vanda.Inacio@ed.ac.uk}). Miguel de Carvalho is Lecturer in Statistics, School of Mathematics, University of Edinburgh, Scotland, UK (\textit{Miguel.deCarvalho@ed.ac.uk}). Adam J. Branscum is Professor, College of Public Health and Human Sciences, Oregon State University, USA (\textit{adam.branscum@oregonstate.edu}).}
\textsc{\footnotesize{key words: Diagnostic test; Bayesian bootstrap; ROC surface; Volume under the surface; Trail Making test} }

\section{\large{\textsf{INTRODUCTION}}}\label{intro}
Evaluating and ranking the performance of medical diagnostic tests is of fundamental importance in health care. Before a test is approved for routine use in practice, its ability to distinguish between different disease stages or conditions must be rigorously evaluated through statistical analysis. The receiver operating characteristic (ROC) curve is a popular tool for evaluating the accuracy of continuous outcome diagnostic tests that classify subjects into two groups: diseased and nondiseased. However, disease progression can be regarded as a dynamic process and, in clinical practice, physicians often face situations that require a decision among three or more diagnostic alternatives. Patients may advance through one or more transitional stages prior to full disease onset, and this is especially true for neurological disorders. For instance, in Section \ref{application} we present an assessment of the discriminatory ability of the Trail Making Test, a widely used test to detect cognitive impairment associated with dementia, to distinguish between Parkinson's disease patients who present normal cognition, mild cognitive impairment, and dementia/severe impairment. As a direct generalisation of ROC curves, the ROC surface has been used to evaluate diagnostic accuracy in ordered three-class classification problems \citep{Yang2000,Nakas2004}; see also \cite{Nakas2014} for an insightful review of three-class ROC surface analysis. The volume under the ROC surface (VUS) has been proposed as a scalar summary measure of diagnostic accuracy, analogous to the area under the ROC curve in the two-class setting.\\
\indent There is a vast literature on parametric, semiparametric, and nonparametric ROC curve analysis; see \cite{Pepe03} and \cite{Zhou2011} for an overview. The amount of existing research on ROC surface analysis is, by comparison, limited.  \cite{Li2009} developed a frequentist nonparametric approach to estimating the ROC surface based on the empirical distribution function and a semiparametric approach that attempts to generalise the parametric (normal) functional form of the ROC surface but that, as pointed out by the authors, relies heavily on the normality assumption.   \cite{Inacio2011} developed a  nonparametric Bayesian method based on finite mixtures of Polya trees to estimate the ROC surface, while \cite{Zhang2011} developed methods for combining multiple biomarkers, \cite{Yu2012} considered rare diseases and high-throughput data, \cite{Li2012} considered ROC surface regression analysis, \cite{Kang2013} developed nonparametric estimators based on kernel methods, and \cite{Tahani2014} considered a frequentist nonparametric predictive approach. An empirical likelihood approach \citep{Wan2012} and inverse probability weighting \citep{Zhang2016} have been used to estimate the volume under the ROC surface, and test statistics were developed by \cite{Hong2015}. \\
\indent In this paper, we develop a computationally appealing, fully nonparametric estimator of the ROC surface based on the Bayesian bootstrap. The Bayesian bootstrap \citep{Rubin1981} is the Bayesian analog of the frequentist bootstrap.  A key difference is the Bayesian bootstrap is based on simulation rather than resampling (details about the Bayesian bootstrap are given in Section \ref{preliminaries}). Our nonparametric estimator of the ROC surface is robust and smooth.  Compared to a kernel approach, it does not depend on a smoothing parameter, the choice of which is a nontrivial issue in practice and has a great impact on inference. Our method is hence a widely applicable approach to inference for the ROC surface that can be used for many populations and for a large number of diseases and continuous diagnostic measures. Moreover, point estimates and credible intervals for the ROC surface and its corresponding VUS are obtained in a single integrated framework. Recent developments of flexible Bayesian models that have been successfully applied in medical diagnostic testing research abound (e.g., \citealt{Erkanli2006,Branscum2008,Branscum2013,Inacio2013,Rodriguez2013,Branscum2015,Hwang2015,Zhao2016,Inacio2017,Inacio2018}).\\
\indent The remainder of the paper is organised as follows. In the next section we introduce background material on ROC surfaces and the Bayesian bootstrap. Section 3 presents our novel nonparametric approach to estimating the ROC surface. The performance of our method is assessed in Section 4 using simulated data, Section 5 applies our approach to data from a Trail Making Test to detect cognitive impairment, and Section 6 concludes the paper. The Trail Making Test data and an \texttt{R} function for implementing our method are provided in the Supplementary Materials.

\section{\large{\textsf{PRELIMINARIES}}}\label{preliminaries}
\subsection{\textsf{ROC surfaces}}
We assume that each subject in the population belongs to one of three ordered diagnostic groups (e.g., no disease, mild disease, advanced disease) and that a diagnostic test with continuous scale outcomes is used for classification into one of the three groups. We further assume that the group to which each subject belongs to is known without error due to the existence of a gold standard test. Without loss of generality, we assume that individuals from group 3 tend to have higher test outcomes than individuals in group 2 who tend to have higher test values than group 1 individuals. Let $Y_1$, $Y_2$, and $Y_3$ be continuous random variables denoting test outcomes in groups 1, 2, and 3, respectively, with $F_1$, $F_2$, and $F_3$ being the corresponding cumulative distribution functions. For any pair of ordered thresholds $(c_1,c_2)$, with $c_1<c_2$, the probabilities of correct classification into each group are given by
\begin{align*}
p_1(c_1,c_2)&=\Pr(Y_1\leq c_1)=F_1(c_1),\\
p_2(c_1,c_2)&=\Pr(c_1<Y_2\leq c_2)=F_2(c_2)-F_2(c_1),\\
p_3(c_1,c_2)&=\Pr(Y_3>c_2)=1-F_3(c_2).
\end{align*}
The ROC surface is then the three-dimensional plot in the unit cube depicting the probabilities of correct classification into each group as the thresholds $c_1$ and $c_2$ vary
\begin{small}
\begin{equation*}
\{(p_1(c_1,c_2),p_2(c_1,c_2),p_3(c_1,c_2)): c_1<c_2 \}=\{(F_1(c_1),F_2(c_2)-F_2(c_1),1-F_3(c_2)): c_1<c_2\}.
\end{equation*}
\end{small}
\noindent For notational simplicity, hereafter we drop the dependence of $p_1$, $p_2$, and $p_3$ on $c_1$ and $c_2$. \\
\indent By writing $c_1=F_{1}^{-1}(p_1)$ and $c_2=F_{3}^{-1}(1-p_3)$, we obtain the functional form of the ROC surface
\begin{equation*}
\text{ROCS}(p_1,p_3)=
\begin{cases}
F_{2}(F_{3}^{-1}(1-p_3))-F_{2}(F_{1}^{-1}(p_1)), \quad \mbox{if } F_{1}^{-1}(p_1)<F_{3}^{-1}(1-p_3),\\
0,\quad \mbox{\hspace{5.4cm}otherwise}.
\end{cases}
\end{equation*}
The volume under the ROC surface is a summary measure of the overall diagnostic accuracy and it is defined as
\begin{align*}
\text{VUS}&=\int_{0}^{1}\int_{0}^{1}\text{ROCS}(p_1,p_3)\text{d}p_3\text{d}p_1\\
&=\int_{0}^{1}\int_{0}^{1-F_3(F_1^{-1}(p_1))}\left\{F_{2}(F_{3}^{-1}(1-p_3))-F_{2}(F_{1}^{-1}(p_1))\right\}\text{d}p_3\text{d}p_1\\
&=\Pr(Y_1<Y_2<Y_3).
\end{align*}
When the three distributions completely overlap, and thus the test has no discriminatory ability, the VUS takes the value $1/6$, whereas a VUS of 1 corresponds to a test that perfectly discriminates between the three groups. Other values of VUS correspond to different degrees of overlap/stochastic ordering between $f_1/F_1$, $f_2/F_2$, and $f_3/F_3$; the closer the VUS is to $1$, the better the classification accuracy (see Figure 1 of the Supplementary Materials).

\subsection{\textsf{Bayesian bootstrap}}
The Bayesian bootstrap (BB) was introduced by \cite{Rubin1981} as a Bayesian counterpart of the original bootstrap proposed by \cite{Efron1979} and it is based on simulation rather than resampling. Let $(y_{1},\ldots,y_{n})$ be a random sample from an unknown distribution $F$ and suppose that the parameter of interest is $F$ itself, which is represented as
\begin{equation*}
F(z)=\sum_{i=1}^{n}\omega_{i}I(y_i\leq z),
\end{equation*}
where $I(\cdot)$ denotes the indicator function, $\omega_i$ is the weight associated to observation $y_i$, with $\omega_i\geq0$ and $\sum_{i=1}^{n}\omega_i=1$. In the classic nonparametric bootstrap, inference about $F$ is obtained by repeatedly generating bootstrap samples, where each bootstrap sample is drawn with replacement from the data. In the $b$th bootstrap replicate, $F^{(b)}$ is computed as
\begin{equation*}
F^{(b)}(z)=\sum_{i=1}^{n}\omega_{i}^{(b)}I(y_i\leq z),
\end{equation*}
where $\omega_{i}^{(b)}$ is the number of times $y_i$ appears in the $b$th bootstrap sample, with $\omega_{i}^{(b)}$ taking values on the discrete set $\{0,1/n,\ldots,n/n\}$. For the Bayesian bootstrap, the weights are considered unknown and their posterior distribution is derived. \cite{Rubin1981} used a diffuse prior, $\prod_{i=1}^{n}\omega_i^{-1}$, which when combined with the (multinomial) likelihood for the data, results in a $\text{Dirichlet}(n;1,\ldots,1)$ distribution for the posterior distribution of the weights. Thus, the weights in the BB are smoother than those from the classical bootstrap. Note that in the BB the data are regarded as fixed, so we do not resample from it. The BB has connections to the Dirichlet Process \citep{Ferguson1973}; specifically, it can be regarded as a non-informative version of the Dirichlet Process \citep[][Theorem 2]{Gasparini1995}. For a further explanation of the different views of the BB we refer the reader to \citet[][p.~971]{Kim2005}.

\section{\large{\textsf{PROPOSED ESTIMATOR}}}
Our estimator extends to the three class-setting the method proposed by \cite{Gu2008} for the ROC curve. It is motivated by a simple, yet interesting and computational appealing representation of the ROC surface that is based on the notion of a placement variable \citep[][Chapter 5]{Pepe03}.  A placement variable is simply a standardisation of test outcomes with respect to a reference population. Consider the following two variables
\begin{equation*}
U_1=F_1(Y_2),\qquad U_3=1-F_3(Y_2),
\end{equation*}
with $U_1$ being the proportion of class 1 subjects with test outcomes less than or equal to $Y_{2}$ and $U_3$ being the proportion of class 3 subjects with test outcomes greater than $Y_2$. Here, group 2 is the reference group. The variables $U_1$ and $U_3$ quantify the degree of separation of the test outcome distributions in the three groups of patients.  Specifically, $U_1$ quantifies the degree of separation between the test outcomes  in groups 1 and 2, whereas $U_3$ quantifies the degree of separation between groups 2 and 3. For instance, if the test outcomes in the three groups are highly separated, the placement of most group 2 subjects is at the upper tail of the group 1 distribution and at the lower tail of the group 3 distribution, so that most group 2 subjects will have large $U_1$ and $U_3$ values. On the other hand, when the three distributions of test outcomes completely overlap, both $U_1$ and $U_3$ will have a $\text{Uniform}(0,1)$ distribution. \\
\indent Interestingly, the ROC surface is the difference between the survival distribution of $U_3$ and the cumulative distribution of $U_1$. Specifically, if $F_1^{-1}(p_1)<F_3^{-1}(1-p_3)$, we have
\begin{align}\label{rocspv}
\text{ROCS}(p_1,p_3)&=F_2(F_3^{-1}(1-p_3))-F_2(F_1^{-1}(p_1)) \nonumber\\
&=\Pr(Y_2\leq F_3^{-1}(1-p_3))-\Pr(Y_2\leq F_1^{-1}(p_1)) \nonumber\\
&=\Pr(1-F_3(Y_2)>p_3)-\Pr(F_1(Y_2)\leq p_1) \nonumber\\
&=\Pr(U_3>p_3)-\Pr(U_1\leq p_1).
\end{align}
Let $(y_{11},\ldots,y_{1n_{1}})$, $(y_{21},\ldots,y_{2n_{2}})$, and $(y_{31},\ldots,y_{3n_{3}})$ be independent (within and between groups) samples of size $n_1$, $n_2$, and $n_3$ from groups 1, 2, and 3, respectively. The result in \eqref{rocspv} provides the rationale for the following computational algorithm.  Let $B$ denote the number of iterations.
\begin{description}[align=left]
\item [\textbf{Step 1:}] \textbf{Computation of placement variables based on the BB.}\\
For $b=1,\ldots,B$, let
\begin{equation*}
U_{1j}^{(b)}=F_1^{(b)}(y_{2j})=\sum_{i=1}^{n_1}v_{1i}^{(b)}I(y_{1i}\leq y_{2j}),
\end{equation*}
and
\begin{equation*}
U_{3j}^{(b)}=1-F_3^{(b)}(y_{2j})=\sum_{\ell=1}^{n_3}v_{3\ell}^{(b)}I(y_{3\ell}>y_{2j}),
\end{equation*}
where $j=1,\ldots,n_2$, $(v_{11}^{(b)},\ldots,v_{1n_{1}}^{(b)})\sim\text{Dirichlet}(n_1;1,\ldots,1)$, and $(v_{31}^{(b)},\ldots,v_{3n_{3}}^{(b)})\sim\text{Dirichlet}(n_3;1,\ldots,1)$.
\item [\textbf{Step 2:}] \textbf{Generate a random realisation of the ROC surface.}\\
Based on \eqref{rocspv}, generate a realisation of $\text{ROCS}^{(b)}(p_1,p_3)$, the difference between the survival function of $(U_{31}^{(b)},\ldots,U_{3n_{2}}^{(b)})$ and the distribution function of $(U_{11}^{(b)},\ldots,U_{1n_{2}}^{(b)})$, i.e.,
\begin{footnotesize}
\begin{equation*}
\text{ROCS}^{(b)}(p_1,p_3)=
\begin{cases}
\sum_{j=1}^{n_2}w_{3j}^{(b)}I(U_{3j}^{(b)}>p_3)-\sum_{j=1}^{n_2}w_{1j}^{(b)}I(U_{1j}^{(b)}\leq p_1),\quad \mbox{if } \sum_{j=1}^{n_2}w_{3j}^{(b)}I(U_{3j}^{(b)}>p_3)>\sum_{j=1}^{n_2}w_{1j}^{(b)}I(U_{1j}^{(b)}\leq p_1),\\
0,\quad \mbox{\hspace{6.8cm}otherwise},
\end{cases}
\end{equation*}
\end{footnotesize}
where $p_1$ and $p_3$ span grids over $[0,1]$, $(w_{11}^{(b)},\ldots,w_{1n_{2}}^{(b)})\sim\text{Dirichlet}(n_2;1,\ldots,1)$, and $(w_{31}^{(b)},\ldots,w_{3n_{2}}^{(b)})\sim\text{Dirichlet}(n_2;1,\ldots,1)$.
\end{description}
The BB estimate of the ROC surface, denoted as $\widehat{\text{ROCS}}(p_1,p_3)$ is obtained by averaging over the ensemble of ROC surfaces $\left\{\text{ROCS}^{(1)}(p_1,p_3),\ldots,\text{ROCS}^{(B)}(p_1,p_3)\right\}$, that is,
\begin{equation*}
\widehat{\text{ROCS}}(p_1,p_3)=\frac{1}{B}\sum_{b=1}^{B}\text{ROCS}^{(b)}(p_1,p_3).
\end{equation*}
Similarly, the posterior mean for the VUS can be computed as
\begin{equation*}
\widehat{\text{VUS}}=\frac{1}{B}\sum_{b=1}^{B}\text{VUS}^{(b)}, \qquad \text{VUS}^{(b)}=\int_{0}^{1}\int_{0}^{1}\text{ROCS}^{(b)}(p_1,p_3)\text{d}p_3\text{d}p_1.
\end{equation*}
A credible interval for the VUS can be obtained from the percentiles of the ensemble $\left(\text{VUS}^{(1)},\ldots,\text{VUS}^{(B)}\right)$.

\section{\large{\textsf{SIMULATION STUDY}}}\label{simulation}
A simulation study was conducted to evaluate the performance of our approach to conduct inference about the ROC surface and its associated VUS.

\subsection{Simulation scenarios}\label{scenarios}
We considered four scenarios as listed in Table \ref{t1}. Scenario 1 corresponds to the case where test outcomes from the three groups follow normal distributions. In Scenario 2, data from the three groups follow non-normal distributions from the same family, namely the family of gamma distributions. In Scenario 3, test outcomes arise from different distributional families.  Lastly, Scenario 4 considers mixtures of distributions for test outcome data, a setting that is common in practice.

\subsection{Models}
For our BB estimator we only need to specify the number of BB iterates $B$, which we set equal to $2000$. For the grid of values for $p_1$ and $p_3$, the probabilities of correct classification in group 1 and 3, respectively, we used 50 equidistant points on $[0,1]$. \\
\indent We compared the performance of our nonparametric BB estimator against its main nonparametric competitors, namely, the frequentist kernel estimator and the empirical estimator. The empirical method simply estimates $F_d$ by its empirical distribution function. For the kernel estimator, the cumulative distribution function in each group is estimated as
\begin{equation*}
\widehat{F}_{d}(y)=\frac{1}{n_{d}}\sum_{i=1}^{n_{d}}\Phi\left(\frac{y-y_{di}}{h_d}\right),\qquad d\in\{1,2,3\},
\end{equation*}
where $\Phi(\cdot)$ stands for the standard normal distribution function. For the bandwidth $h_d$, which controls the amount of smoothing, we considered two options. The first option was
\begin{equation}\label{bwdef}
h_{d}=0.9\min\{\text{SD}(\mathbf{y}_d),\text{IQR}(\mathbf{y}_d)/1.34\}n_{d}^{-0.2},
\end{equation}
where $\text{SD}(\mathbf{y}_d)$ and $\text{IQR}(\mathbf{y}_d)$ are the standard deviation and interquantile range, respectively, of $\mathbf{y}_d=(y_{d1},\ldots,y_{dn_{d}})$.  This is the default choice in the \texttt{R} statistical software \citep{R17} and it is implemented in the  function \texttt{bw.nrd0}. It is well-known \citep[e.g.,][p.~61]{Wand1994} that this `rule', although providing reasonable bandwidths for non-normal data, is `optimal' when the data distribution is normal. For this reason, we have also considered a bandwidth selected by least squares cross-validation \citep[][Chapter 3]{Wand1994}, which is implemented in \texttt{R} by the function \texttt{bw.ucv}. Estimation of VUS for the empirical and kernel approaches used the following  closed form expressions \citep[see e.g.,][]{Kang2013}:
\begin{align*}
\widehat{\text{VUS}}_{e}&=\frac{1}{n_1n_2n_3}\sum_{i=1}^{n_1}\sum_{j=1}^{n_2}\sum_{\ell=1}^{n_3}I(y_{1i}<y_{2j}<y_{3\ell}),\\
\widehat{\text{VUS}}_{k}&=\frac{1}{n_1n_2n_3}\sum_{i=1}^{n_1}\sum_{j=1}^{n_2}\sum_{\ell=1}^{n_3}\Phi\left(\frac{y_{2j}-y_{1i}}{\sqrt{h_1^2+h_2^2}}\right)\Phi\left(\frac{y_{3\ell}-y_{2j}}{\sqrt{h_2^2+h_3^2}}\right),
\end{align*}
where $\text{VUS}_{e}$ and $\text{VUS}_{k}$ stand, respectively, for the empirical and kernel VUS.\\
\indent For Scenario 1 (where test outcomes in each group follow a normal distribution), we also included a comparison to a model involving independent parametric normal distributions, in order to assess the efficiency of our nonparametric estimator in this context. \\
\subsection{Results}
\begin{figure}[H]
\begin{center}
\subfigure{\includegraphics[page = 1, width=5.75cm]{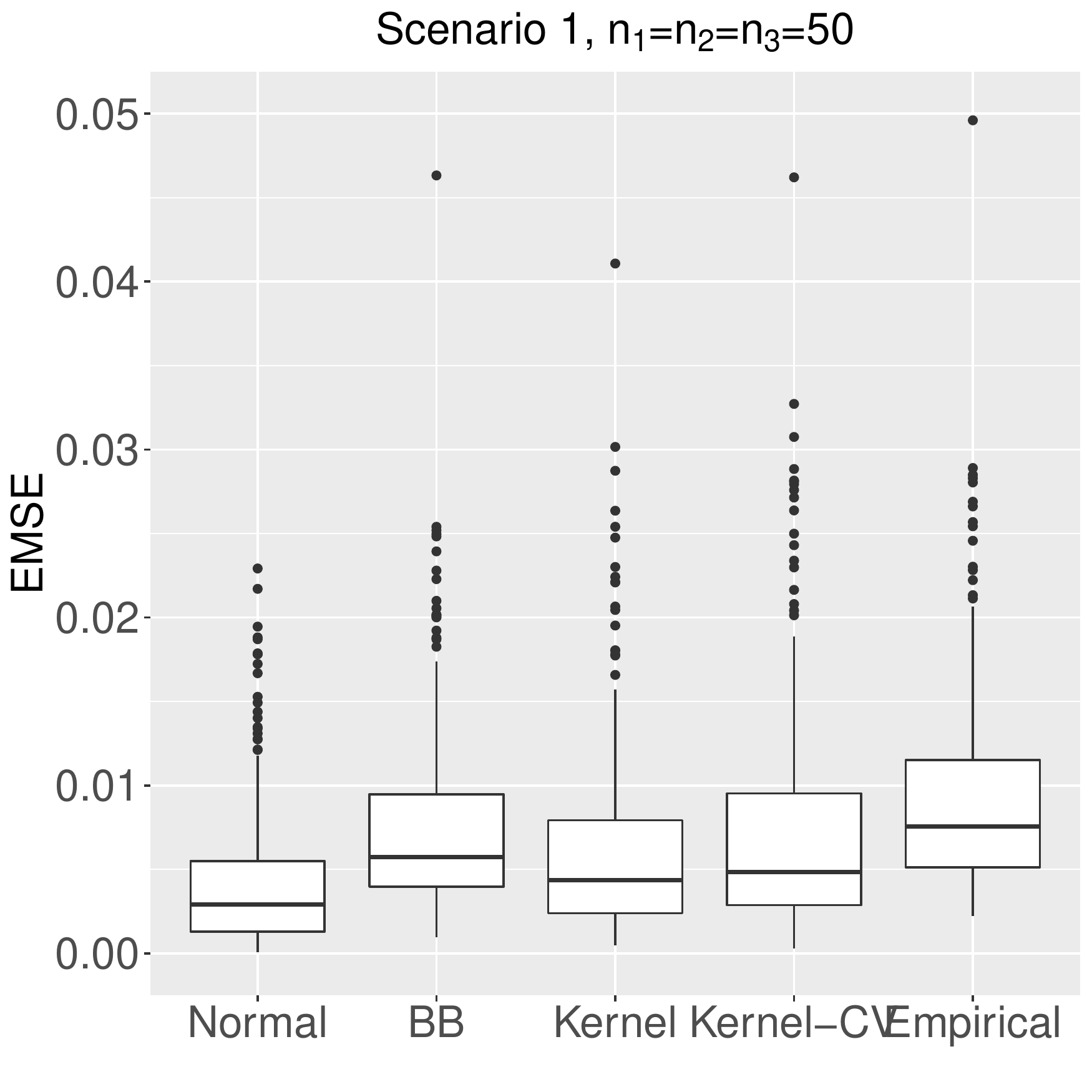}}\hspace{0.3cm}
\subfigure{\includegraphics[page = 1, width=5.75cm]{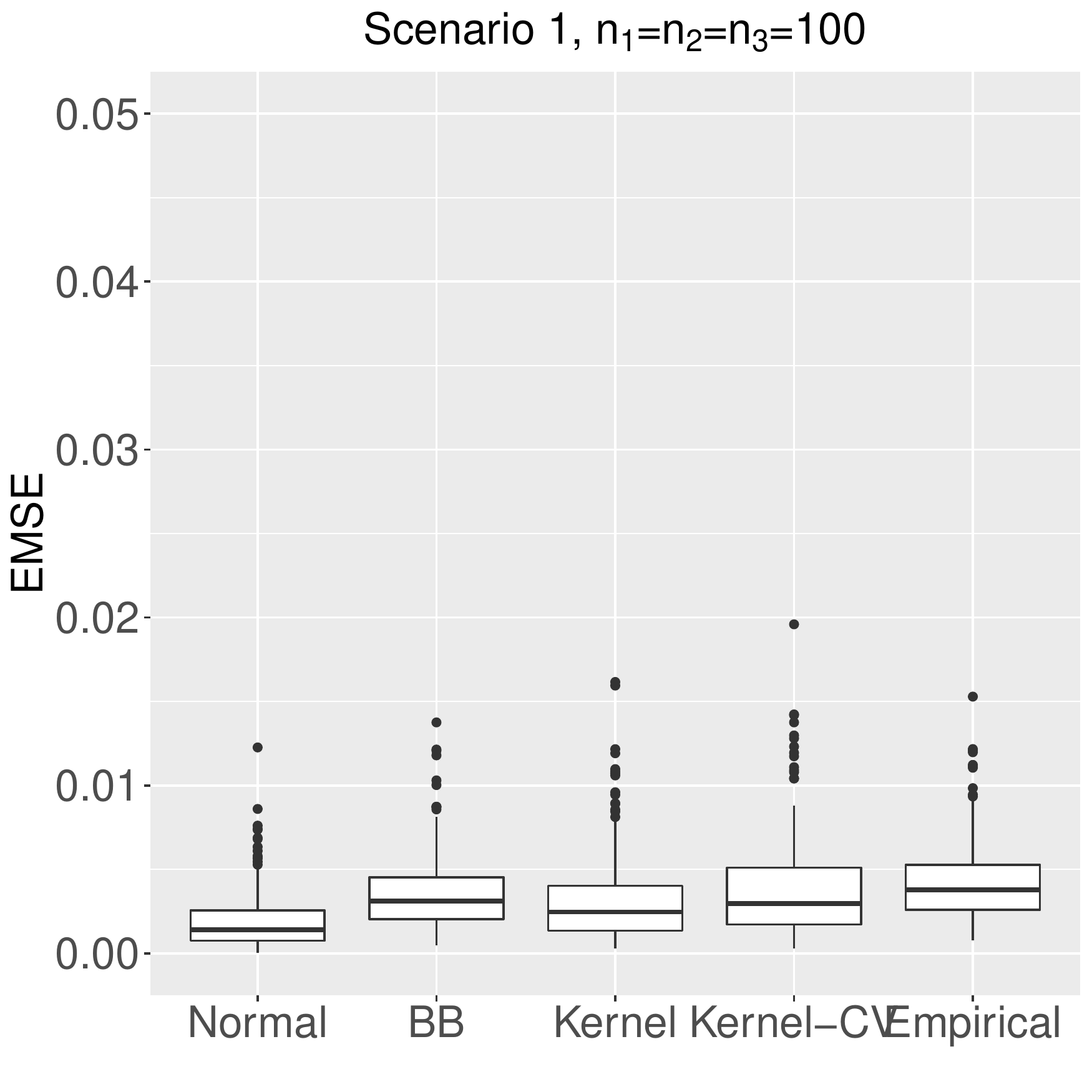}}\hspace{0.3cm}
\subfigure{\includegraphics[page = 1, width=5.75cm]{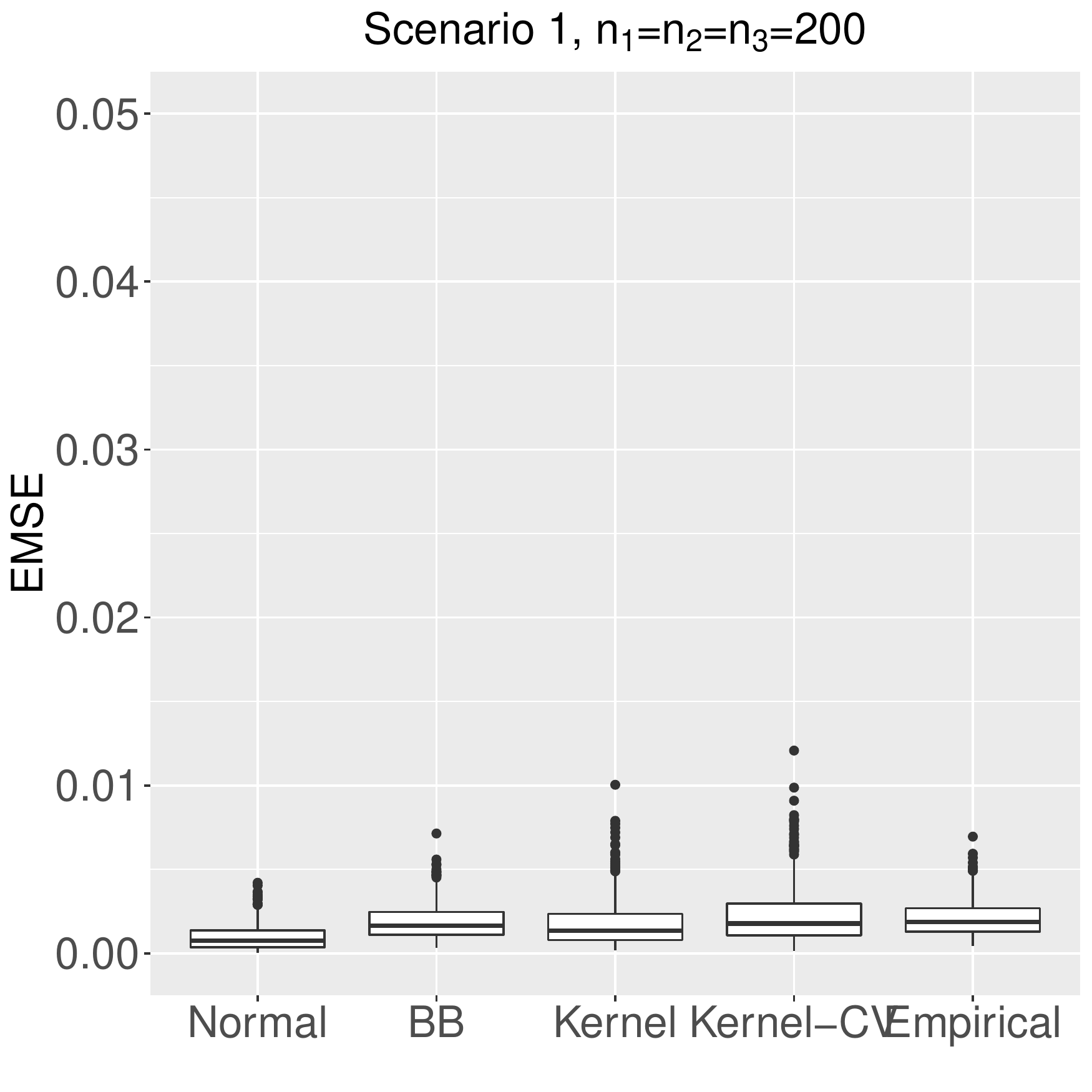}}\\ \vspace{0.3cm}
\subfigure{\includegraphics[page = 1, width=5.75cm]{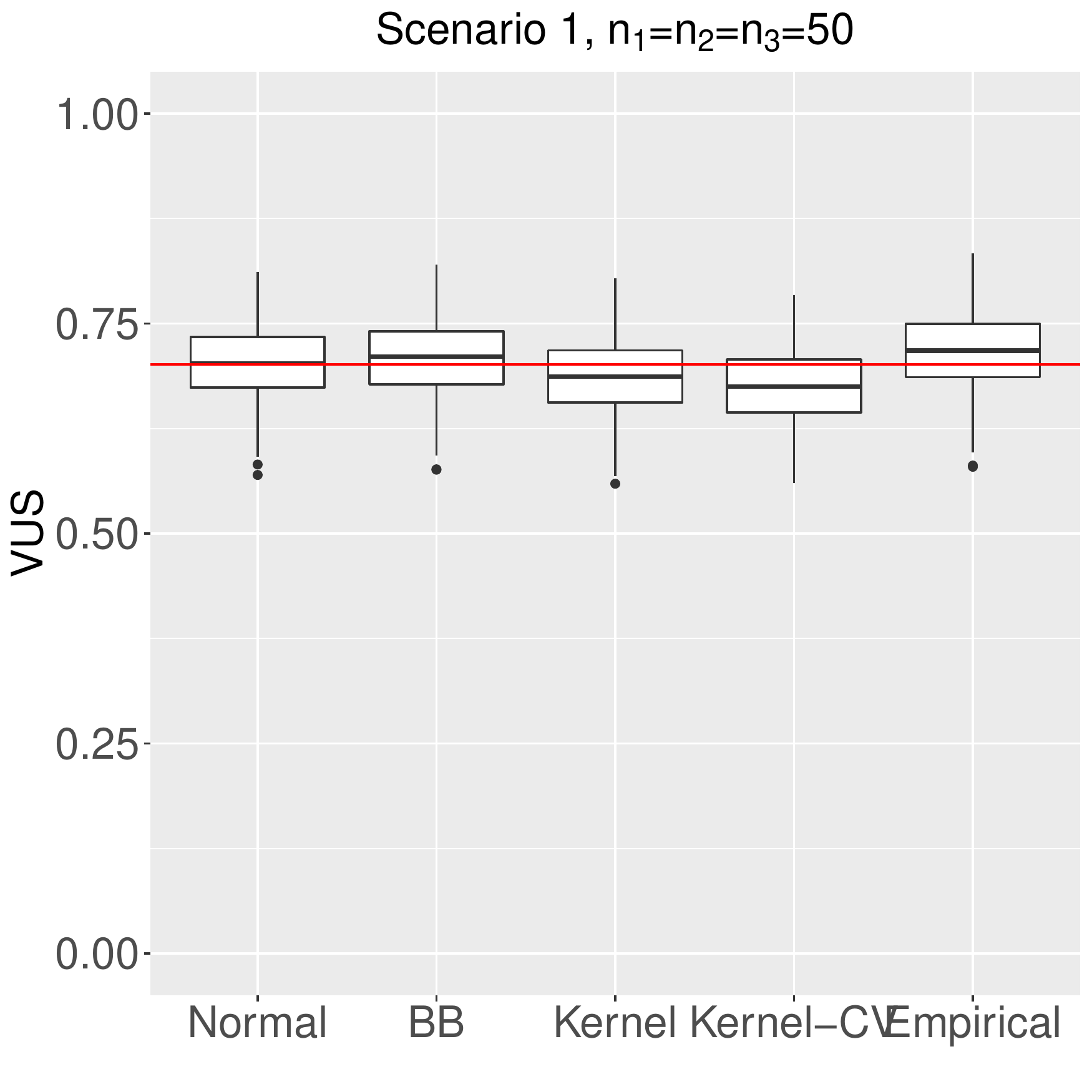}}\hspace{0.3cm}
\subfigure{\includegraphics[page = 1, width=5.75cm]{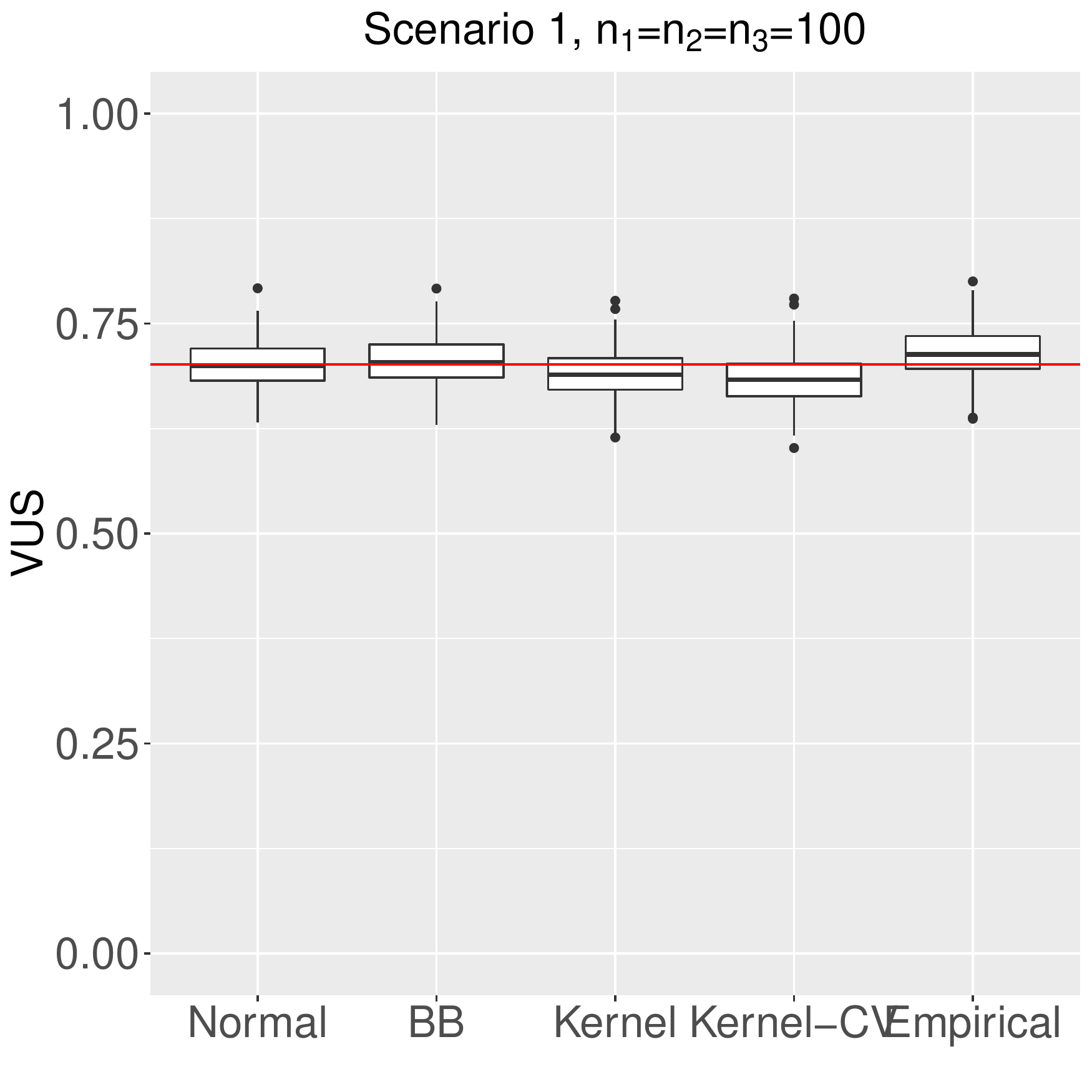}}\hspace{0.3cm}
\subfigure{\includegraphics[page = 1, width=5.75cm]{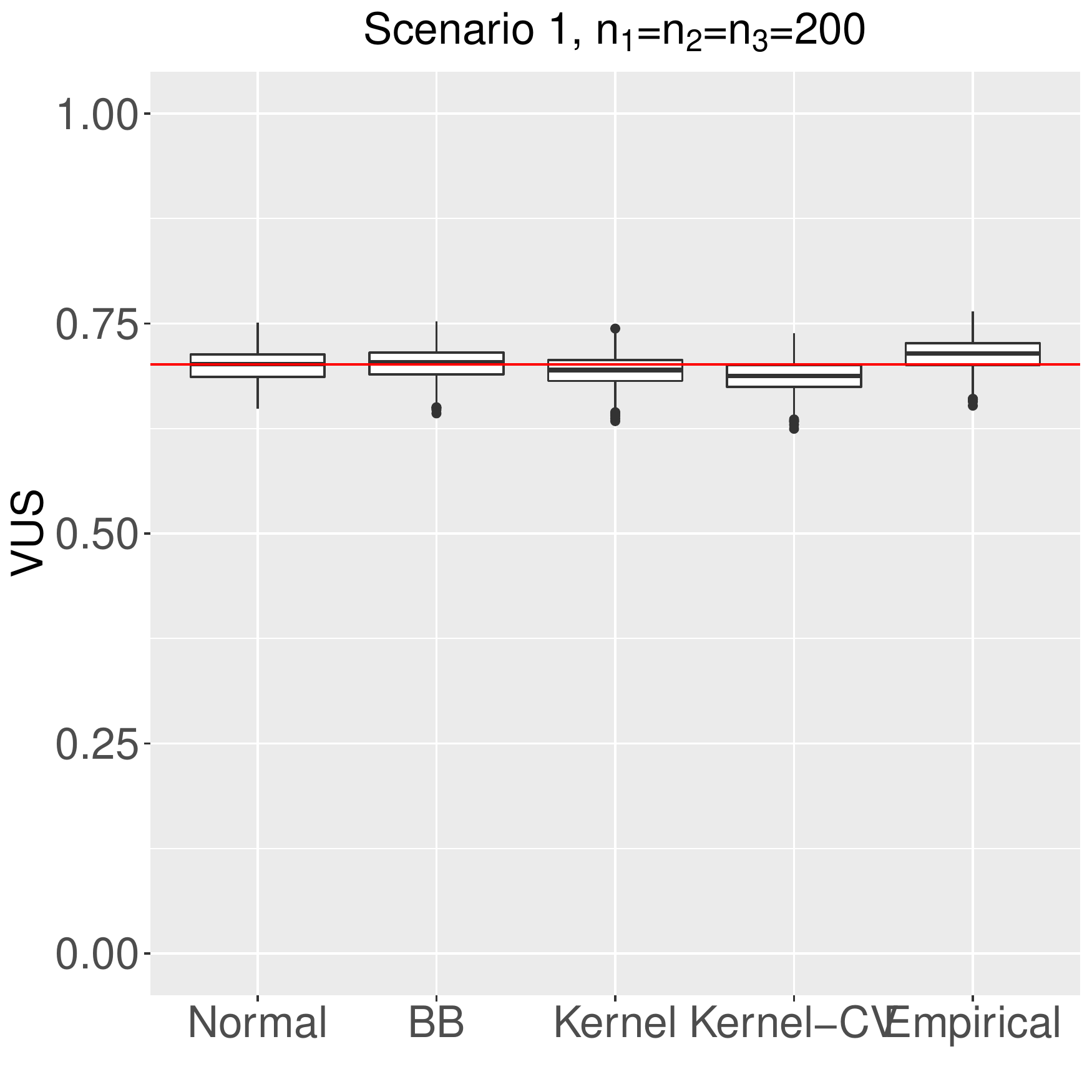}}
\caption{\footnotesize{Scenario 1. Boxplots summarising simulation results for the EMSE (top row) and estimates of the VUS (bottom row). The solid red line corresponds to the true VUS. Here Kernel denotes the kernel estimate with bandwidth calculated using equation \eqref{bwdef} and Kernel-CV stands for the kernel estimate with the bandwidth selected by least squares cross-validation.}}
\label{sc1}
\end{center}
\end{figure}

\begin{figure}[H]
\begin{center}
\subfigure{\includegraphics[page = 1, width=5.75cm]{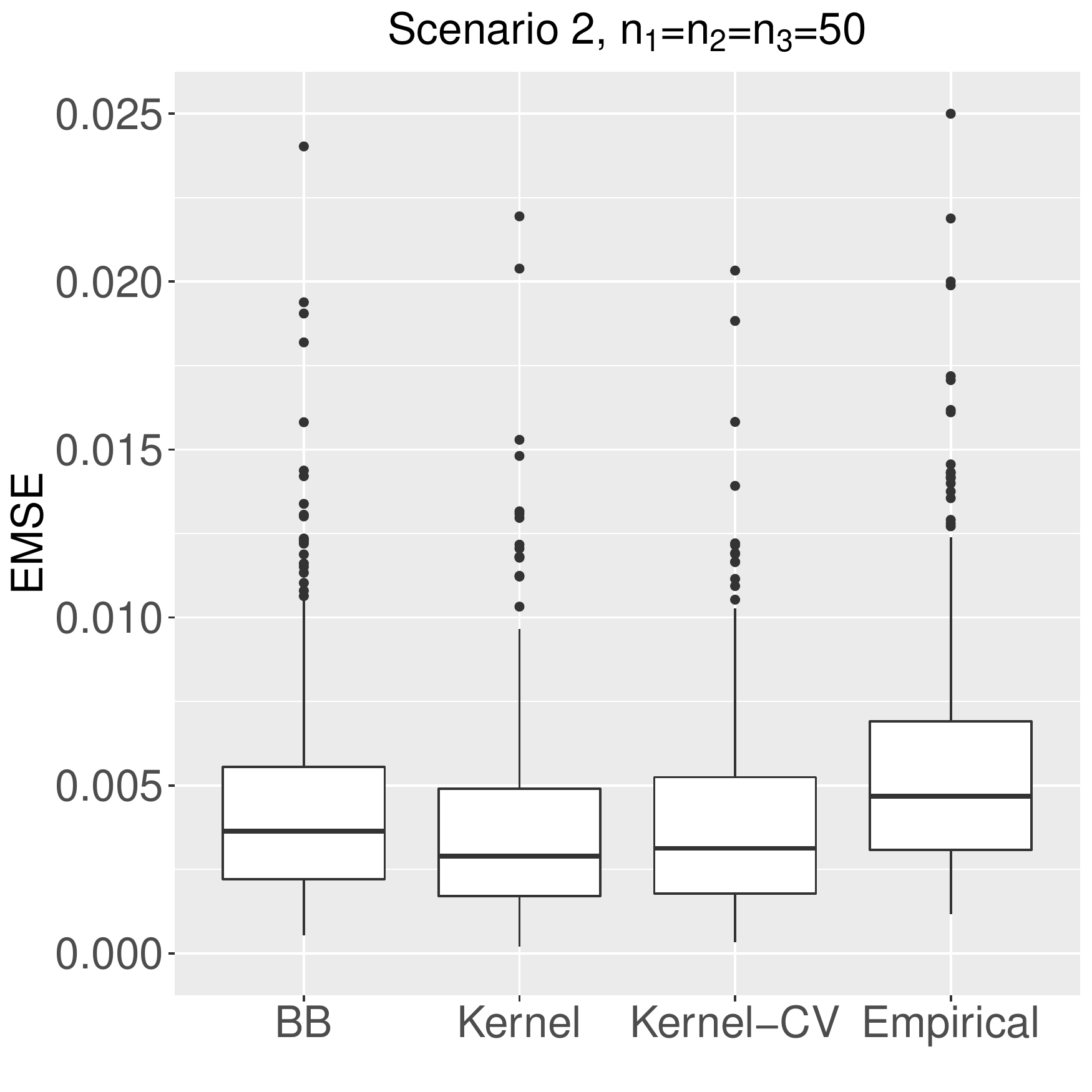}}\hspace{0.3cm}
\subfigure{\includegraphics[page = 1, width=5.75cm]{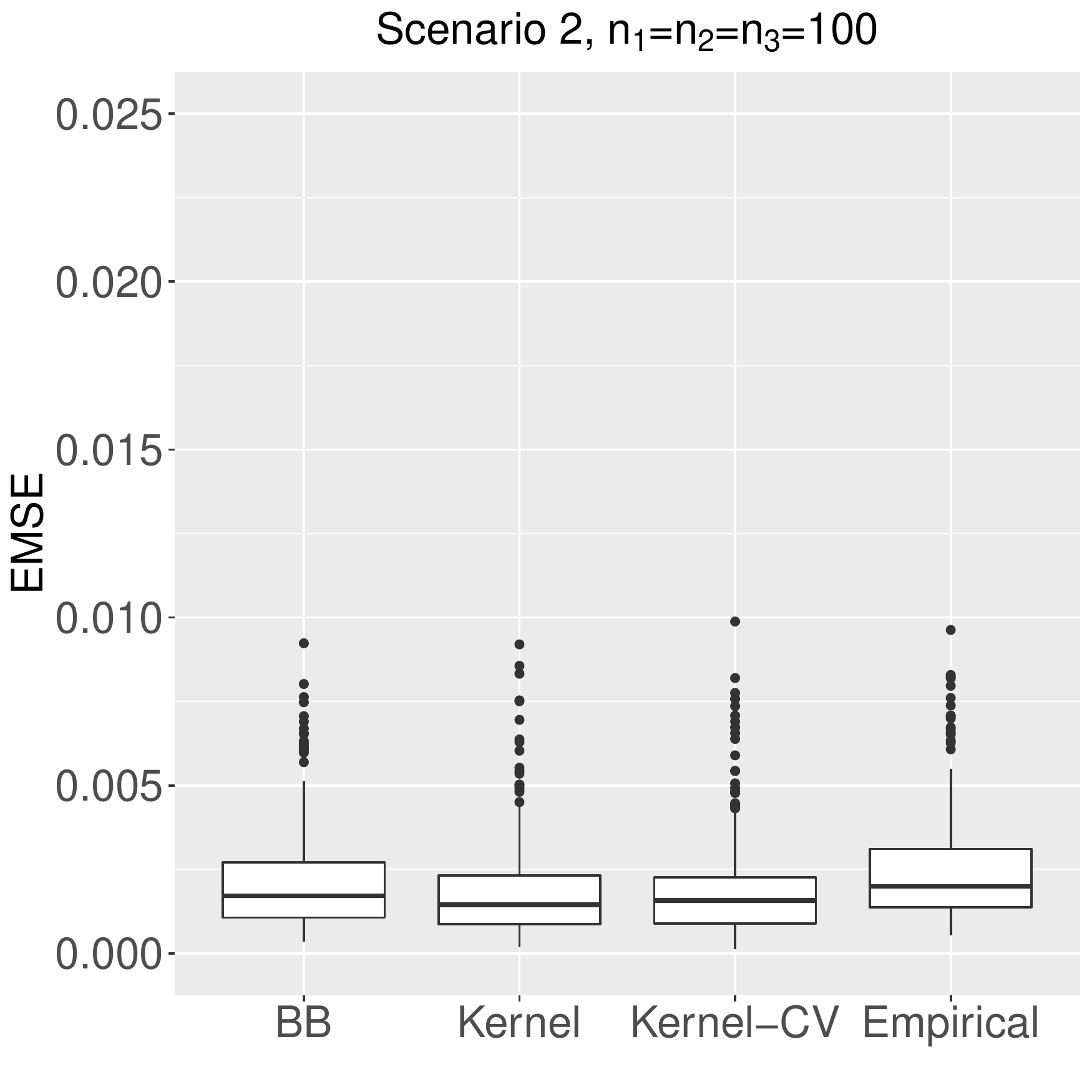}}\hspace{0.3cm}
\subfigure{\includegraphics[page = 1, width=5.75cm]{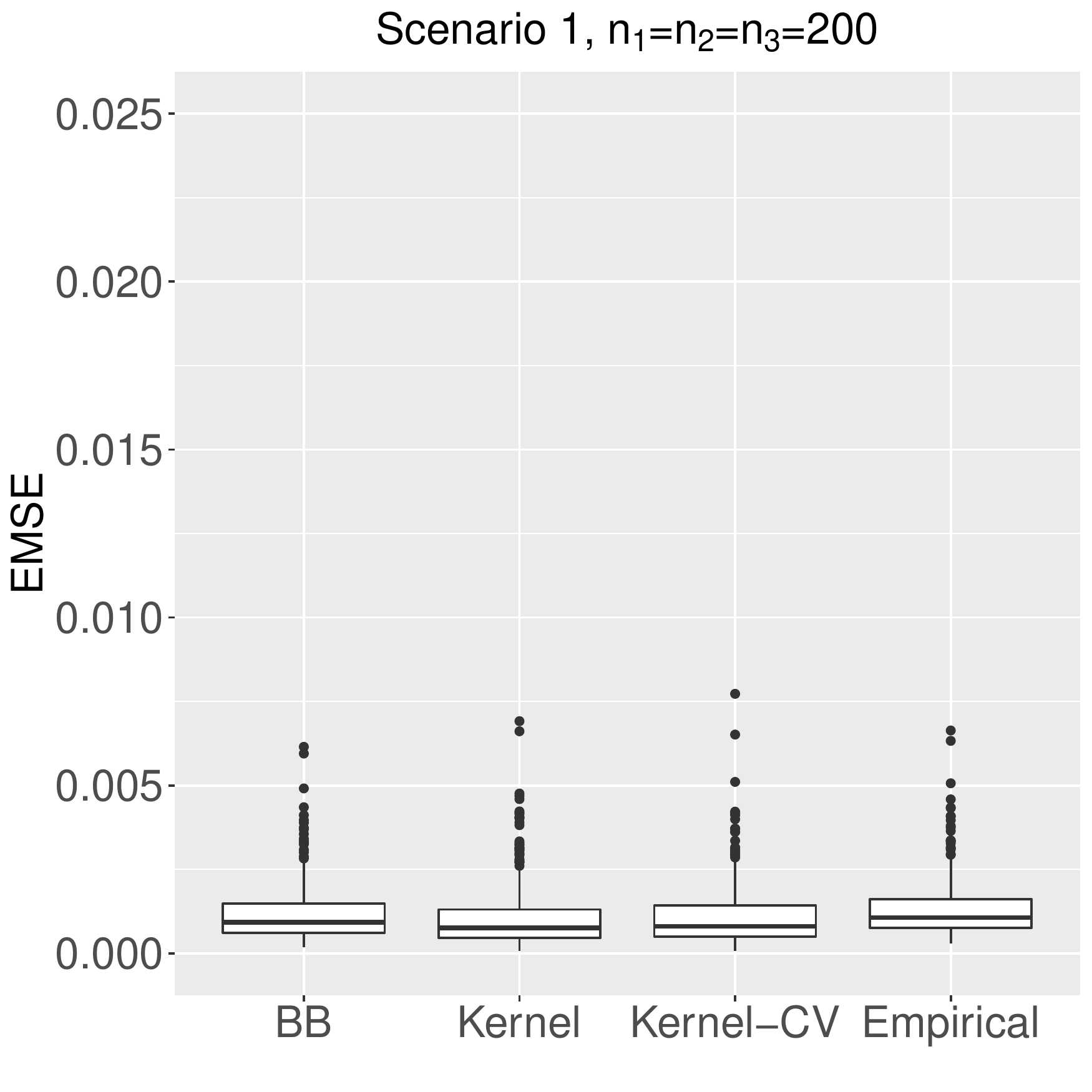}}\\ \vspace{0.3cm}
\subfigure{\includegraphics[page = 1, width=5.75cm]{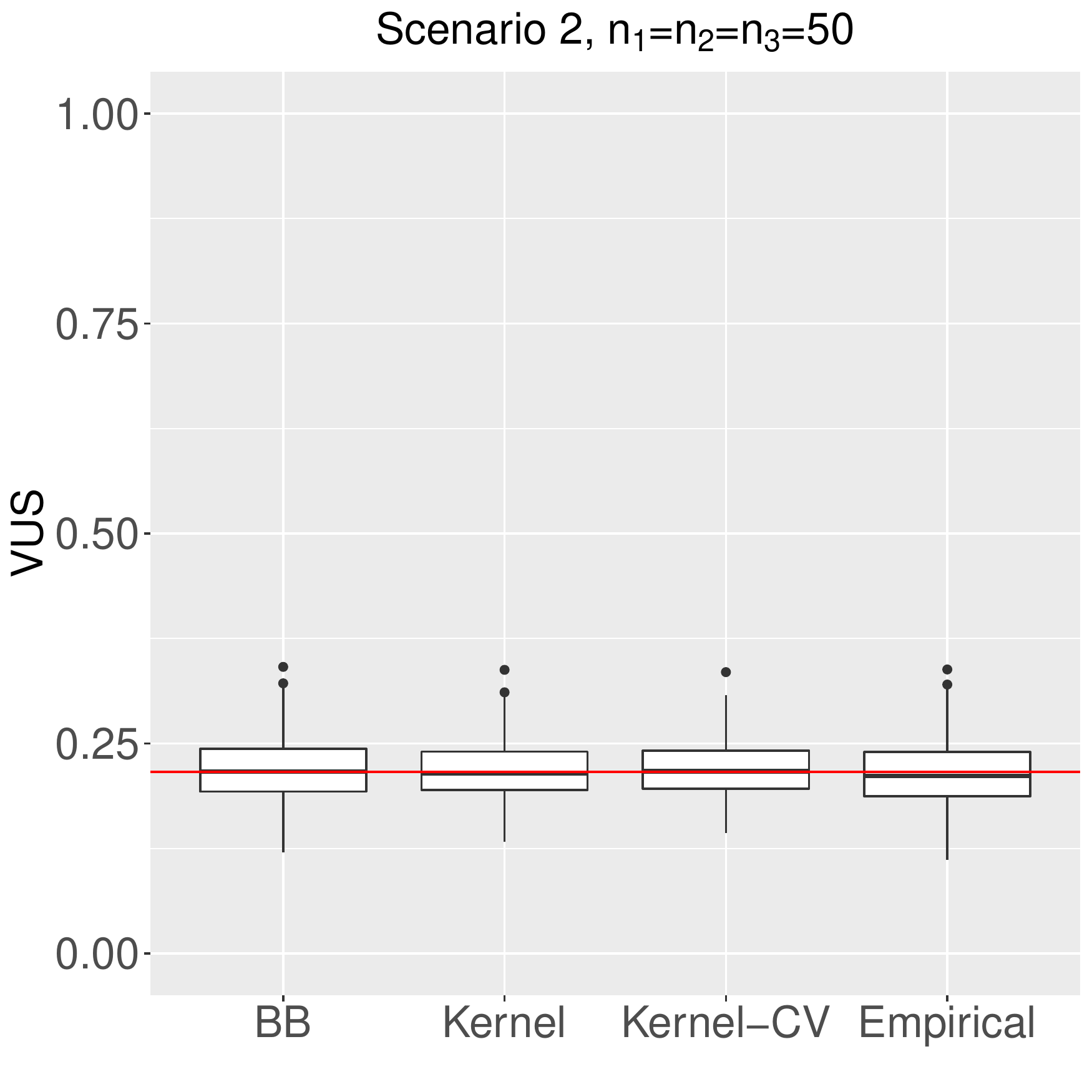}}\hspace{0.3cm}
\subfigure{\includegraphics[page = 1, width=5.75cm]{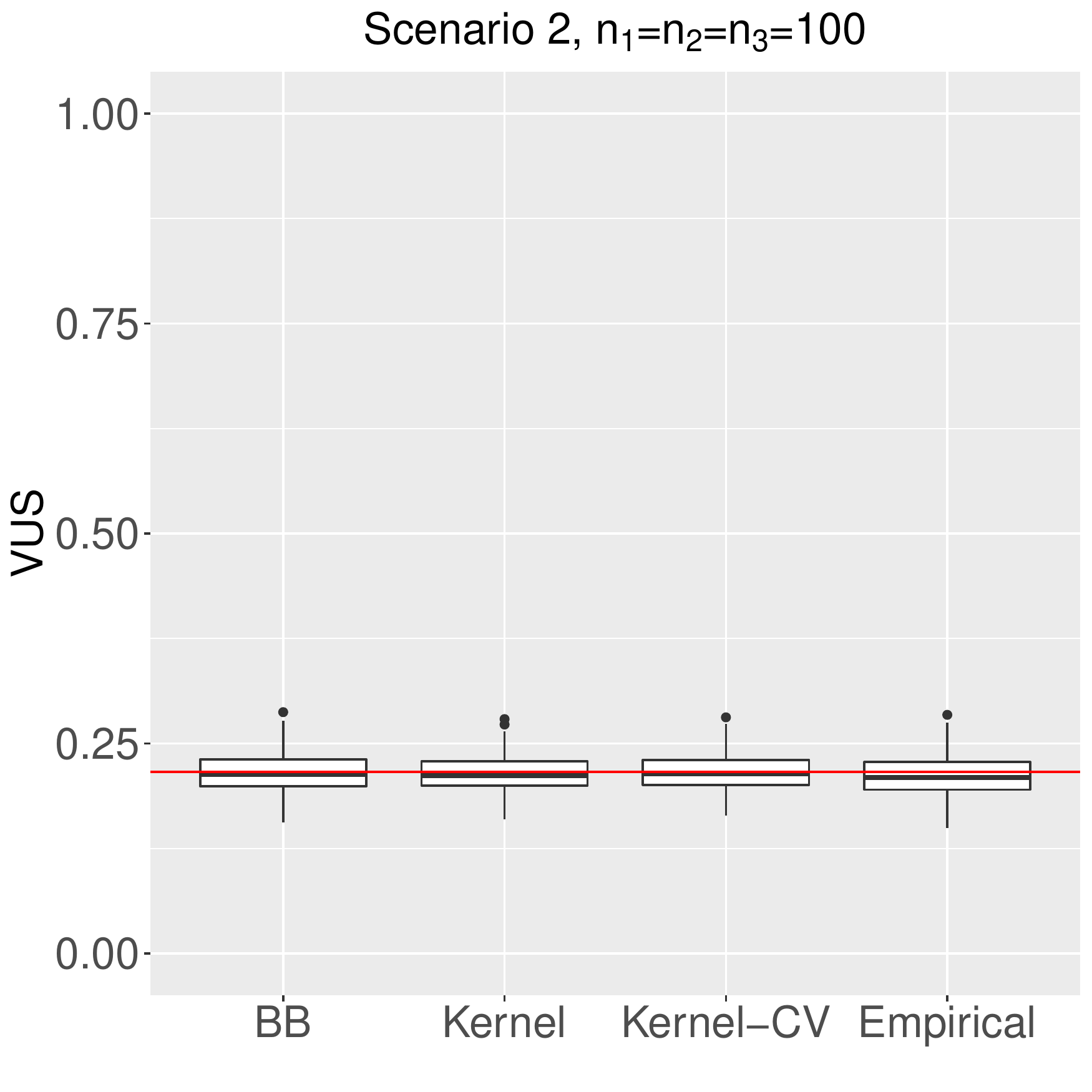}}\hspace{0.3cm}
\subfigure{\includegraphics[page = 1, width=5.75cm]{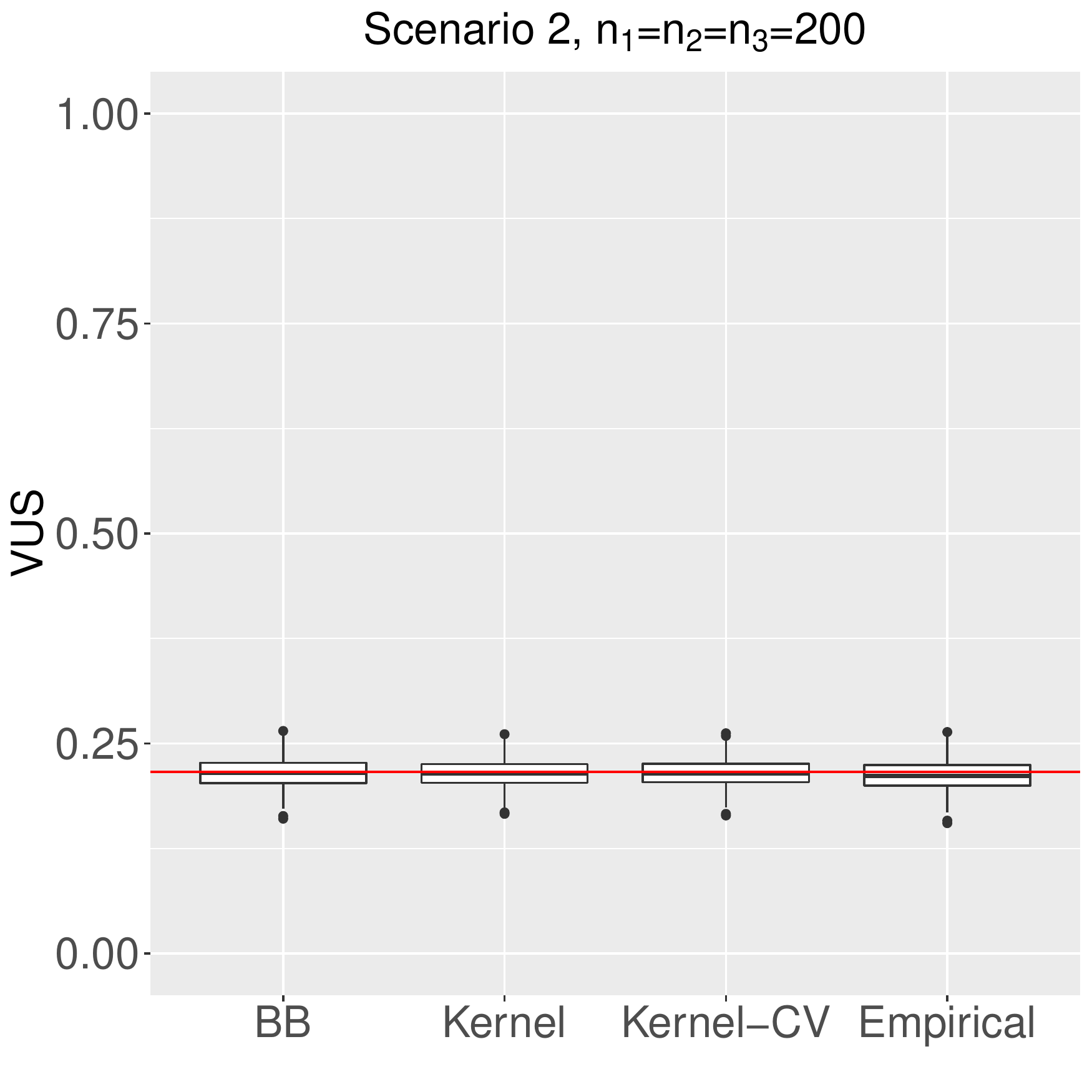}}
\caption{\footnotesize{Scenario 2. Boxplots summarising simulation results for the EMSE (top row) and estimates of the VUS (bottom row). The solid red line corresponds to the true VUS. Here Kernel denotes the kernel estimate with bandwidth calculated using equation \eqref{bwdef} and Kernel-CV stands for the kernel estimate with the bandwidth selected by least squares cross-validation.}}
\label{sc2}
\end{center}
\end{figure}

\begin{figure}[H]
\begin{center}
\subfigure{\includegraphics[page = 1, width=5.75cm]{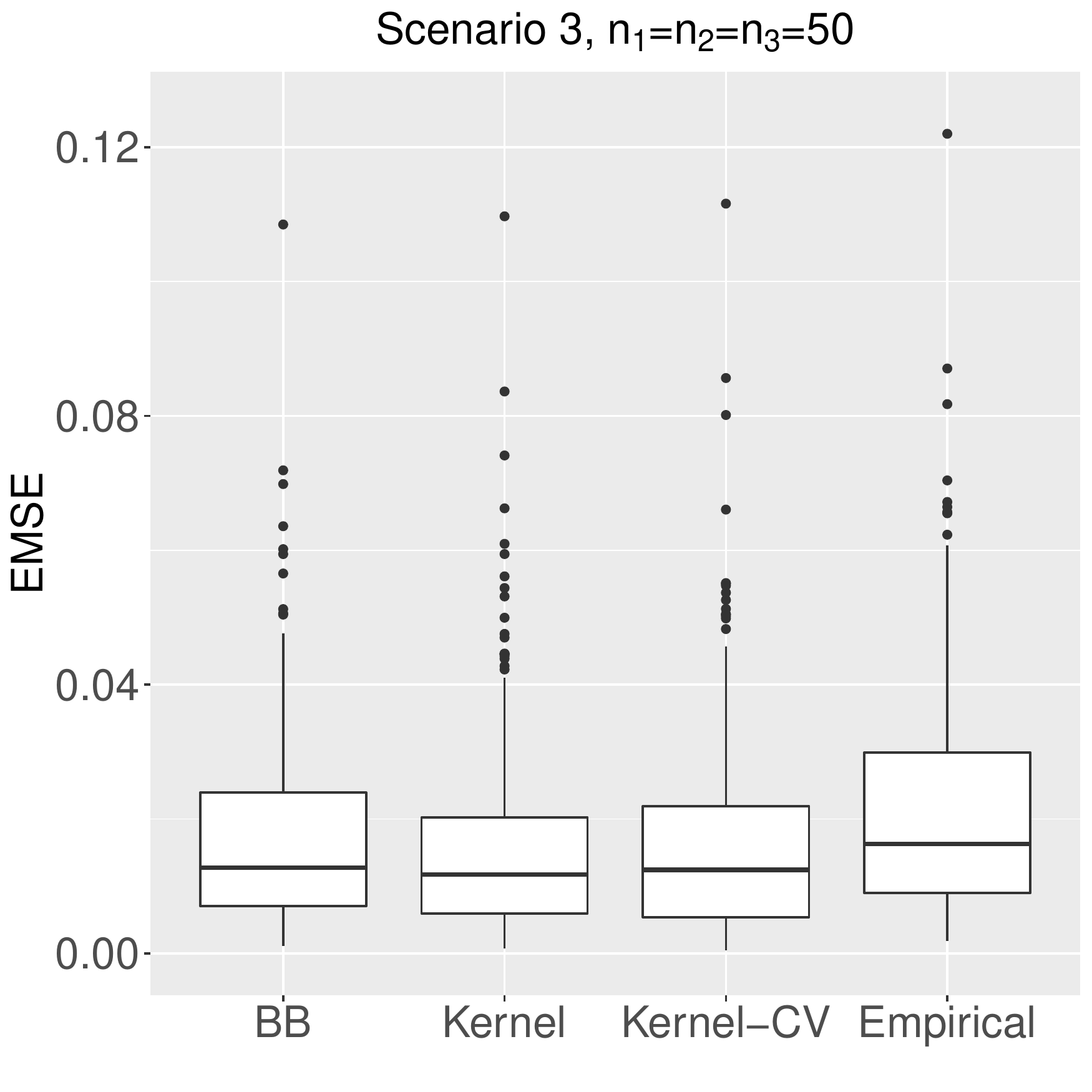}}\hspace{0.3cm}
\subfigure{\includegraphics[page = 1, width=5.75cm]{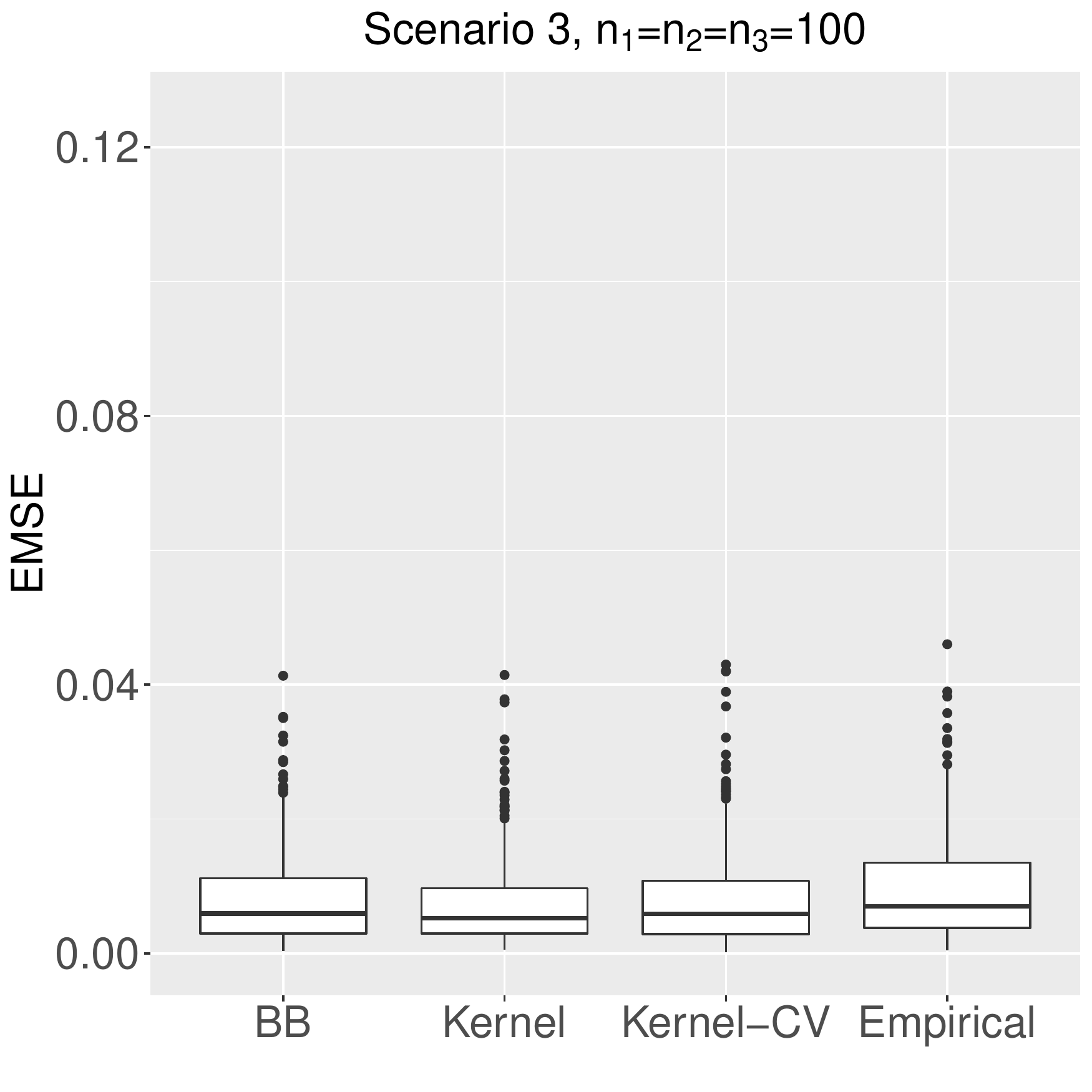}}\hspace{0.3cm}
\subfigure{\includegraphics[page = 1, width=5.75cm]{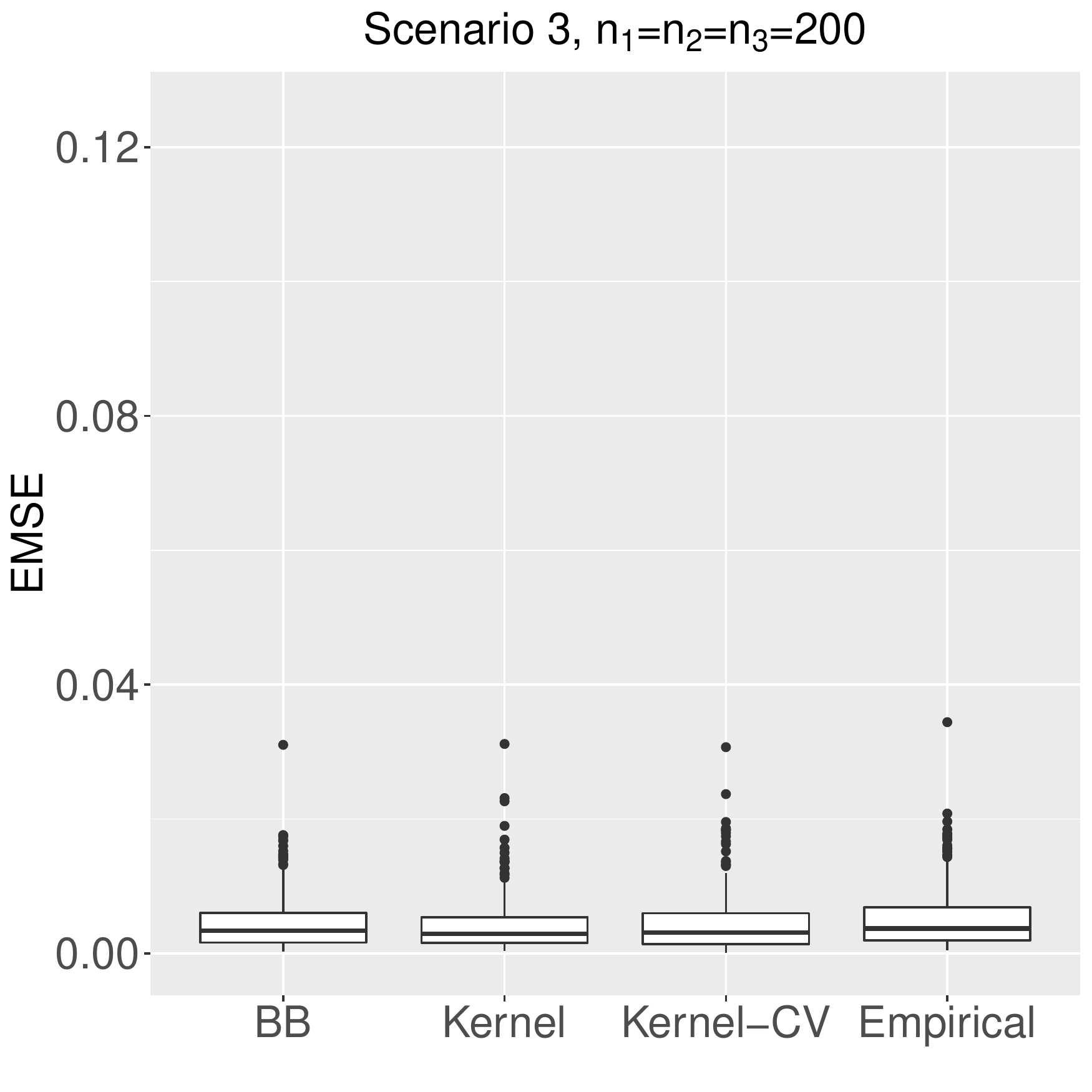}}\\ \vspace{0.3cm}
\subfigure{\includegraphics[page = 1, width=5.75cm]{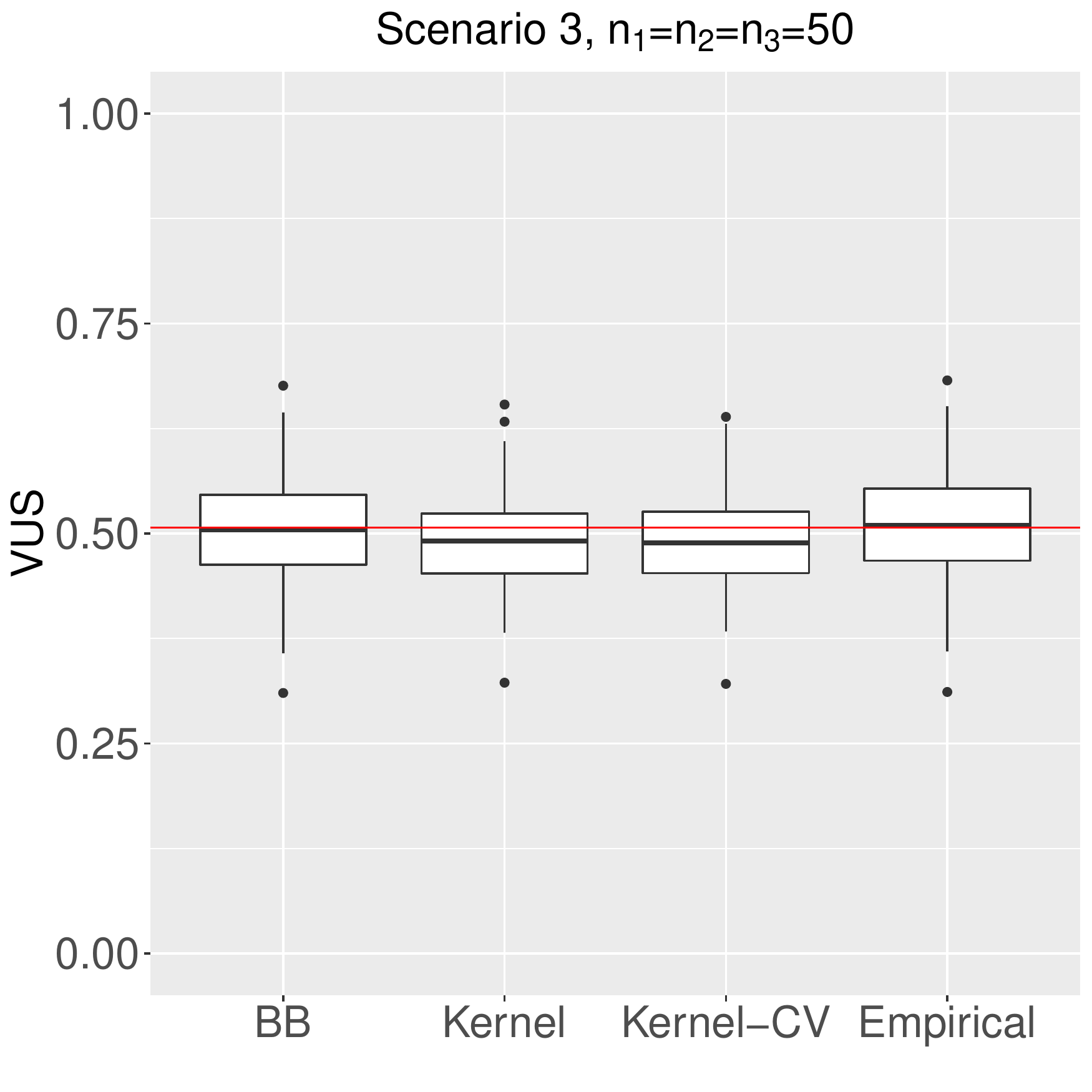}}\hspace{0.3cm}
\subfigure{\includegraphics[page = 1, width=5.75cm]{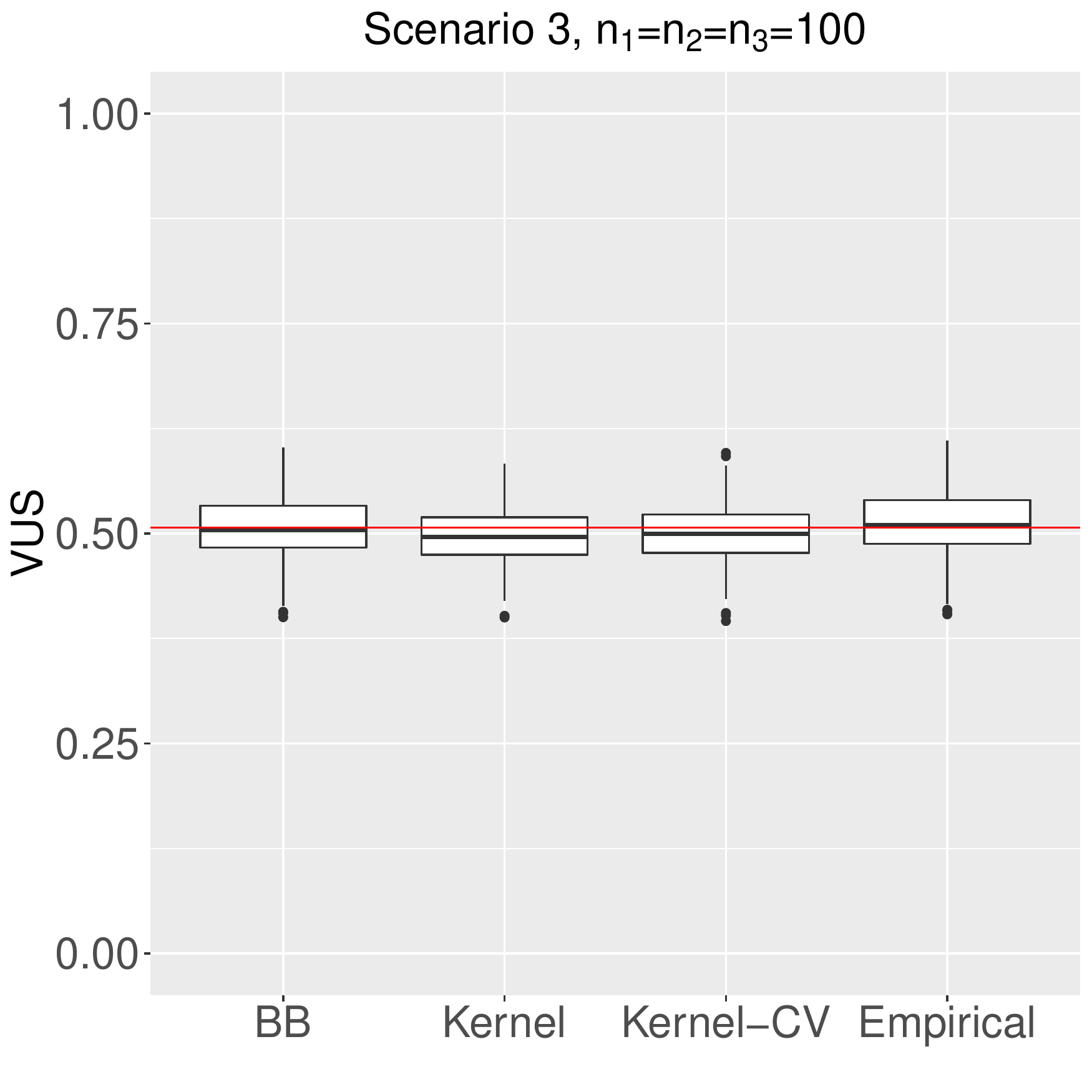}}\hspace{0.3cm}
\subfigure{\includegraphics[page = 1, width=5.75cm]{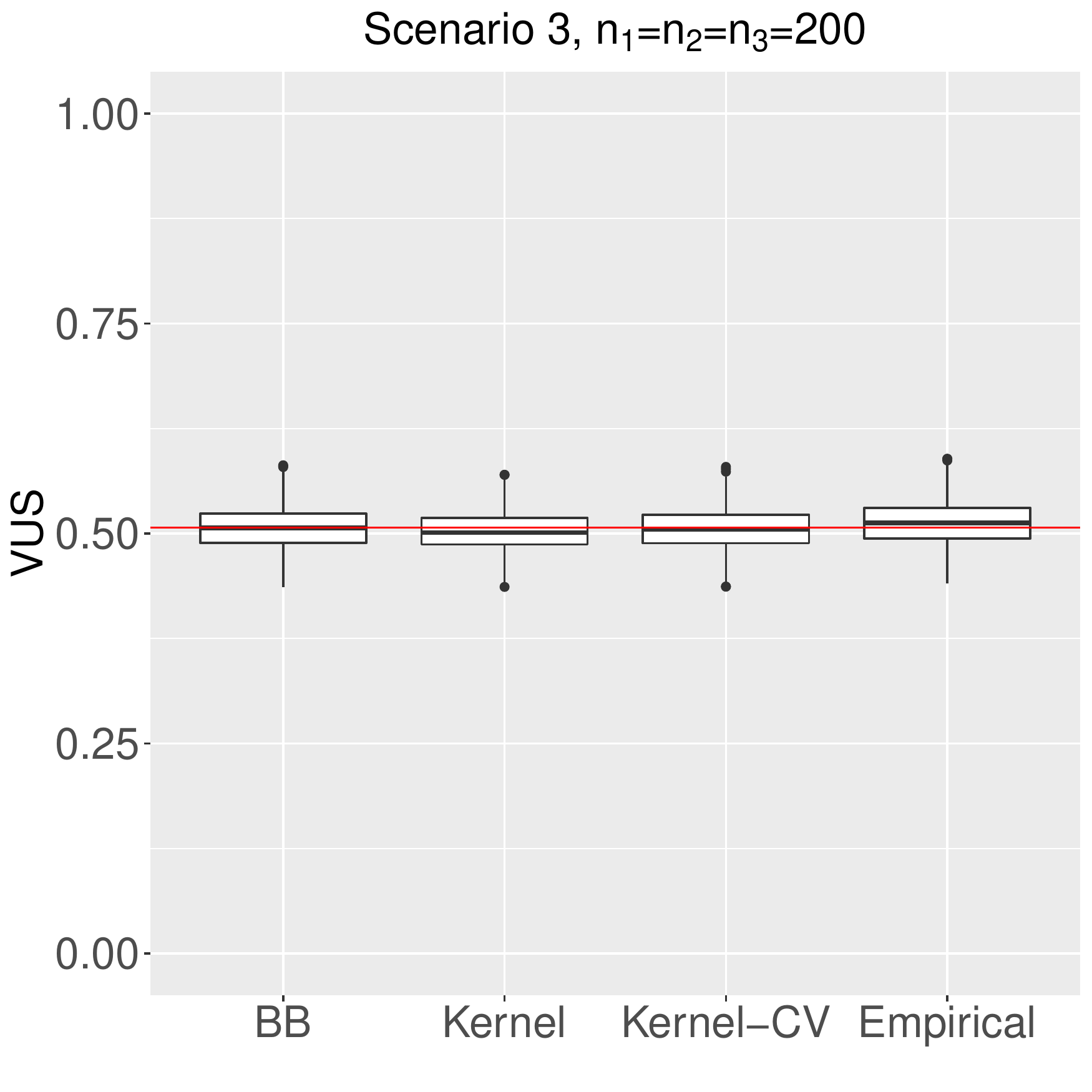}}
\caption{\footnotesize{Scenario 3. Boxplots summarising simulation results for the EMSE (top row) and estimates of the VUS (bottom row). The solid red line corresponds to the true VUS. Here Kernel denotes the kernel estimate with bandwidth calculated using equation \eqref{bwdef} and Kernel-CV stands for the kernel estimate with the bandwidth selected by least squares cross-validation.}}
\label{sc3}
\end{center}
\end{figure}

\begin{figure}[H]
\begin{center}
\subfigure{\includegraphics[page = 1, width=5.75cm]{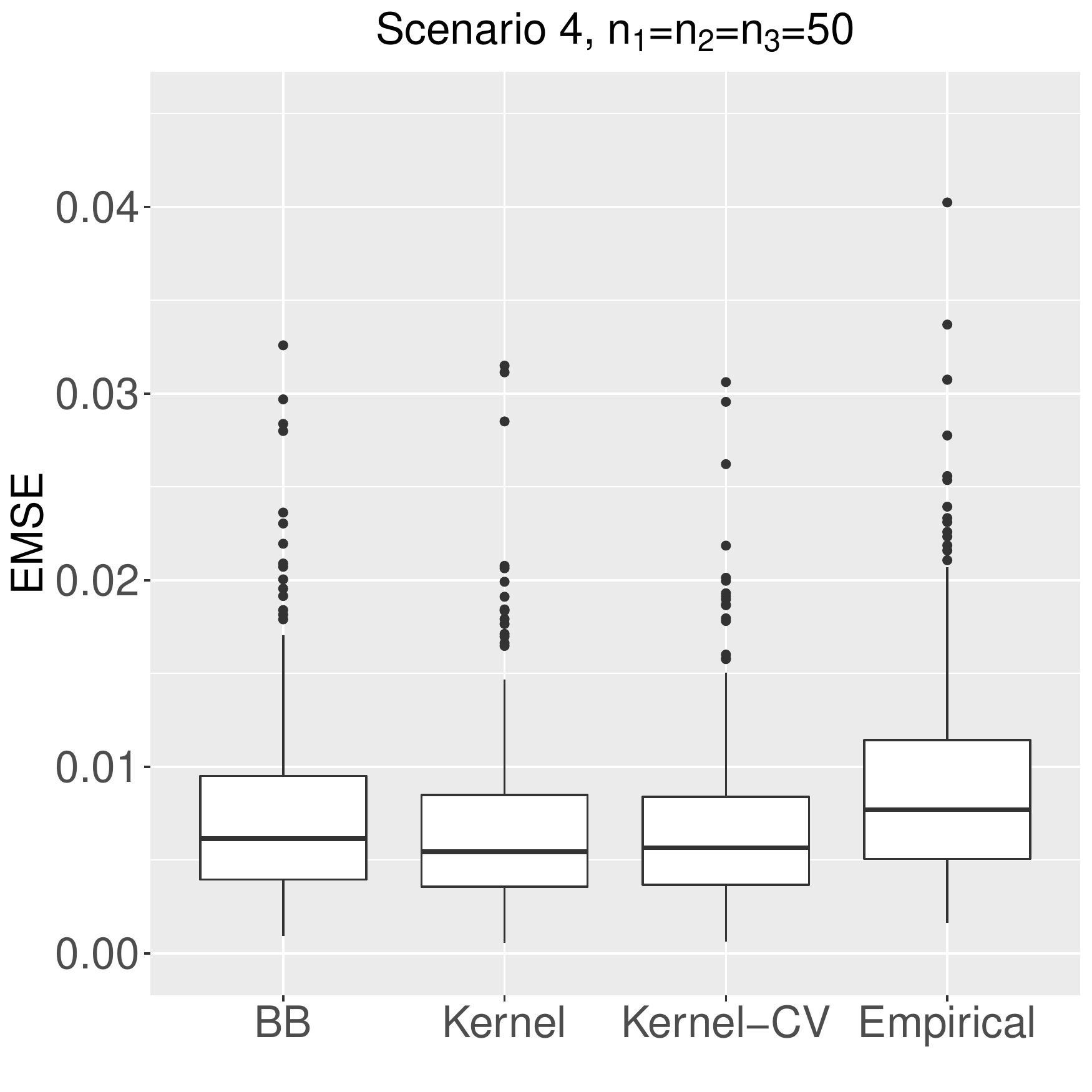}}\hspace{0.3cm}
\subfigure{\includegraphics[page = 1, width=5.75cm]{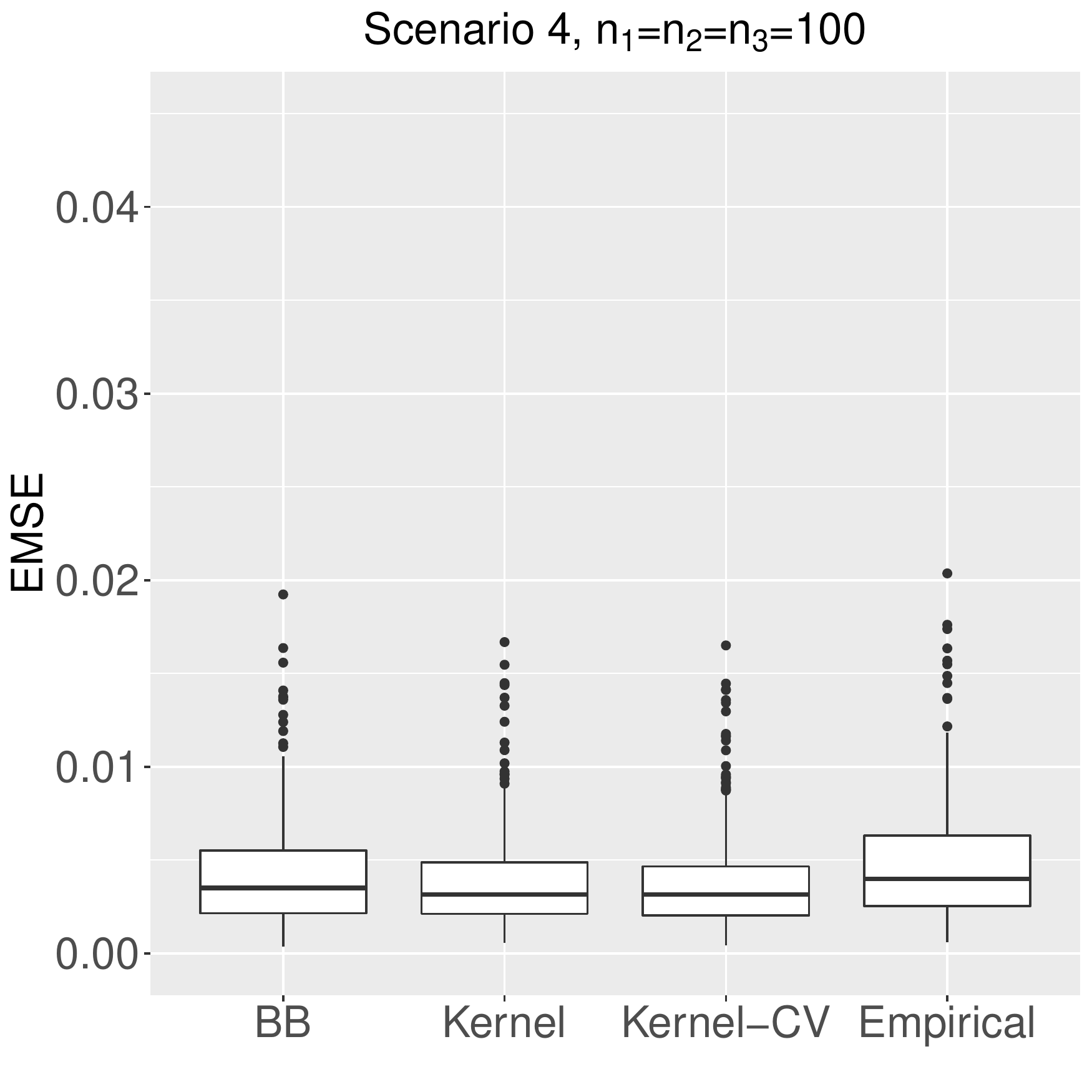}}\hspace{0.3cm}
\subfigure{\includegraphics[page = 1, width=5.75cm]{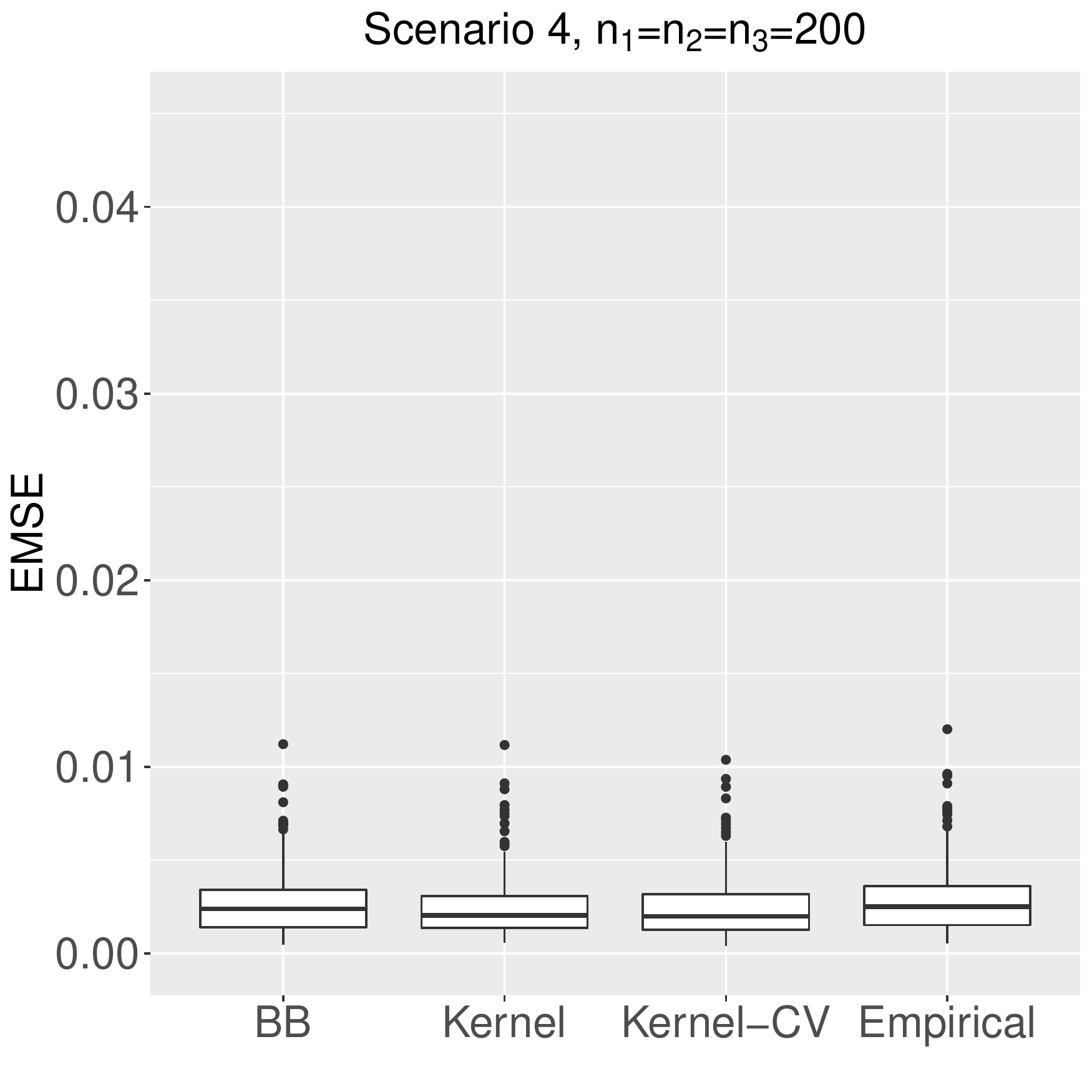}}\\ \vspace{0.3cm}
\subfigure{\includegraphics[page = 1, width=5.75cm]{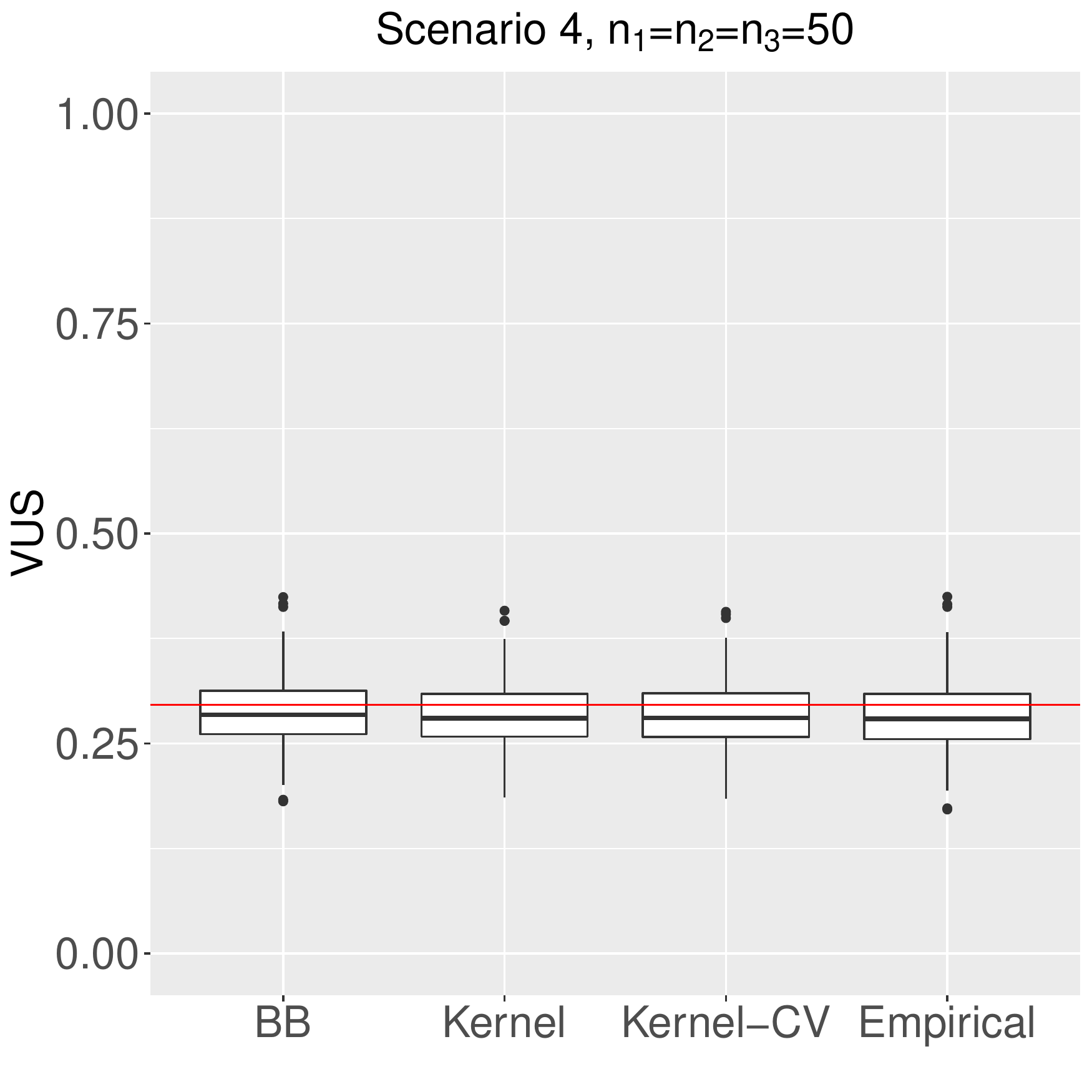}}\hspace{0.3cm}
\subfigure{\includegraphics[page = 1, width=5.75cm]{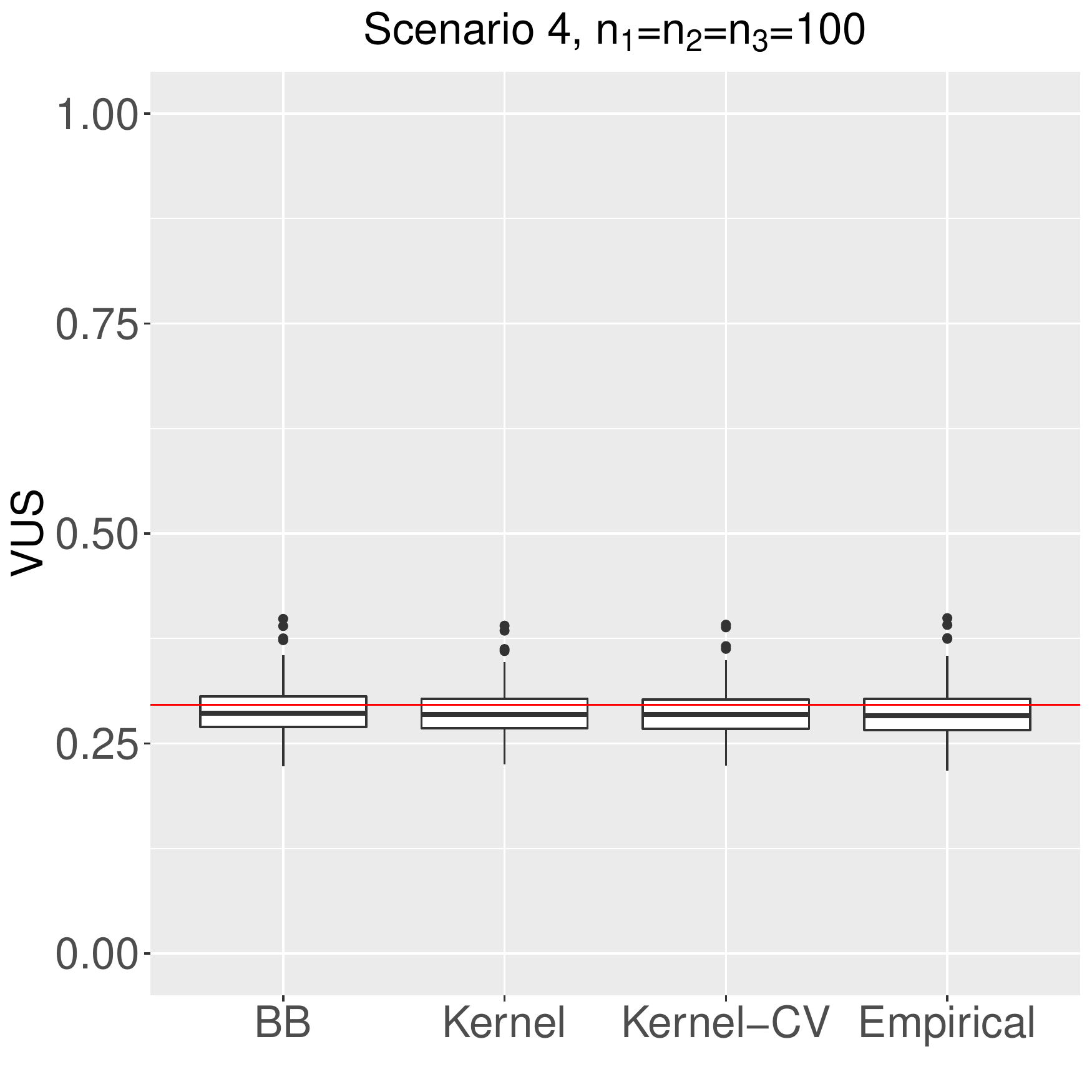}}\hspace{0.3cm}
\subfigure{\includegraphics[page = 1, width=5.75cm]{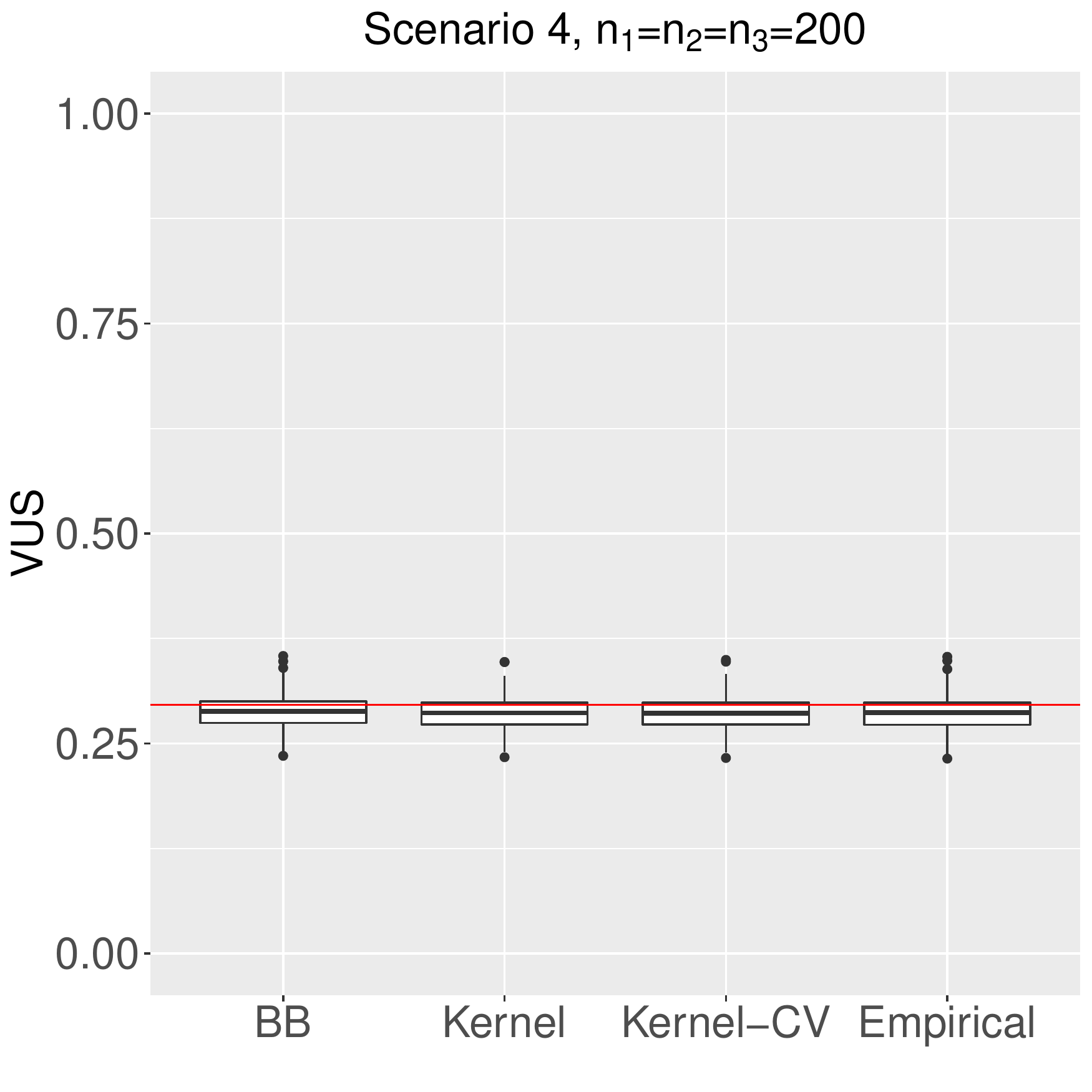}}
\caption{\footnotesize{Scenario 4. Boxplots summarising simulation results for the EMSE (top row) and estimates of the VUS (bottom row). The solid red line corresponds to the true VUS. Here Kernel denotes the kernel estimate with bandwidth calculated using equation \eqref{bwdef} and Kernel-CV stands for the kernel estimate with the bandwidth selected by least squares cross-validation.}}
\label{sc4}
\end{center}
\end{figure}

For each of the four scenarios described in Section \ref{scenarios}, 300 datasets were generated using sample sizes of $(n_1,n_2,n_3)=(50,50,50)$,  $(n_1,n_2,n_3)=(100,100,100)$, and $(n_1,n_2,n_3)=(200,200,200)$.
The discrepancy between the estimated and true ROC surface was measured by the empirical mean squared error
\begin{align*}
\text{EMSE}&=\frac{1}{n_{p_1}}\frac{1}{n_{p_3}}\sum_{u_{1}=1}^{n_{p_1}}\sum_{u_{3}=1}^{n_{p_3}}\left\{\widehat{\text{ROCS}}(p_{1u_{1}},p_{3u_{3}})-\text{ROCS}(p_{1u_{1}},p_{3u_{3}})\right\}^2\\
&\approx\int_{0}^{1}\int_{0}^{1}\left\{\widehat{\text{ROCS}}(p_{1},p_{3})-\text{ROCS}(p_{1},p_{3})\right\}^2,\
\end{align*}
where $n_{p_1}=n_{p_3}=50$. The estimated VUS and the EMSEs for Scenarios 1--4 are presented in Figures \ref{sc1} to \ref{sc4}. Specifically, for each scenario and sample size considered, we present a boxplot of the VUS estimates (along with the true value) and EMSEs produced by each method. In addition, the estimated ROC surfaces, along with the true surfaces, are shown in Figures 2--5 of the Supplementary Materials. In Scenario 1, we can appreciate a minor loss in efficiency of our BB estimator, which is a small price to pay for the benefit of the robustness that leads to accurate data driven estimates under increasingly complex scenarios (Figures \ref{sc2} to \ref{sc4}). The BB estimator outperformed, in terms of the empirical mean squared error, the empirical estimator for most of the scenarios, especially for the sample size $(n_1,n_2,n_3)=(50,50,50)$. The BB estimator was on par with the kernel estimator, as measured both in terms of the EMSE and the computational time, with the additional advantage of not needing to select a smoothing parameter and of providing simultaneously both point and interval estimates. As expected, uncertainty associated with our BB estimator decreased as the sample size increased. Lastly, the frequentist coverage of the $95\%$ credible intervals for the VUS are presented in Table \ref{t2}. We found the coverages to be between $0.95$ and $1$, which shows the validity of our inferences.

\section{\large{\textsf{APPLICATION}}}\label{application}
\begin{figure}[H]
\begin{center}
\subfigure{\includegraphics[page = 1, width=5.85cm]{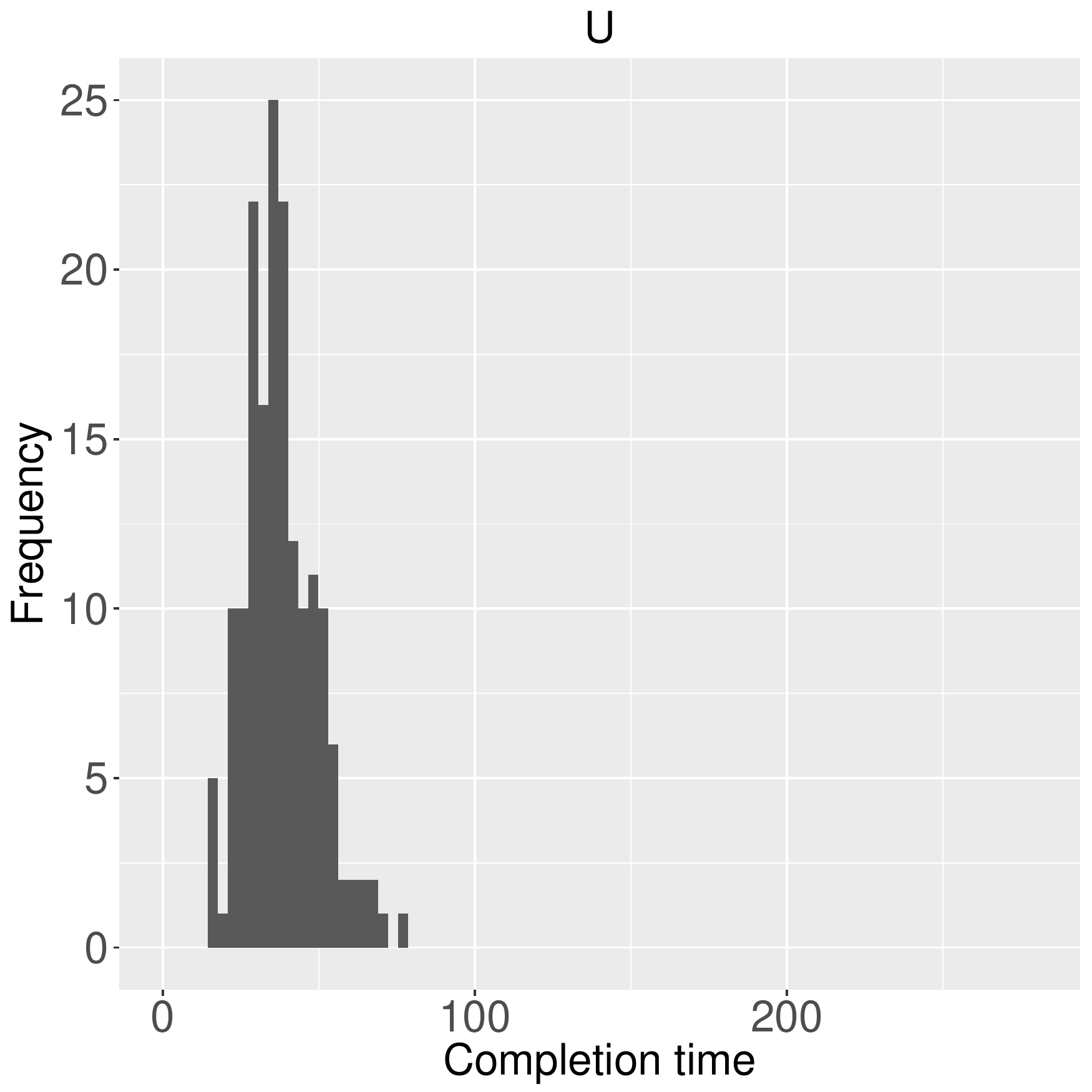}}\hspace{0.4cm}
\subfigure{\includegraphics[page = 1, width=5.85cm]{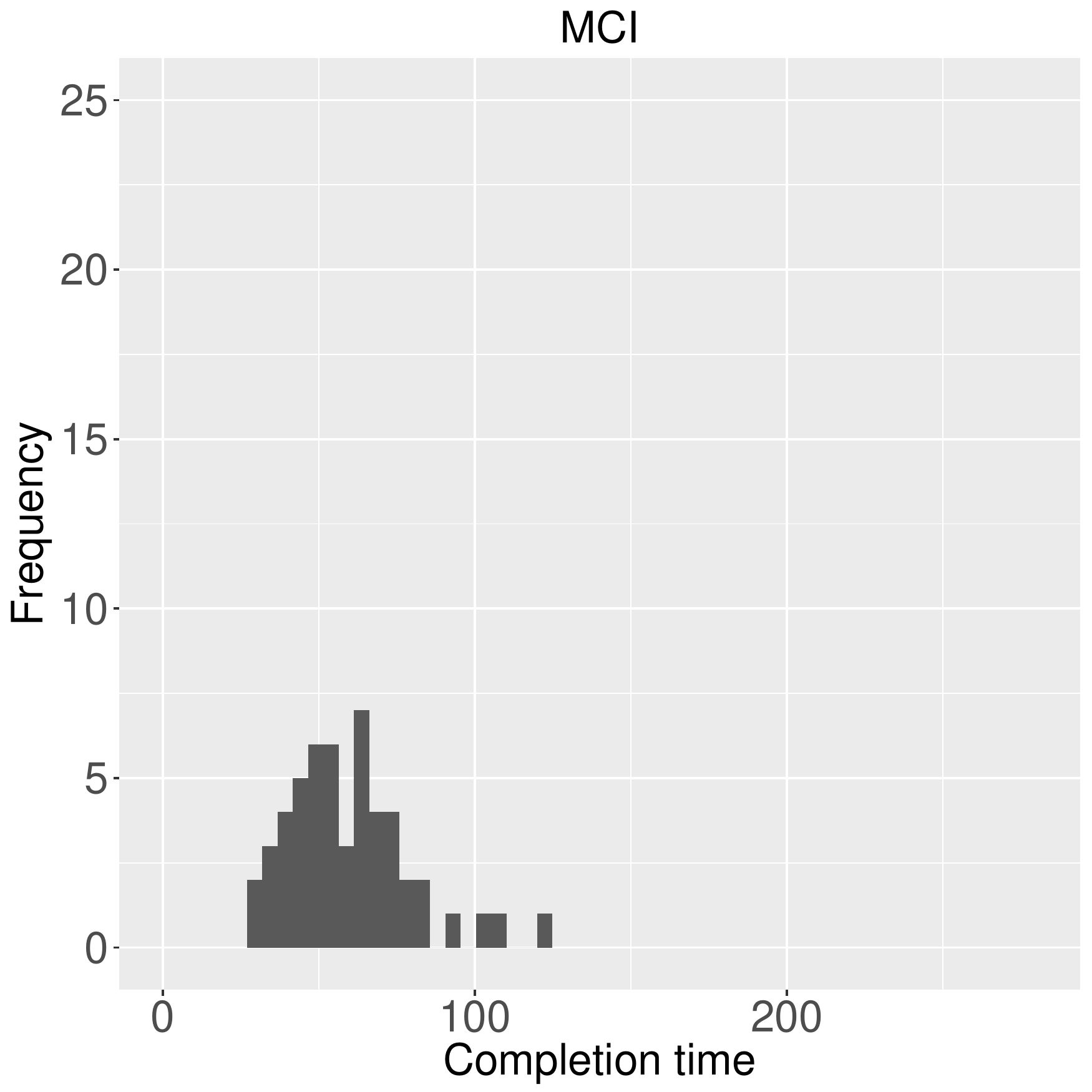}}\\
\subfigure{\includegraphics[page = 1, width=5.85cm]{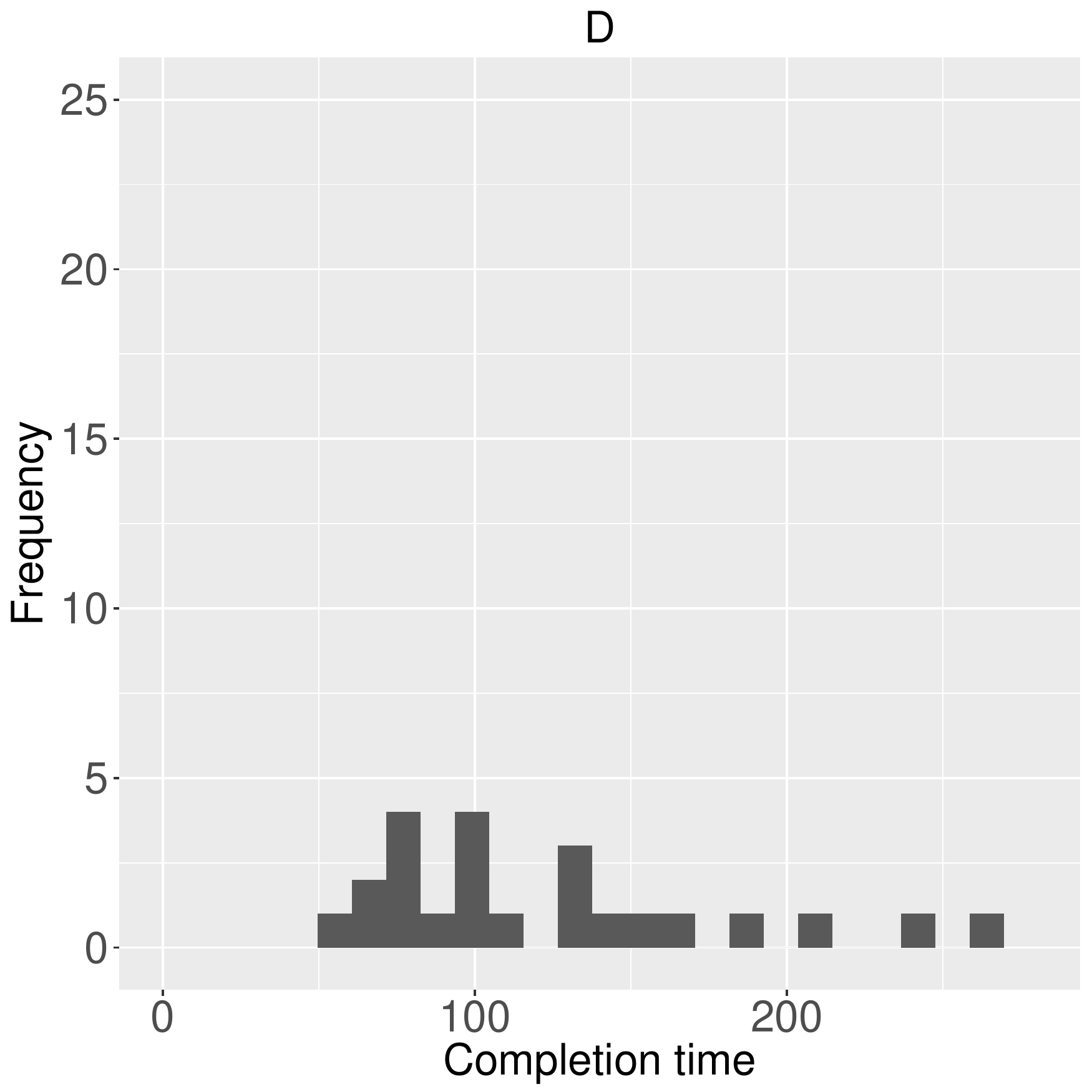}}\hspace{0.4cm}
\subfigure{\includegraphics[page = 1, width=5.85cm]{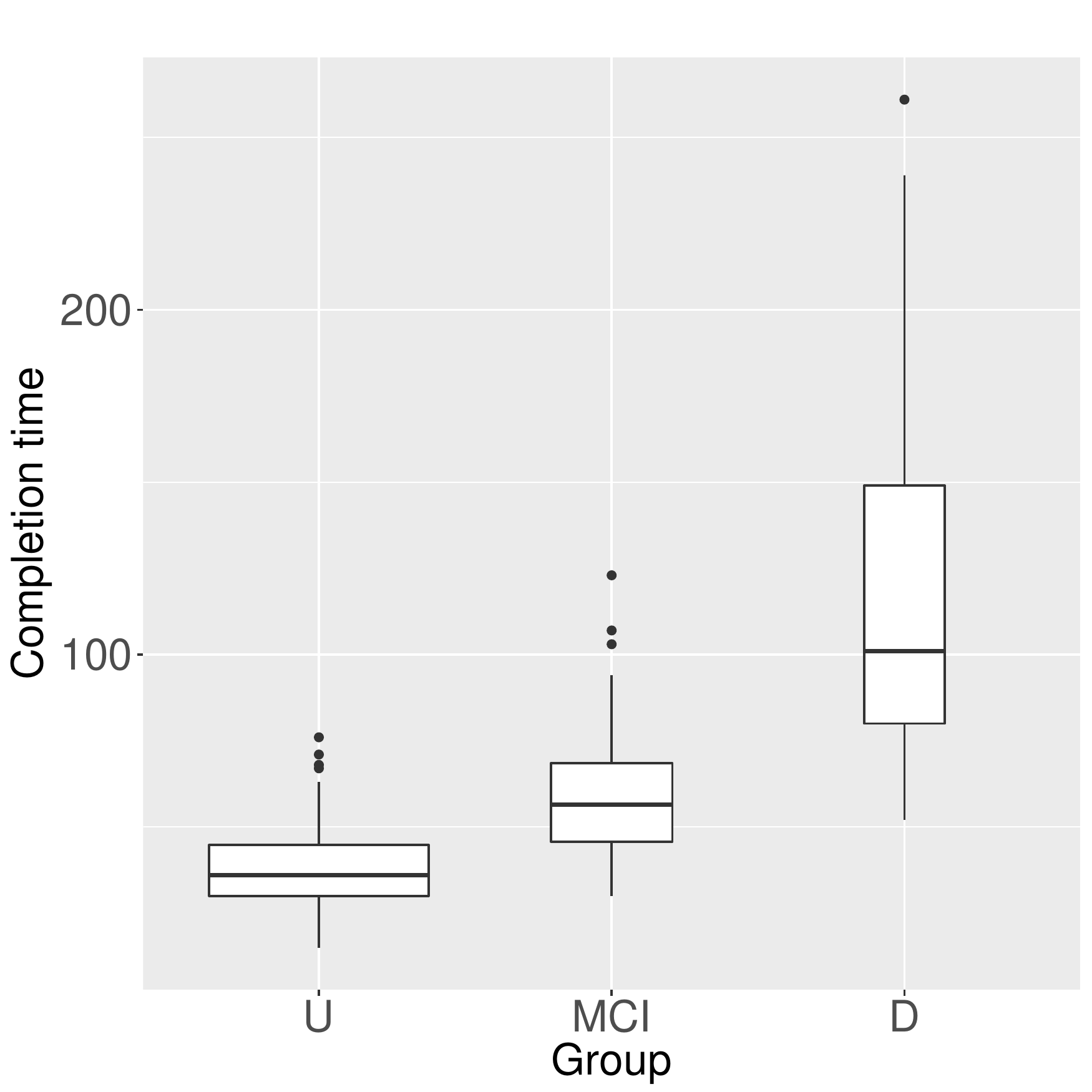}}
\caption{\footnotesize{Trail Making Test Part A data: histograms and variable-width boxplots of the completion times (in seconds) in each group.}}
\label{hists}
\end{center}
\end{figure}

\begin{figure}[H]
\begin{center}
\hspace{0.6cm}
\subfigure[]{\includegraphics[page = 1, width=6.5cm]{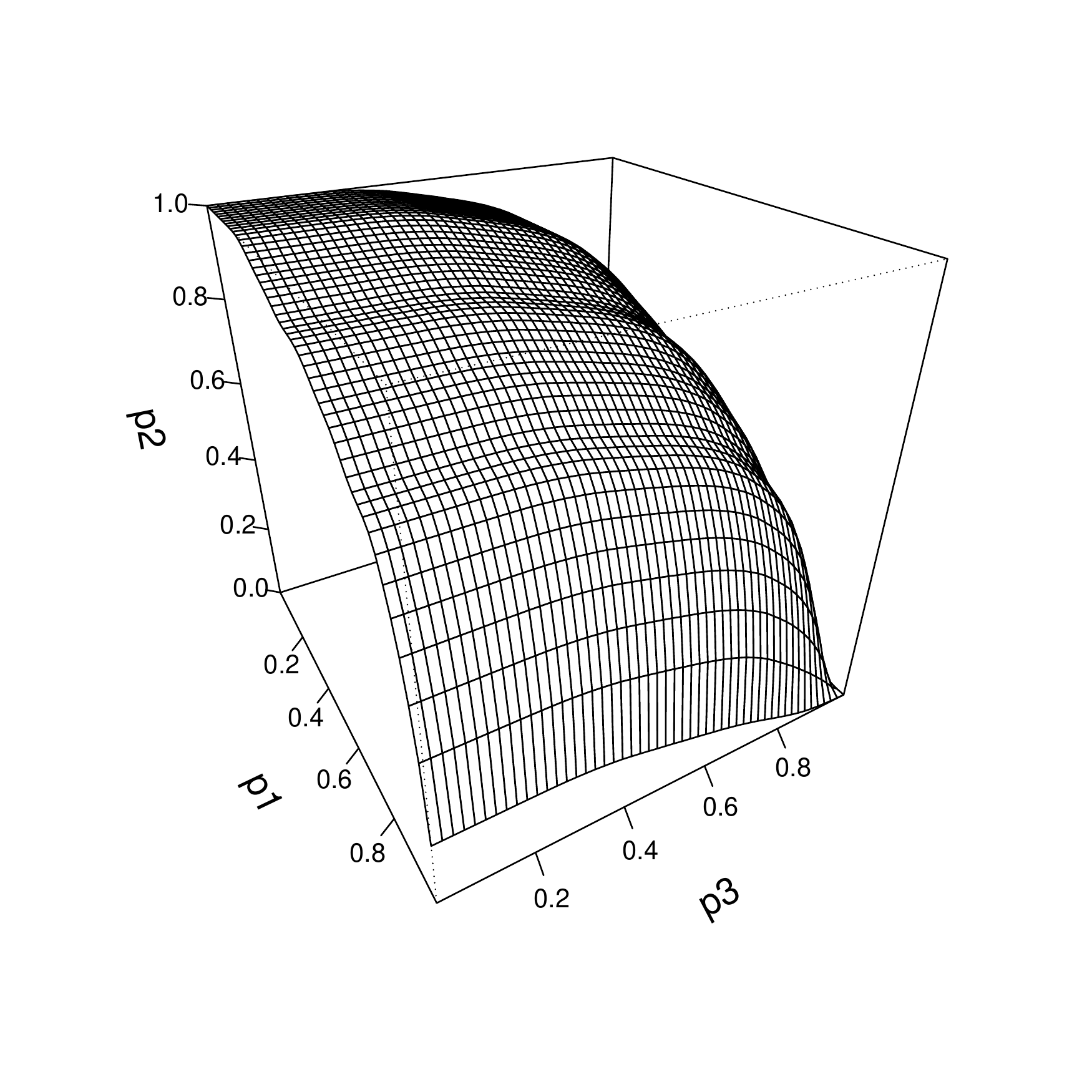}}\hspace{0.6cm}
\subfigure[]{\includegraphics[page = 1, width=5.85cm]{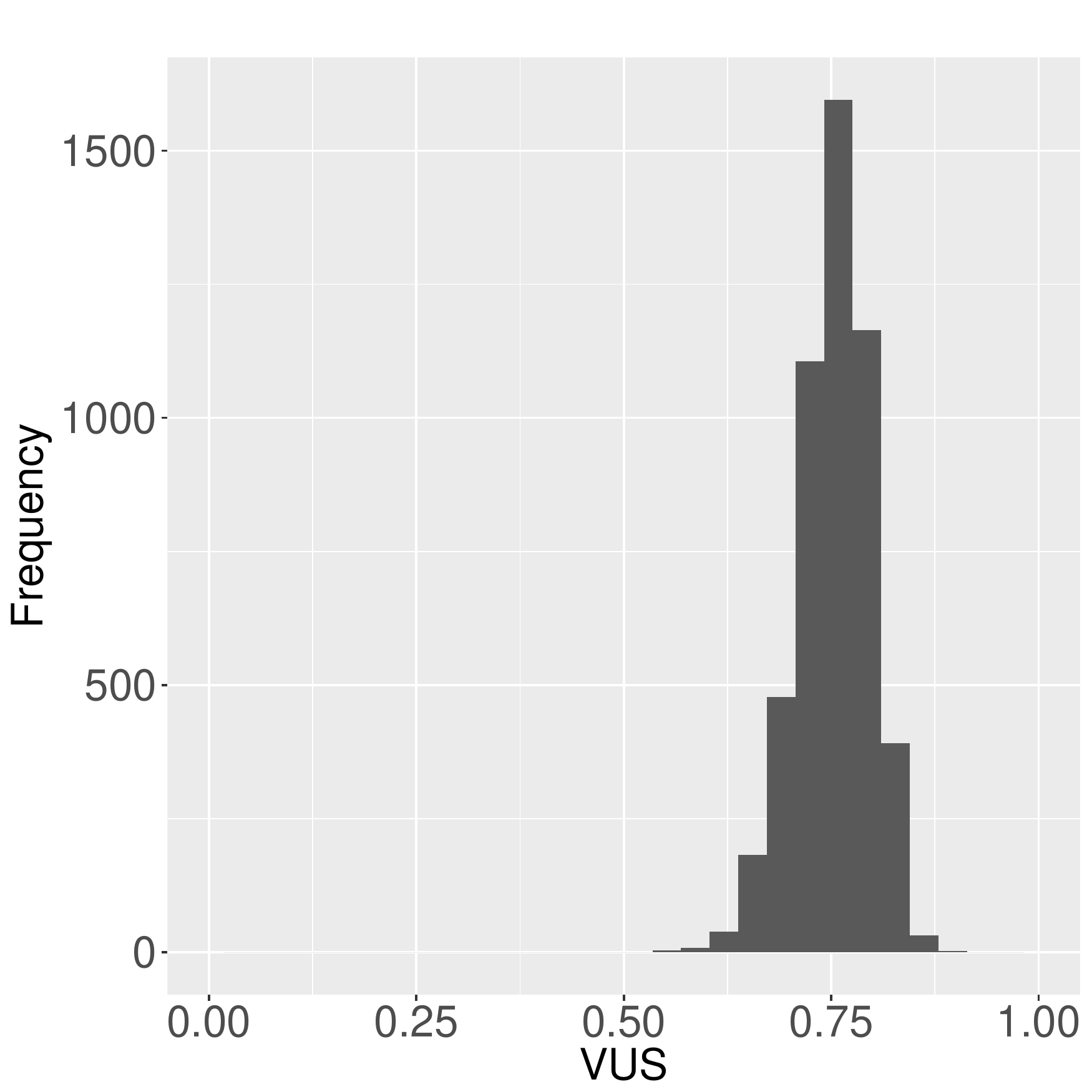}}\\
\subfigure[]{\includegraphics[page = 1, width=5.65cm]{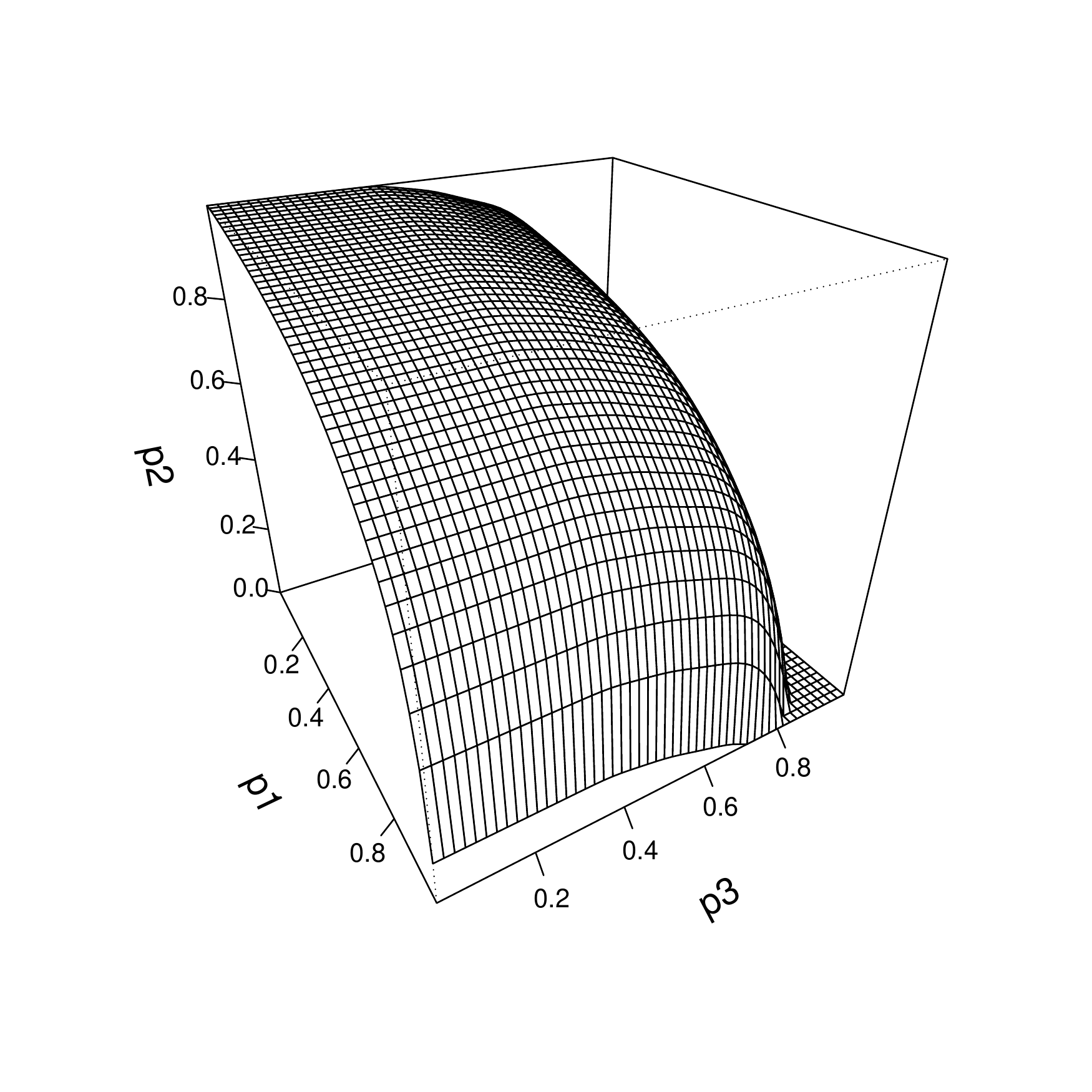}}\hspace{0.2cm}
\subfigure[]{\includegraphics[page = 1, width=5.65cm]{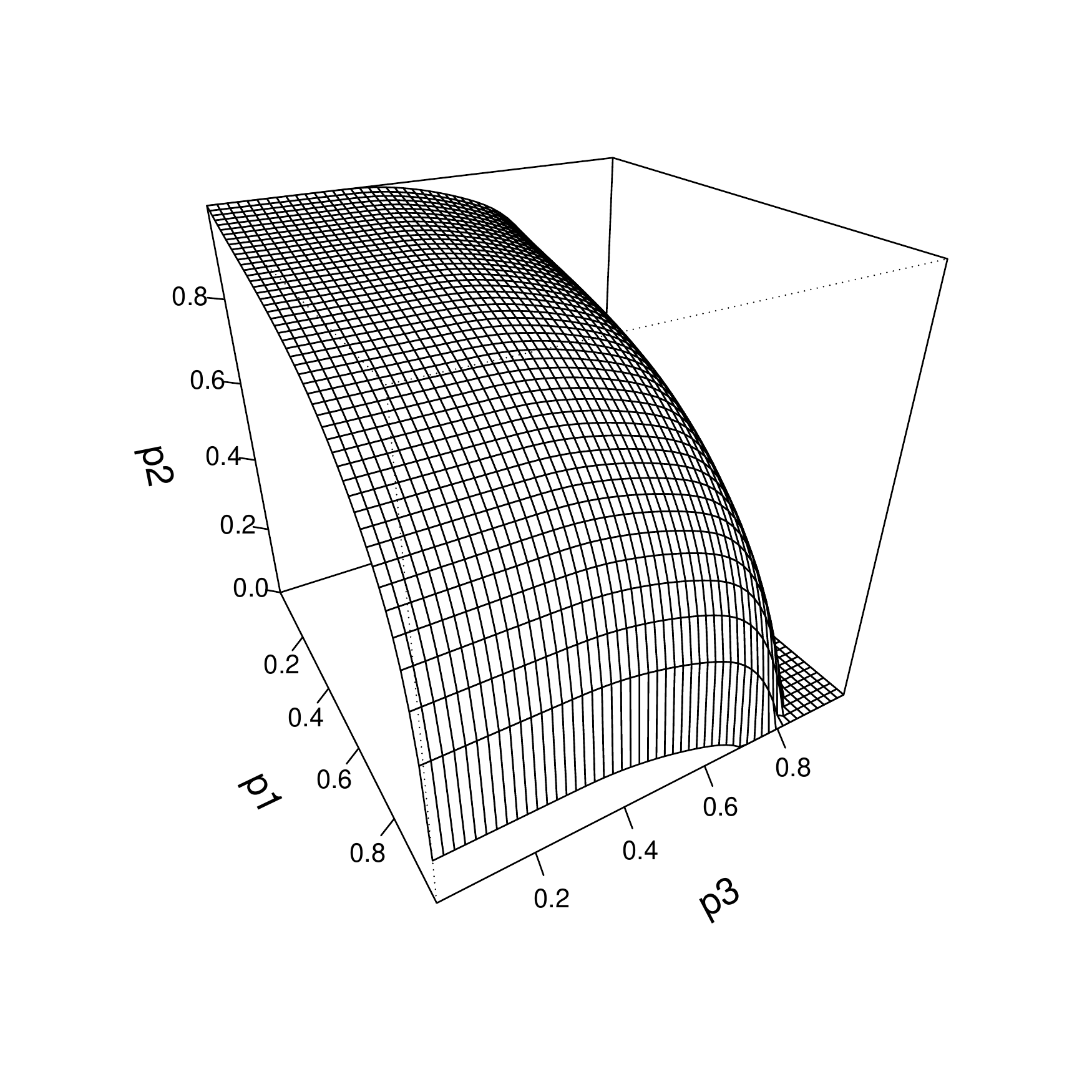}}\hspace{0.2cm}
\subfigure[]{\includegraphics[page = 1, width=5.65cm]{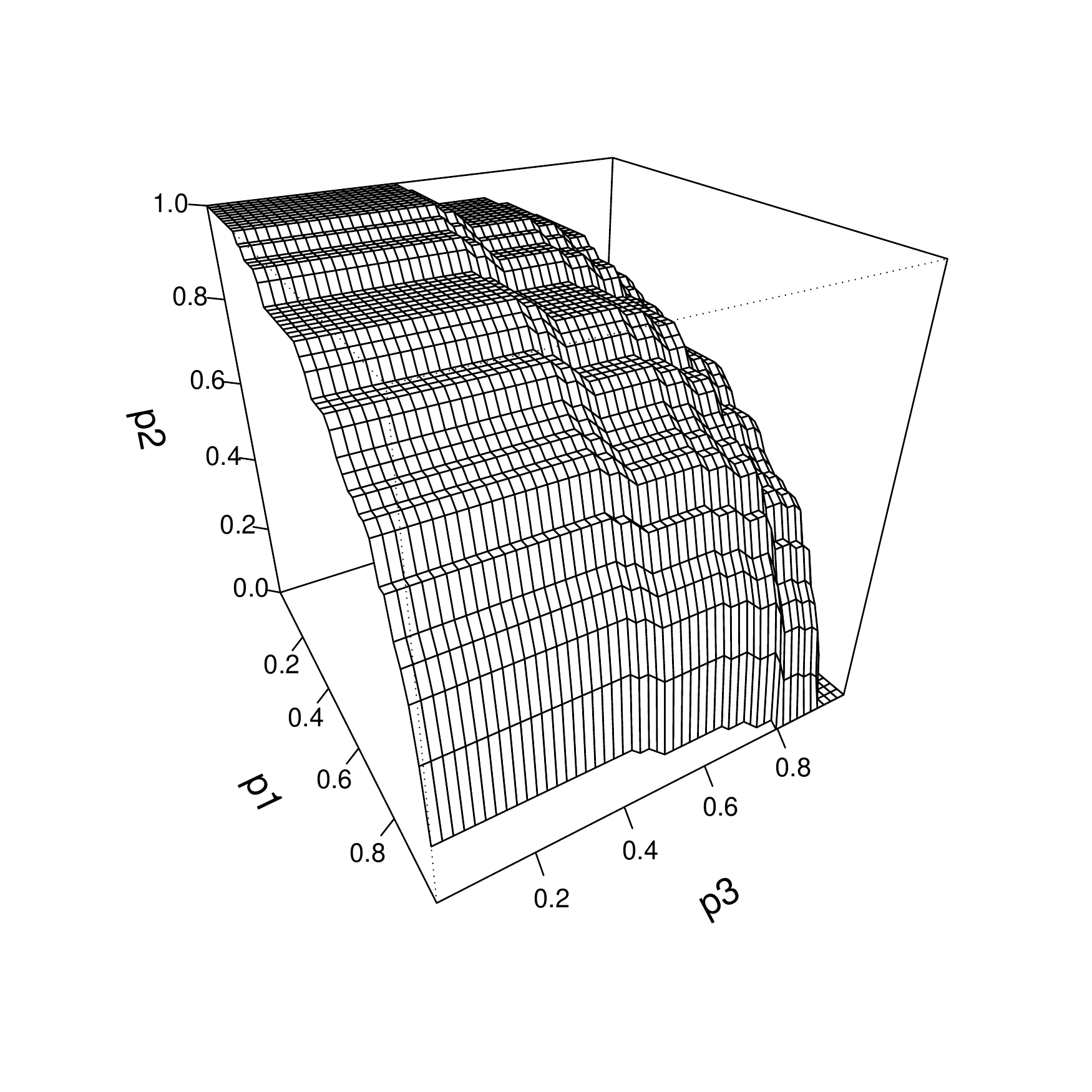}}
\caption{\footnotesize{ Trail Making Test Part A data. Top row: BB estimate of the ROC surface (a) and histogram of the 5000 BB sampled VUS (b). Bottom row: estimated ROC surfaces using the kernel approach with bandwidths computed as in \eqref{bwdef} (c), kernel approach with bandwidths selected by cross-validation (d), and the empirical approach (e).}}
\label{surfaces}
\end{center}
\end{figure}

The Trail Making Test (TMT) is a neuropsychological test that provides information about visual search speed, scanning, speed of processing, as well as, executive functioning. The TMT test is commonly used as a diagnostic test of cognitive impairment associated with dementia. The TMT comprises two parts, both consisting of 25 circles distributed over a sheet of paper or on a computer screen. In Part A, the circles are numbered from 1 to 25 and patients are tasked with connecting them in a sequential order (1--2--3, etc). In part B, the patient alternates between numbers and letters (1--A--2--B, etc). The goal is to finish both parts of the test as quickly as possible, and completion times are used as the primary performance metrics. While Part A is primarily used to assess cognitive processing speed, Part B is used to examine executive functioning.\\
\indent We analysed TMT Part A completion times for 245 patients with Parkinson's disease \citep{Bantis2017}. Based on a battery of tests for characterising cognitive impairment, $170$ patients were diagnosed as unimpaired (U), 52 patients were diagnosed as having mild cognitive impairment (MCI), and 23 patients were diagnosed as having dementia (D). Parkinson's disease patients who have dementia were expected to have slower completion times than those with MCI, and patients with MCI were expected to have slower completion times than those with no cognitive impairment, that is, the anticipated ordering of completion times is $\text{U}<\text{MCI}<\text{D}$. Figure \ref{hists} shows histograms and boxplots of the completion times for each group. We can observe a very reasonable separation between completion times in the three groups, with an almost non-existing overlap between completion times in the unimpaired and dementia group.\\
\indent We applied our BB methodology to the TMT Part A data. We used $5000$ iterations and, as in Section \ref{simulation}, $p_1$ and $p_3$ lie on a grid of $50$ even points on $[0,1]$. The BB estimates of the cumulative distribution function in each group along with $95\%$ pointwise credible bands presented in Figure 6 of the Supplementary Materials show, consistent with the histograms and boxplots of the completion times, considerable separation between the distributions of completion times in the unimpaired and mild cognitive impairment groups compared to the severe dementia group. The BB estimate of the ROC surface has the appealing feature of being smooth (without the need for specifying a smoothing parameter), therefore allowing for useful interpretation of diagnostic performance at all threshold values (Figure \ref{surfaces}a). Figure \ref{surfaces}b presents a histogram of the $5000$ sampled VUS values; the BB estimate ($95\%$ credible interval) is $0.75$ $(0.65,0.83)$, which indicates that completion time on Part A of the TMT accurately discerns between U, MCI, and D in Parkinson disease patients.\\
\indent We also applied the kernel and empirical estimators in the same manner as described in Section \ref{simulation}. Confidence intervals for the VUS were obtained through a nonparametric bootstrap consisting of $1000$ resamples. The estimated surfaces are shown in Figure \ref{surfaces} (c)--(e). Note that the empirical surface lacks the smoothness property, while the kernel approach achieves it but at the cost of using a bandwidth parameter. The corresponding VUS estimates are $0.70$ $(0.62,0.79)$, $0.67$ $(0.63,0.80)$, and $0.74$ $(0.66,0.83)$ for the kernel method with bandwidth calculated using equation \eqref{bwdef}, the kernel method with bandwidth selected by cross validation, and the empirical method, respectively. The empirical VUS is similar to our BB estimate, whereas the kernel VUS, for both bandwidths, is slightly lower. This is in agreement to what has been reported by \cite{Bantis2017}. Overall, all estimates are similar and suggest that TMT Part A completion time is an accurate test for dementia stage in Parkinson's disease patients.

\section{\large{\textsf{CONCLUDING REMARKS}}}
We have developed a flexible, nonparametric method based on the Bayesian bootstrap and on the notion of placement value for conducting inference about the ROC surface and its functionals. In addition to providing point and interval estimates in a single integrated framework, our method is computationally easy to implement and very fast. A simulation study illustrated the ability of our approach to produce accurate estimates for a variety of data-generating distributions, and it demonstrated that our estimator is a viable alternative to current nonparametric surface estimators. Furthermore, the validity of our inferences, in terms of frequentist probability of coverage, was demonstrated. The TMT data analysis revealed high accuracy for distinguishing between Parkinson's disease patients who present normal cognition, mild cognitive impairment and dementia.  An interesting avenue for future research is the potential use of the Bayesian bootstrap for learning about the ROC surface of tests subject to a limit of detection.

\bibliographystyle{hapalike}
\bibliography{References_ROCSBB}

\newpage

\begin{table}[H]
\centering
\begin{tabular}{cccc}
Scenario & $Y_1$ & $Y_2$ & $Y_3$  \\ \hline
1 & $\text{N}(0,1^2)$ & $\text{N}(1.5,1^2)$ & $\text{N}(3,1^2)$\\
2 & $\text{Gamma}(2,1)$ & $\text{Gamma}(3,1)$ & $\text{Gamma}(5,2)$\\
3 & $t_2$ & $\text{Beta}(2,2)$ & $\chi^2_2$\\
4 & $0.5\text{N}(0,1^2)+0.5\text{N}(3,1^2)$  & $0.5\text{N}(1,1^2)+0.5\text{N}(4,1.5^2)$ & $0.5\text{N}(2,1^2)+0.5\text{N}(5,2^2)$
\end{tabular}
\caption{\footnotesize{Scenarios considered for the simulation study.}}
\label{t1}
\end{table}

\begin{table}[H]
\centering
\begin{tabular}{cccc}
Scenario & $(n_1,n_2,n_3)=(50,50,50)$ & $(n_1,n_2,n_3)=(100,100,100)$ &  $(n_1,n_2,n_3)=(200,200,200)$   \\ \hline
1 & $0.95$ & $0.98$ & $0.97$\\
2 & $0.99$ & $1$ & $0.99$\\
3 & $0.95$ & $0.97$ & $0.95$\\
4 & $0.98$ & $0.98$ & $0.98$
\end{tabular}
\caption{\footnotesize{VUS $95\%$ coverage probabilities.}}
\label{t2}
\end{table}

\newpage

\section{\large{\textsf{SUPPLEMENTARY MATERIALS}}}
\setcounter{figure}{0}    

\begin{figure}[H]
\begin{center}
\subfigure{\includegraphics[page = 1, width=4.25cm]{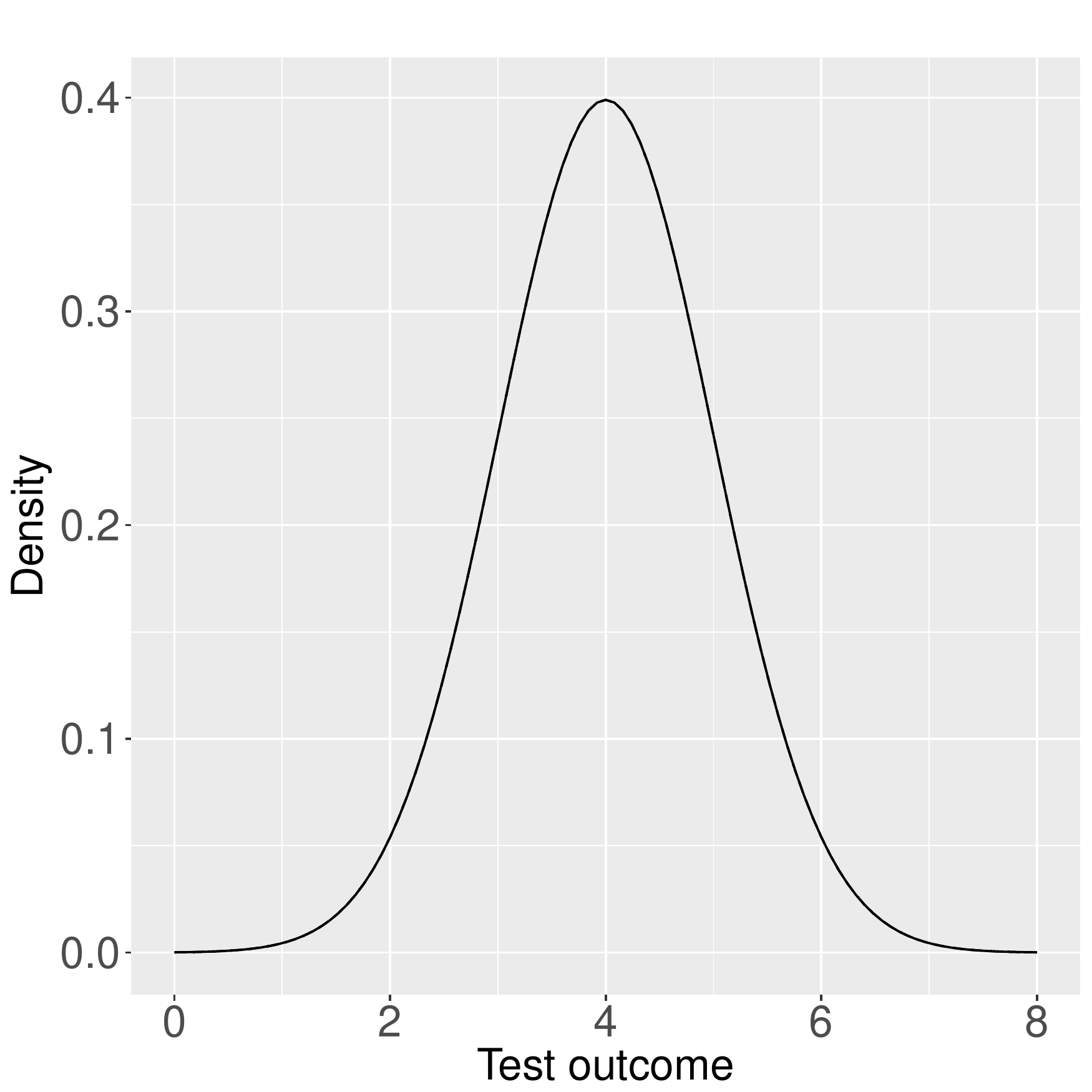}}
\subfigure{\includegraphics[page = 1, width=4.25cm]{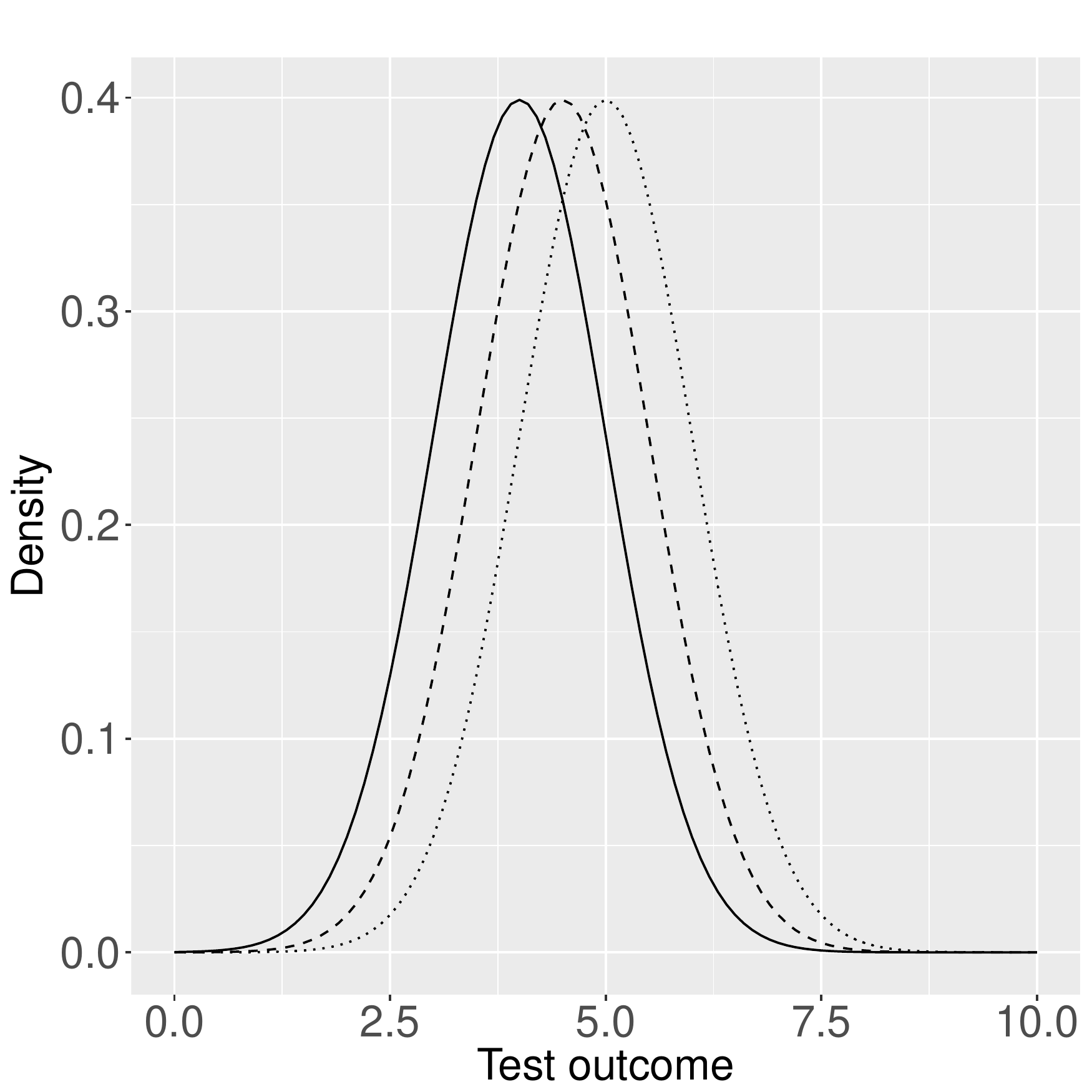}}
\subfigure{\includegraphics[page = 1, width=4.25cm]{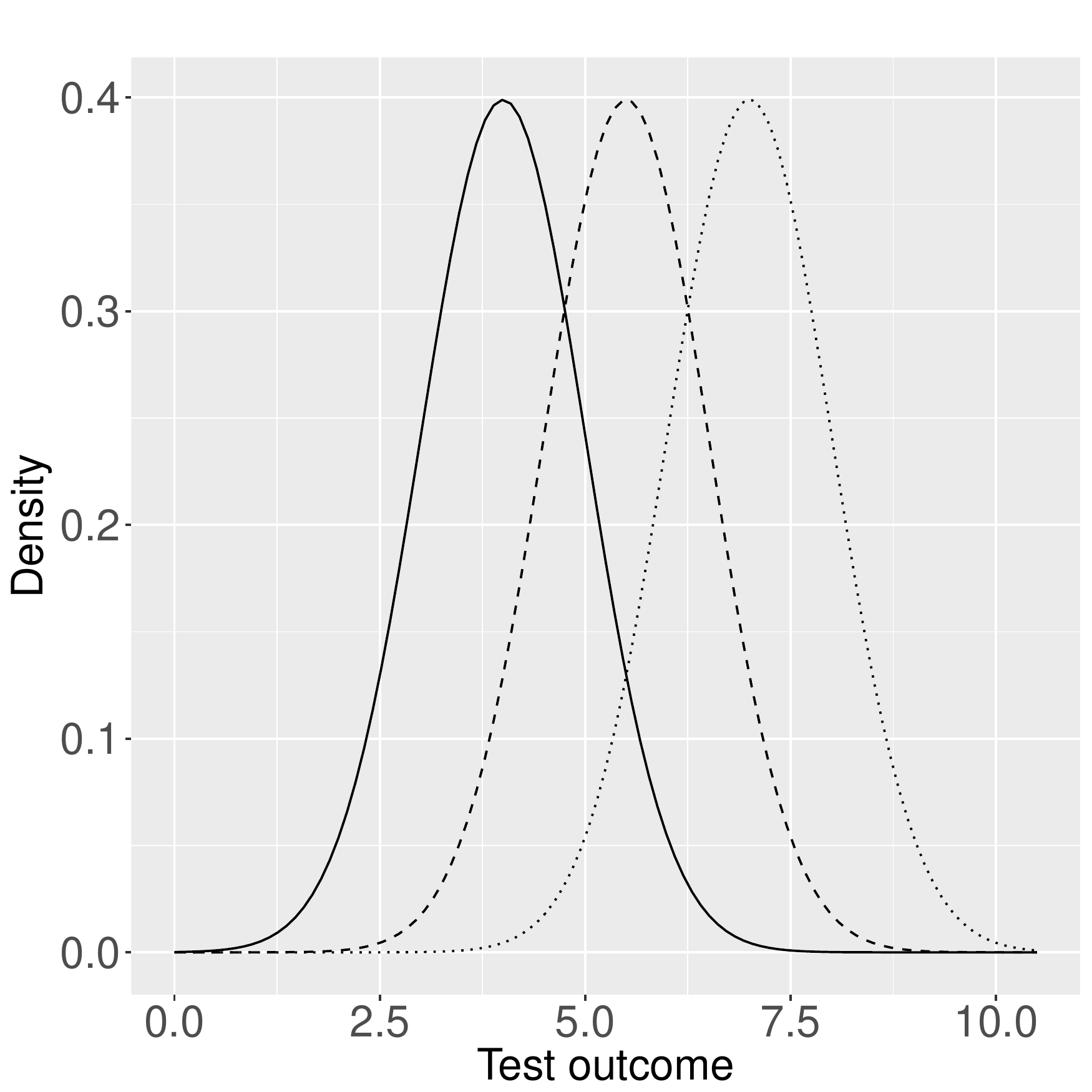}}
\subfigure{\includegraphics[page = 1, width=4.25cm]{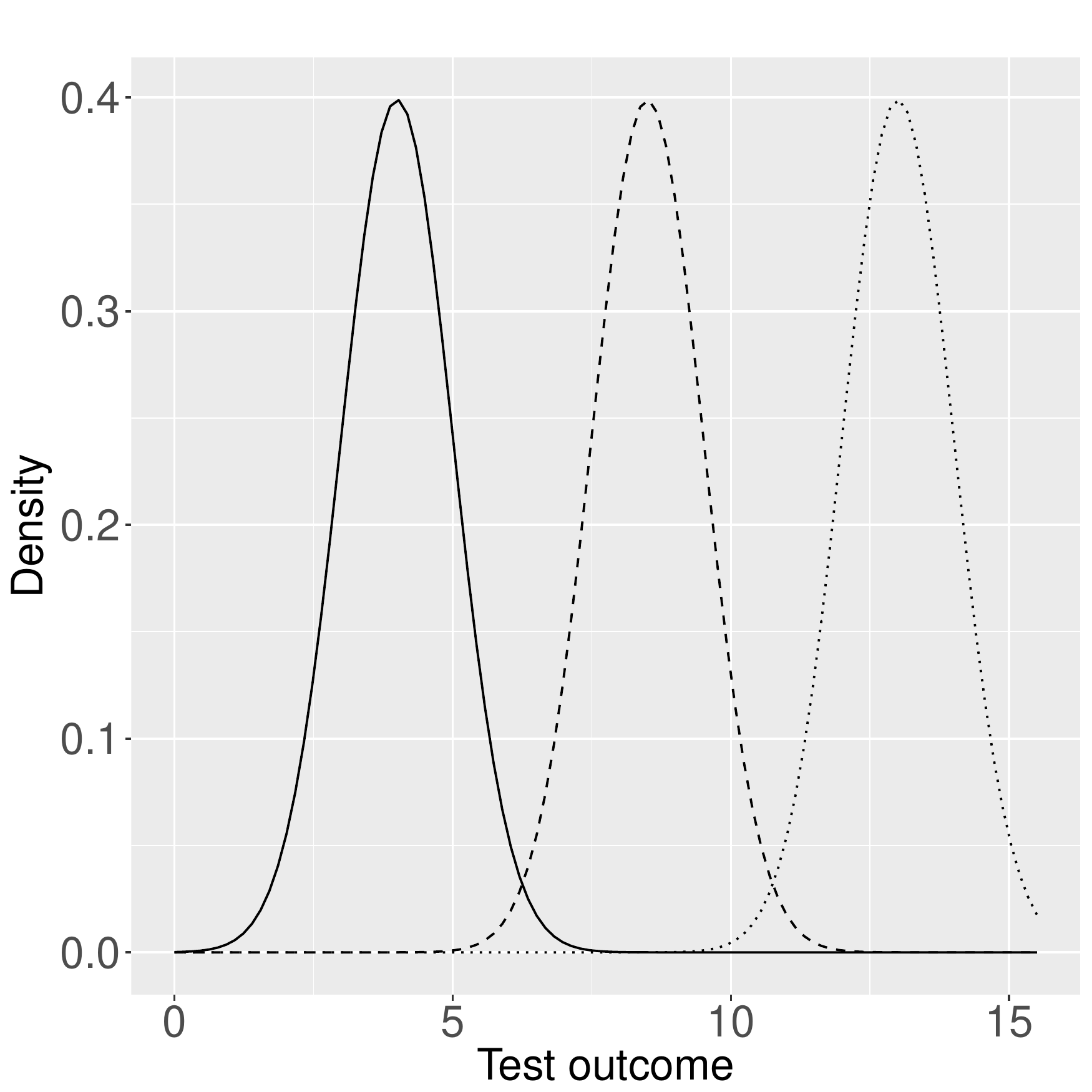}}\\
\subfigure{\includegraphics[page = 1, width=4.25cm]{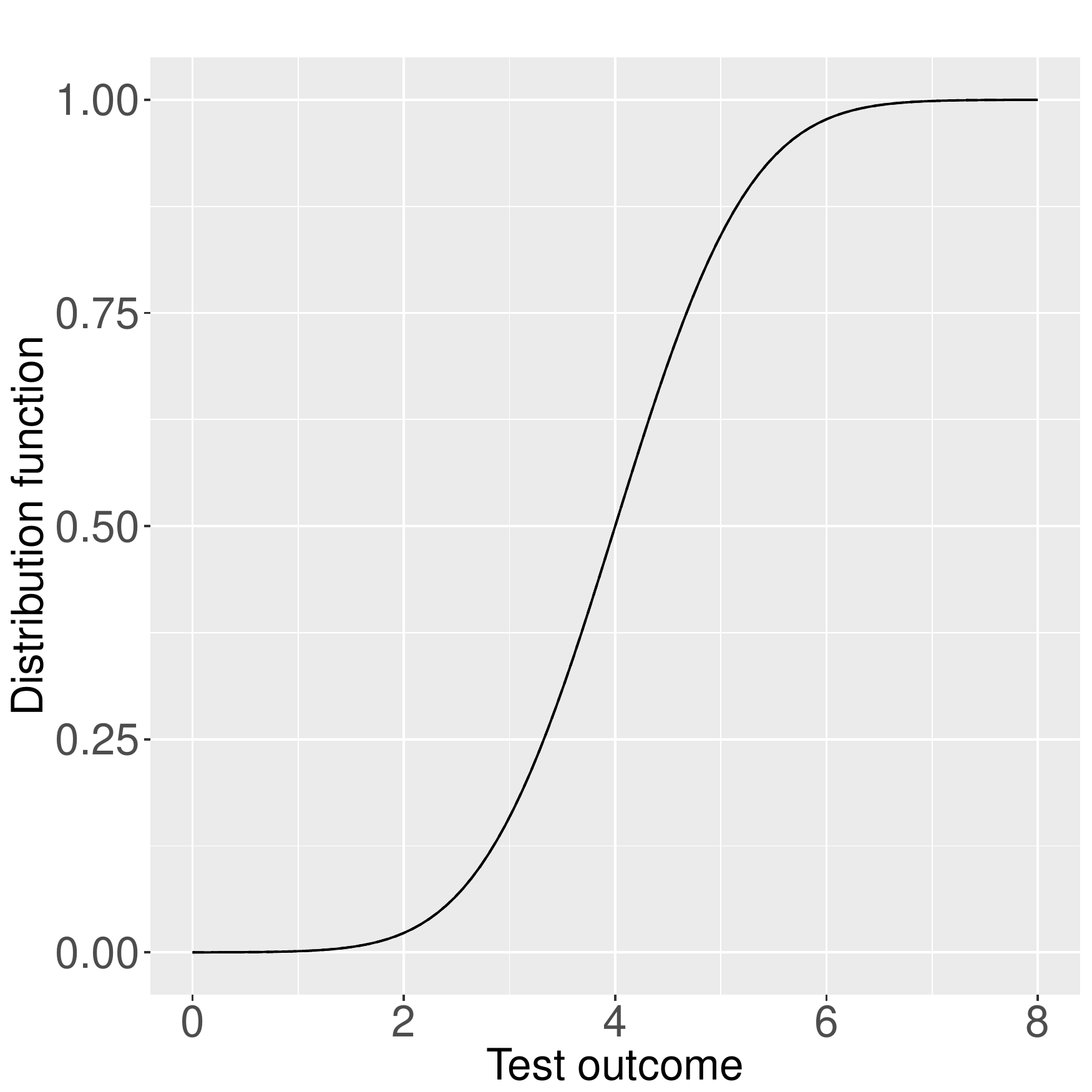}}
\subfigure{\includegraphics[page = 1, width=4.25cm]{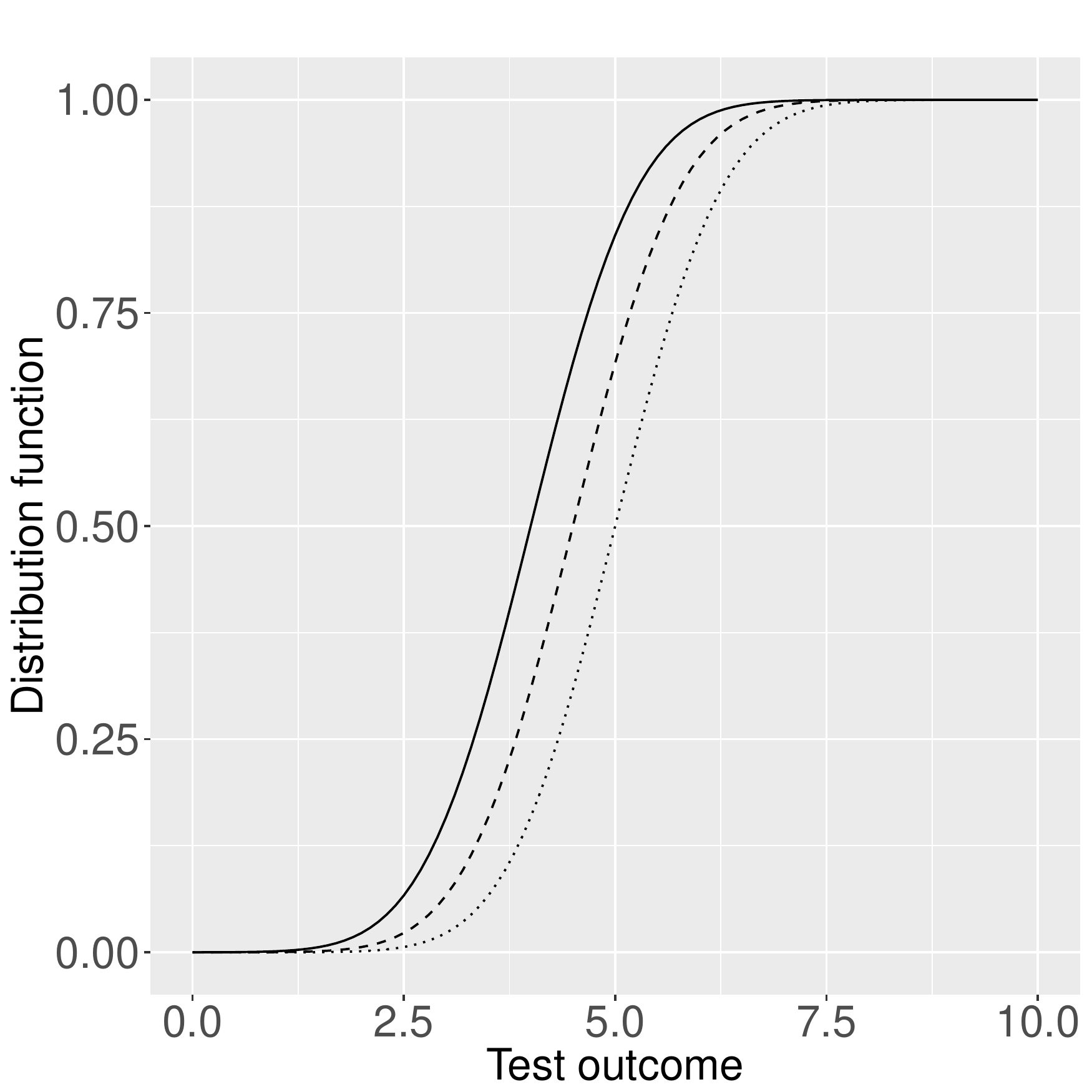}}
\subfigure{\includegraphics[page = 1, width=4.25cm]{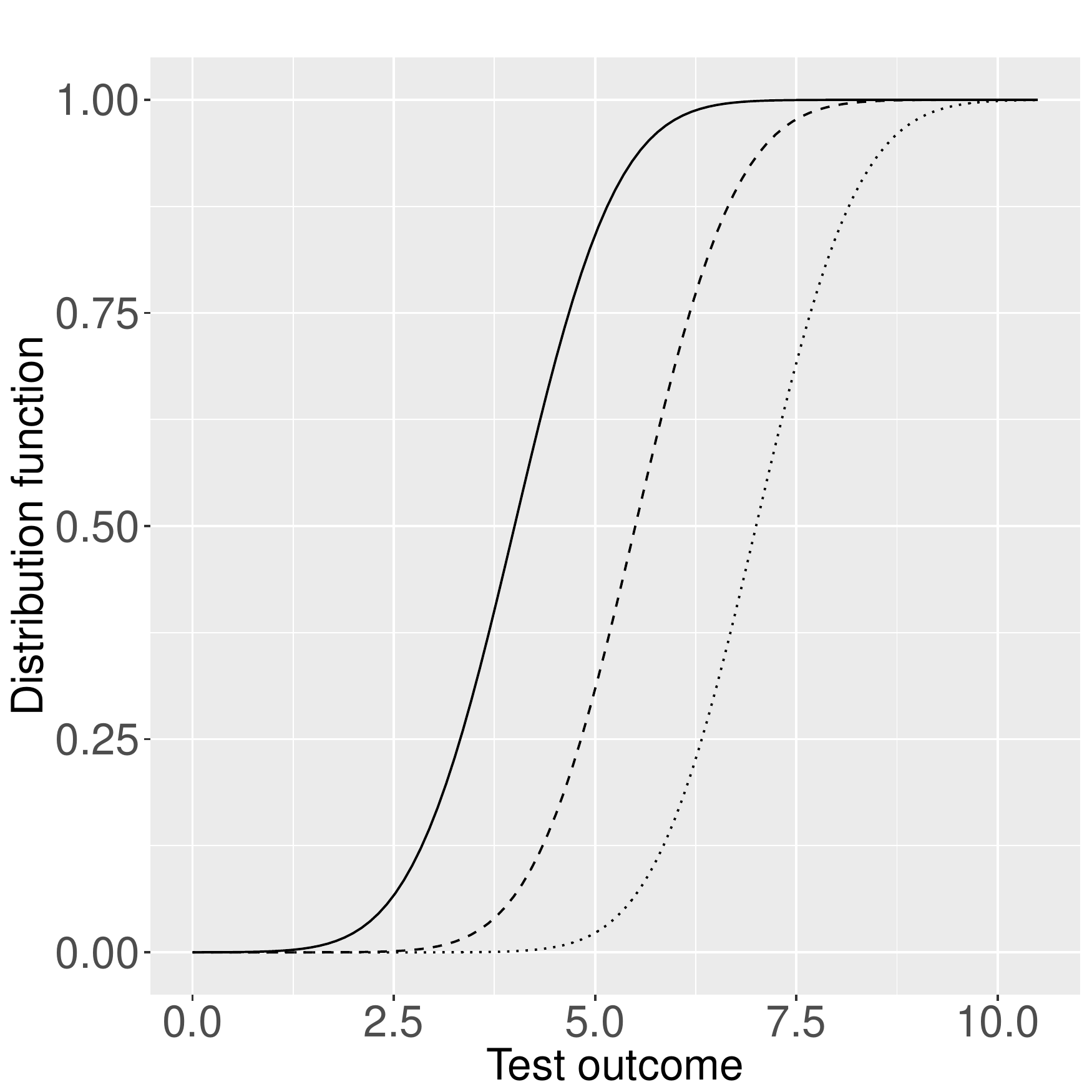}}
\subfigure{\includegraphics[page = 1, width=4.25cm]{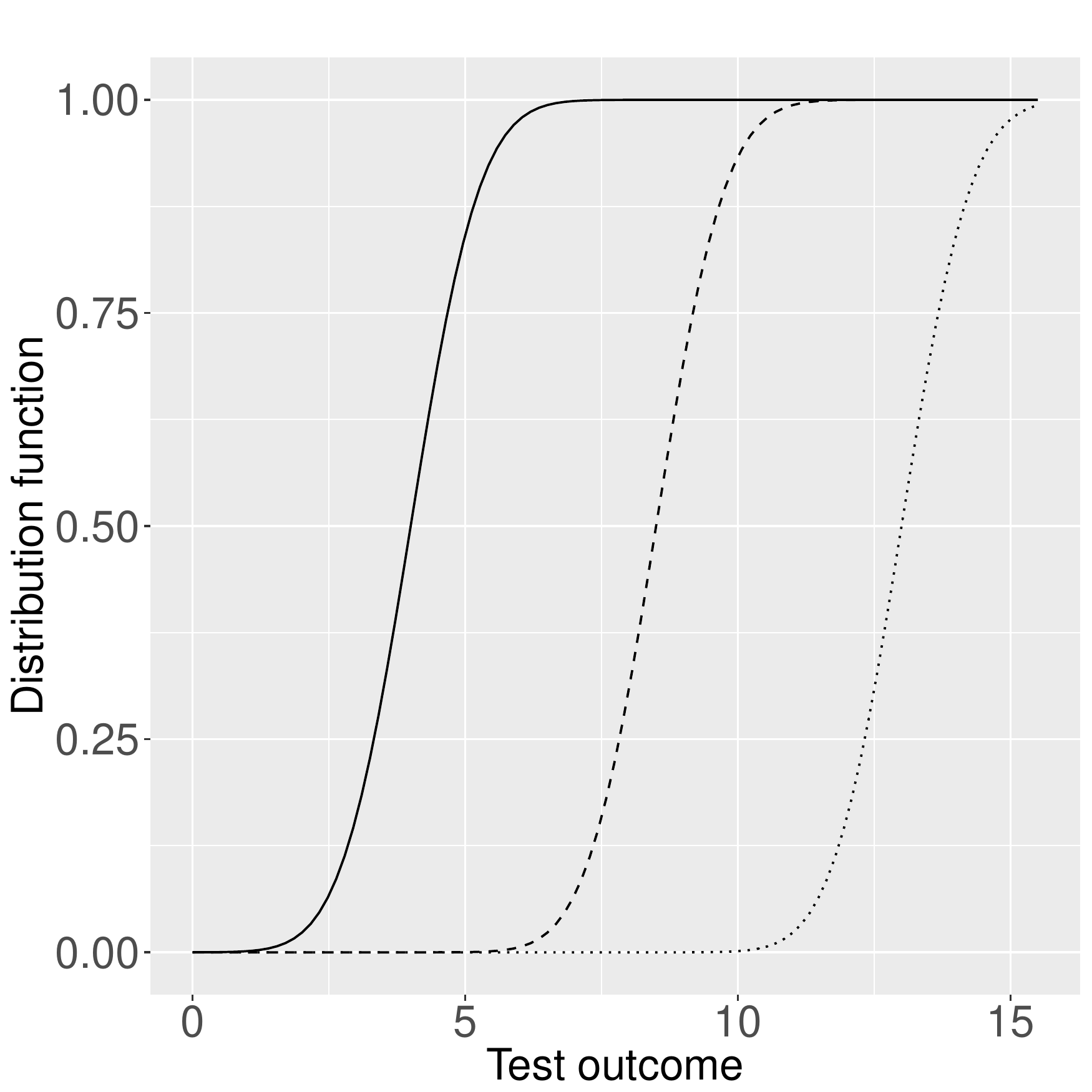}}\\ \vspace{0.2cm}
\subfigure{\includegraphics[page = 1, width=4.25cm]{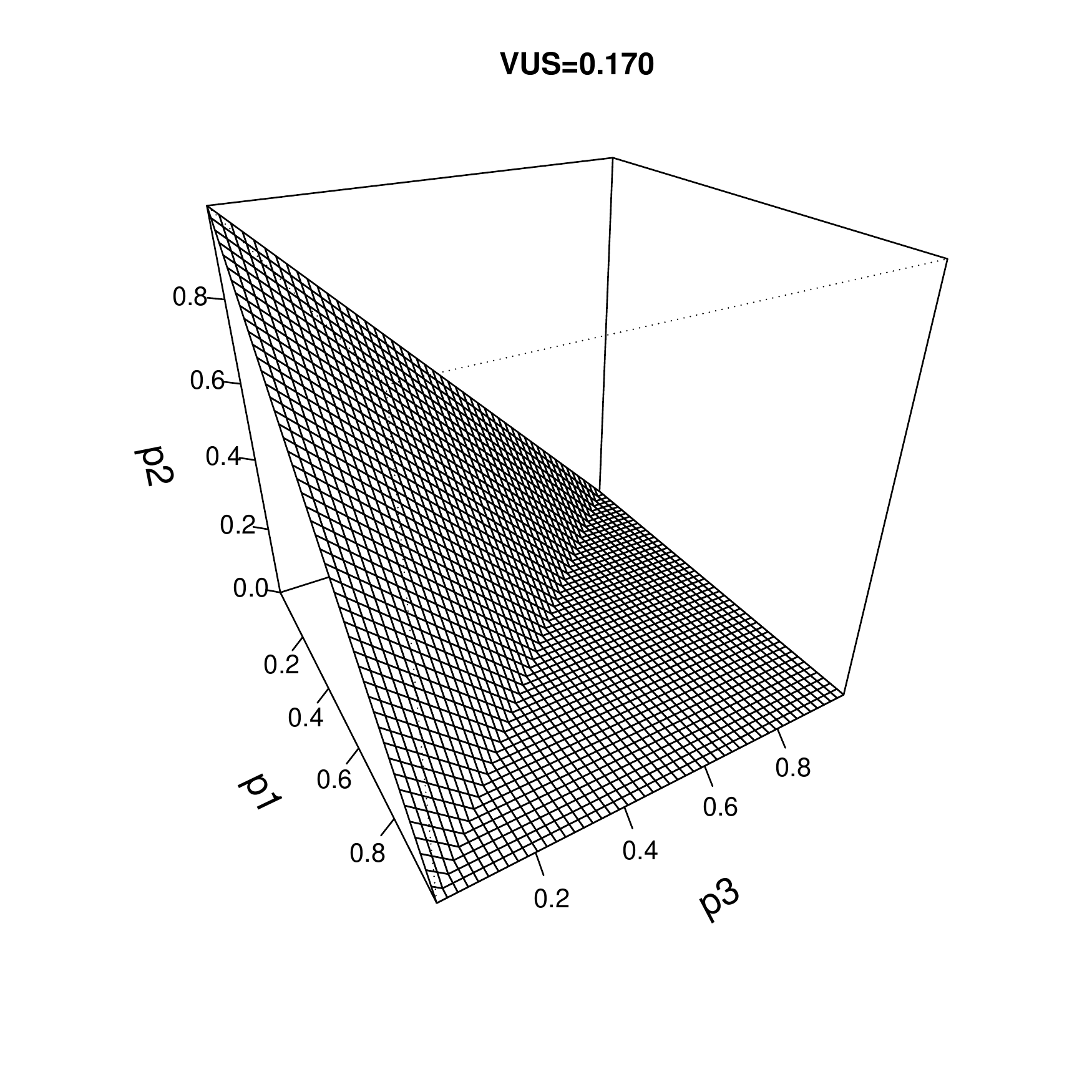}}
\subfigure{\includegraphics[page = 1, width=4.25cm]{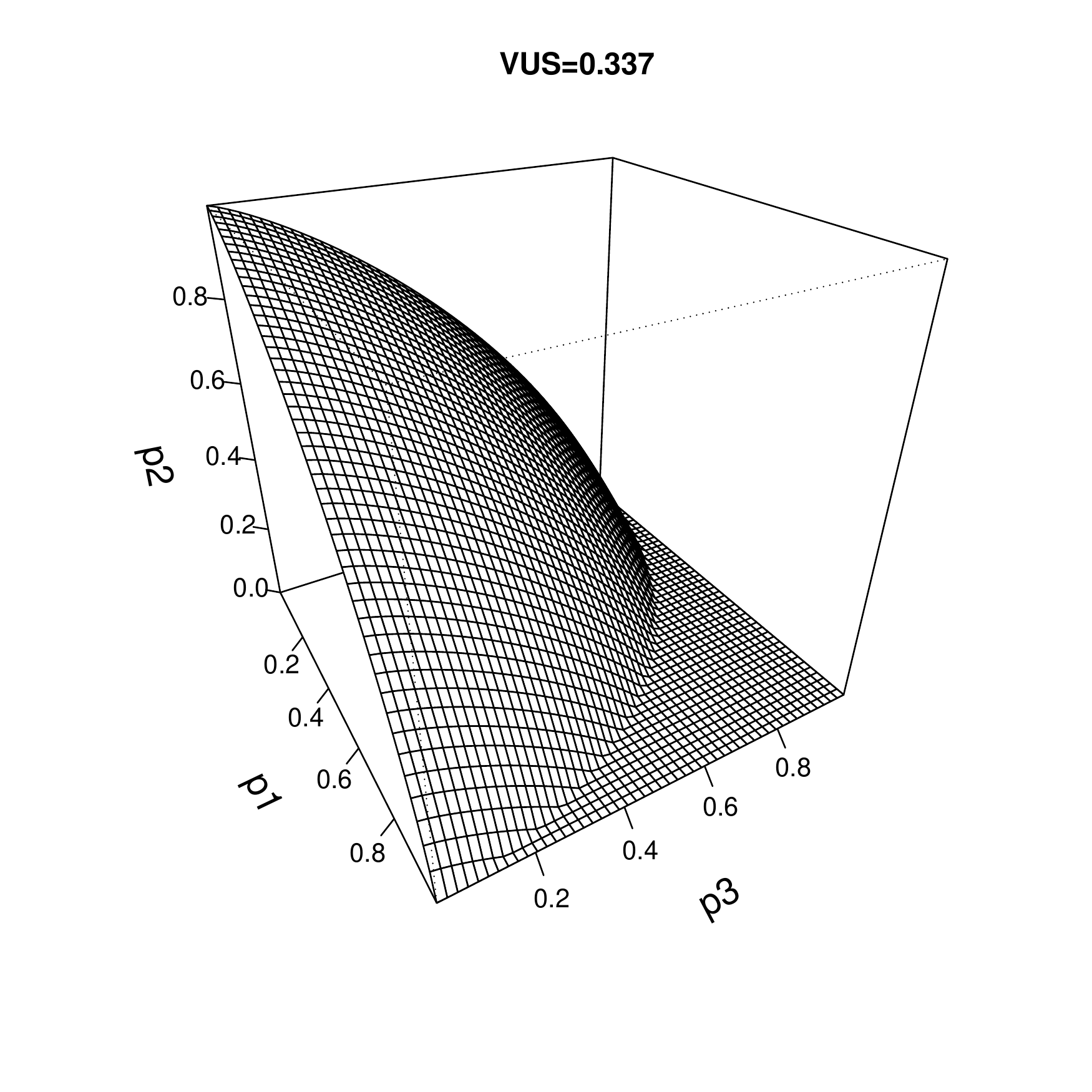}}
\subfigure{\includegraphics[page = 1, width=4.25cm]{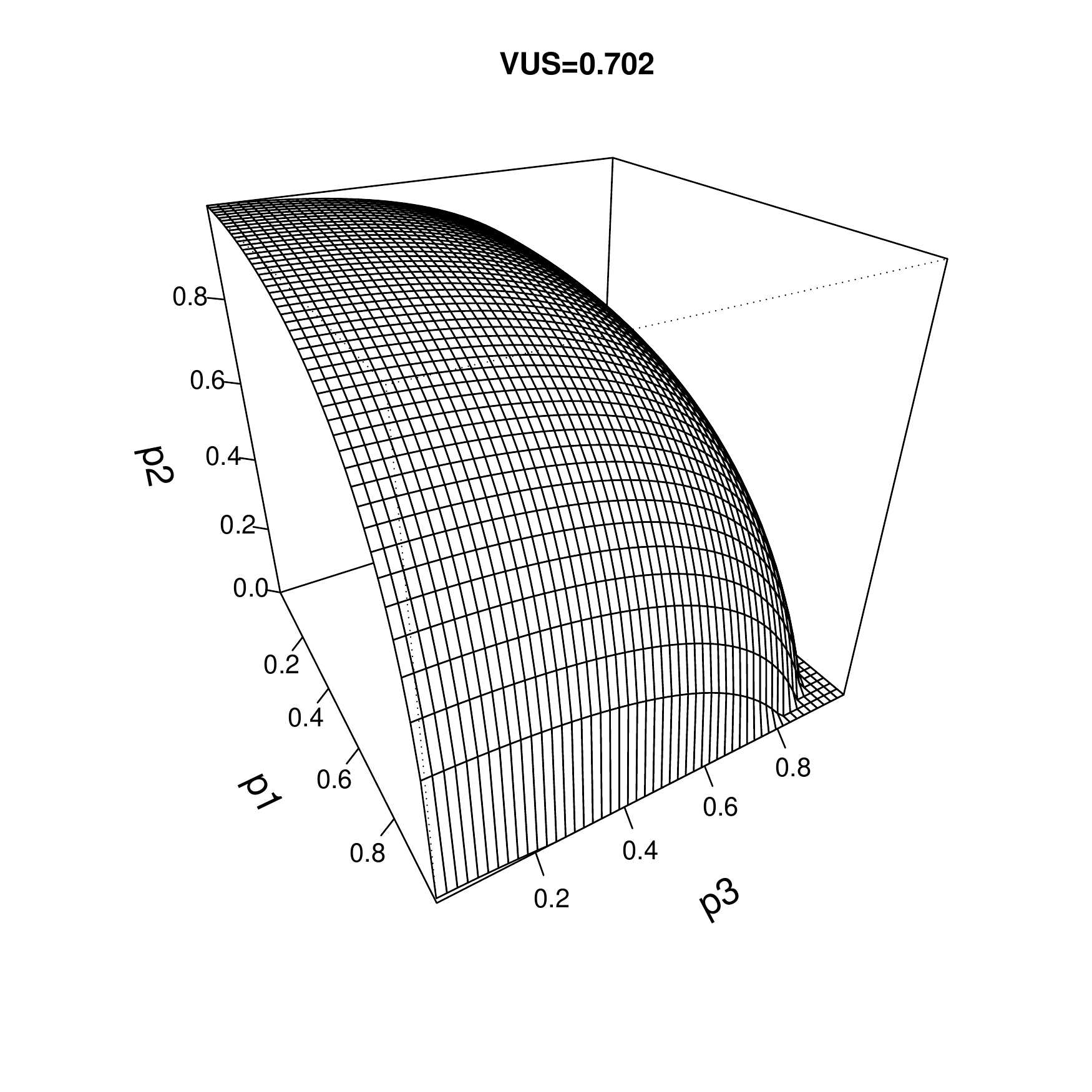}}
\subfigure{\includegraphics[page = 1, width=4.25cm]{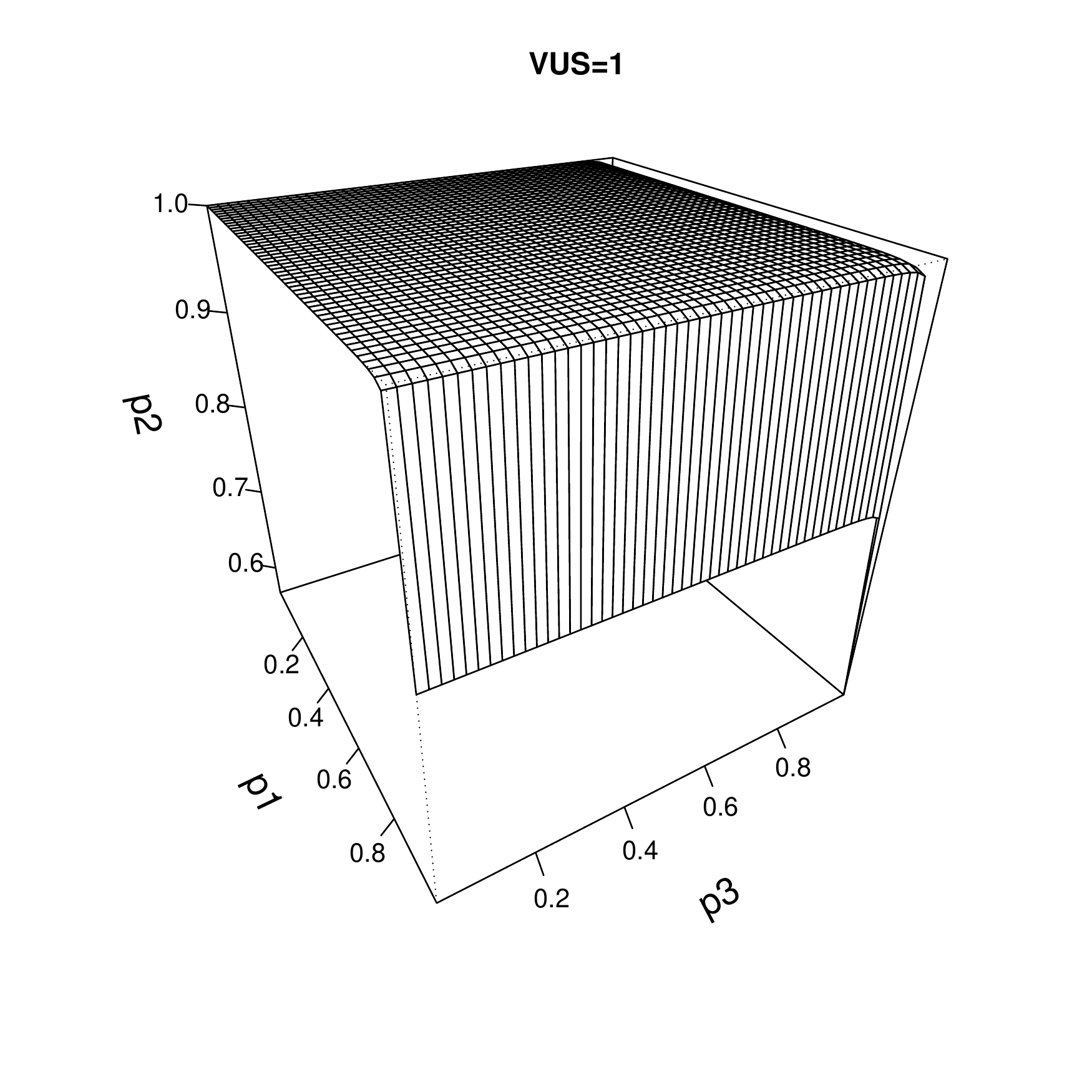}}
\caption{\footnotesize{Different degrees of overlap/stochastic ordering between f1/ F1, f2/ F2, and f3/ F3 for hypothetical test outcomes distributions and corresponding ROC surfaces. Solid lines represent test outcomes in group 1, dashed lines in group 2, and dotted lines in group 3.}}
\label{figsintro}
\end{center}
\end{figure}

\begin{figure}[H]
\begin{center}
\subfigure{\includegraphics[page = 1, width=4.75cm]{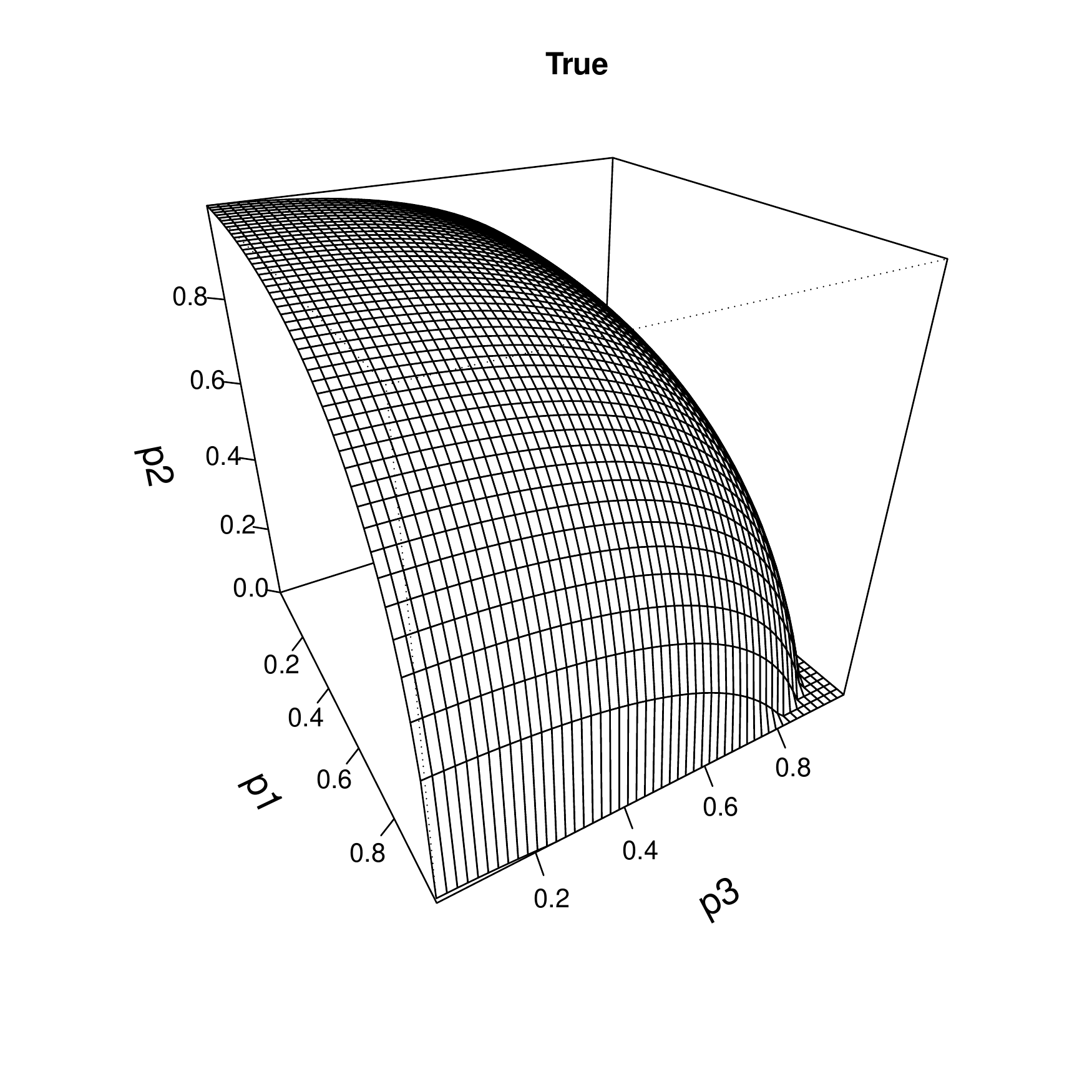}}\\
\subfigure{\includegraphics[page = 1, width=4.5cm]{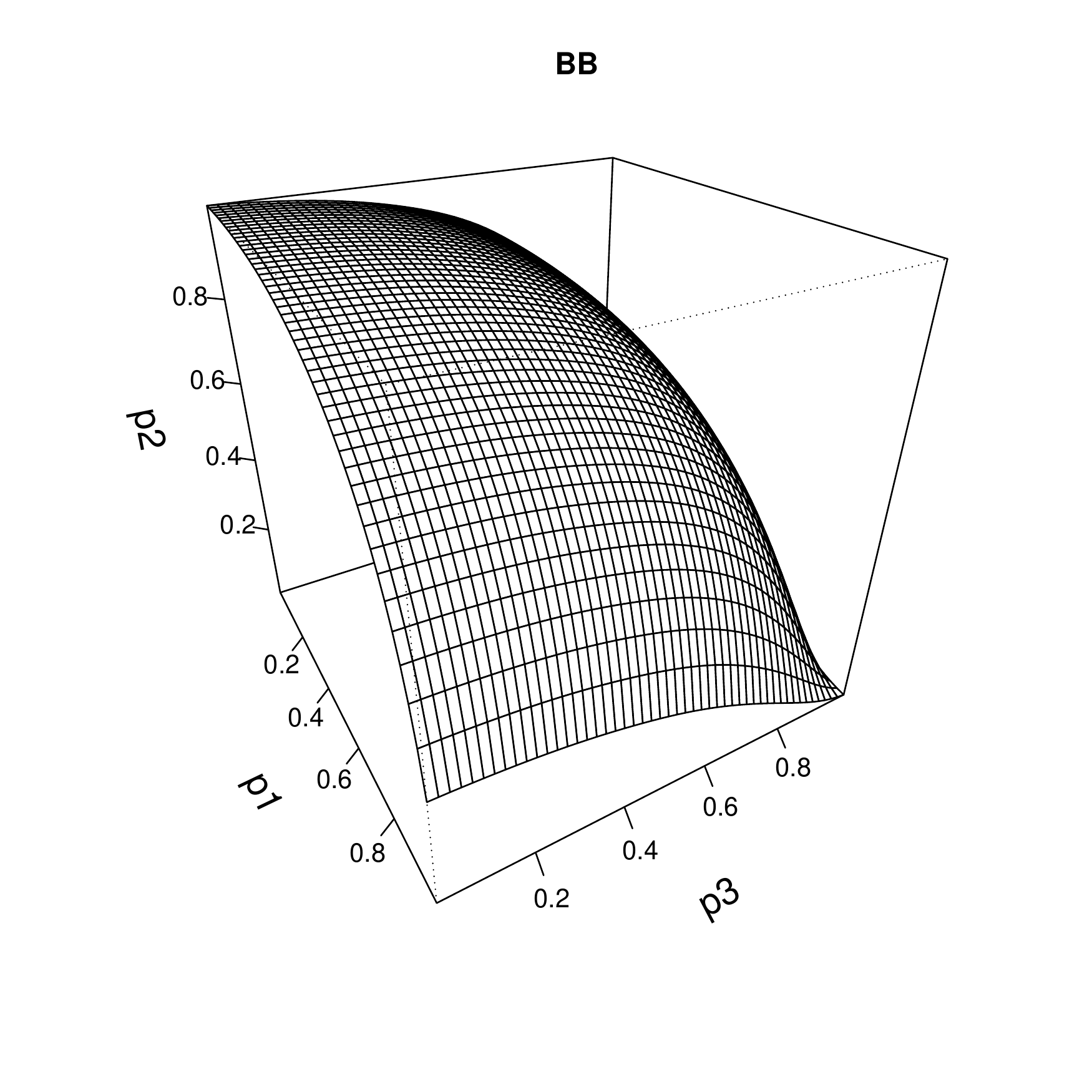}}
\subfigure{\includegraphics[page = 1, width=4.5cm]{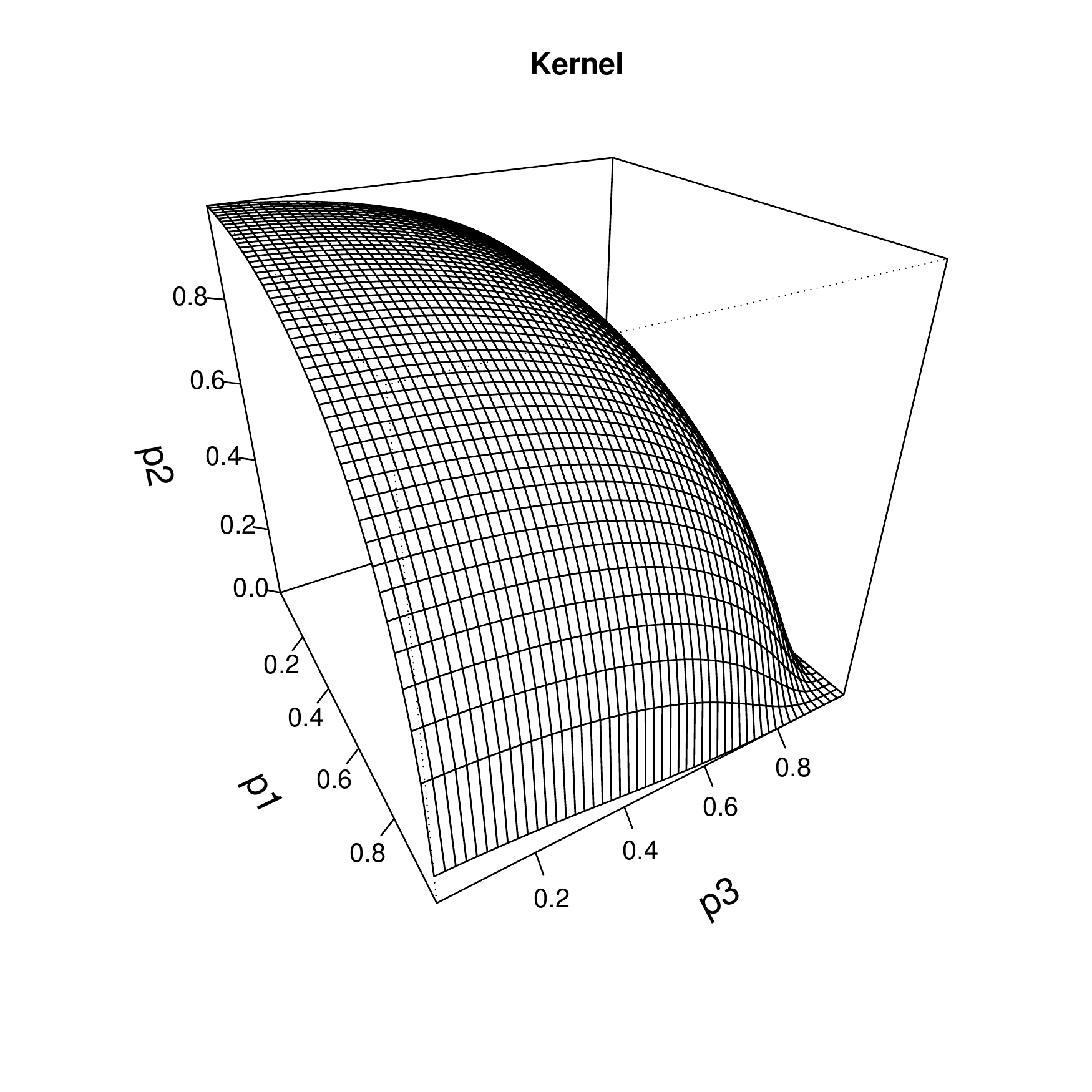}}
\subfigure{\includegraphics[page = 1, width=4.5cm]{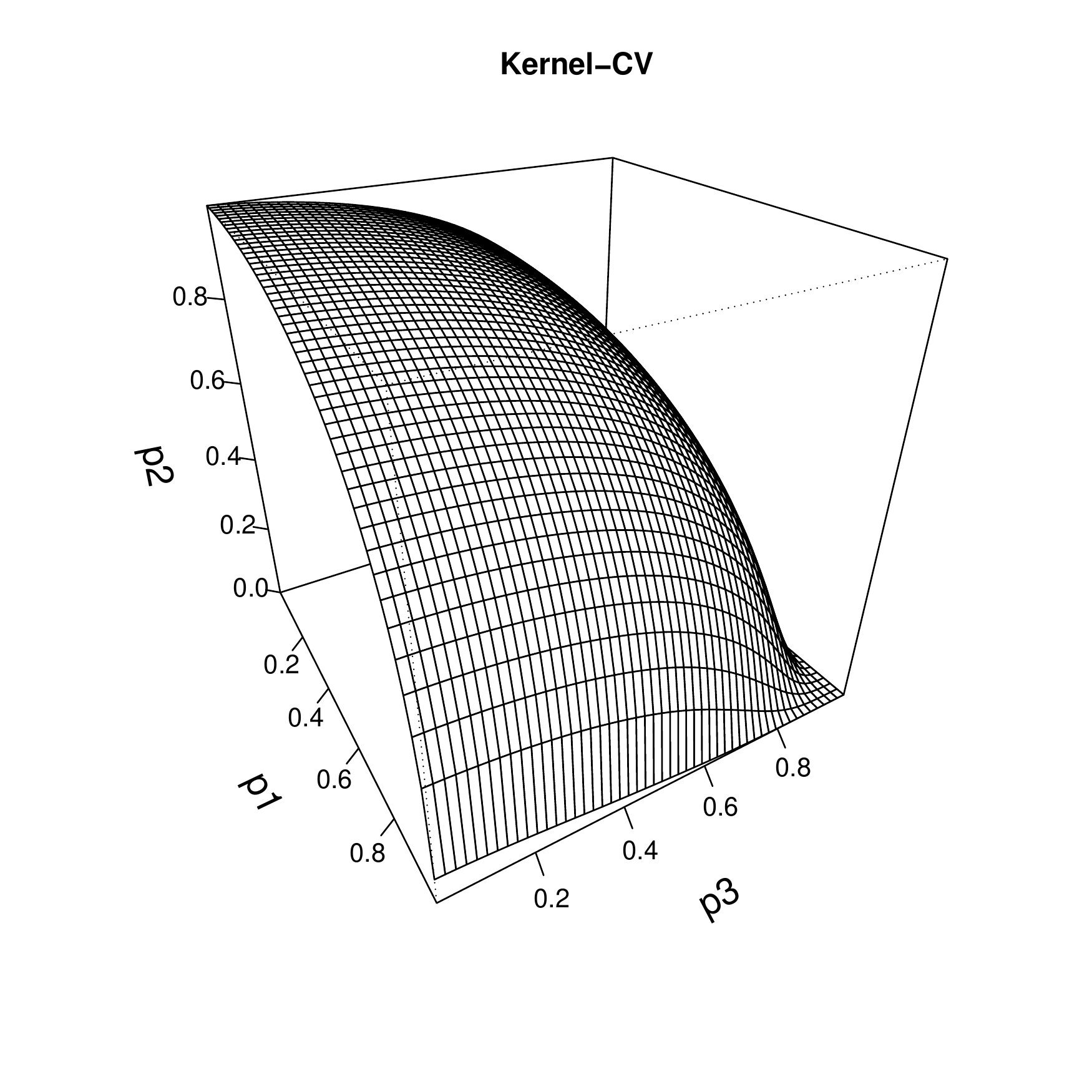}}
\subfigure{\includegraphics[page = 1, width=4.5cm]{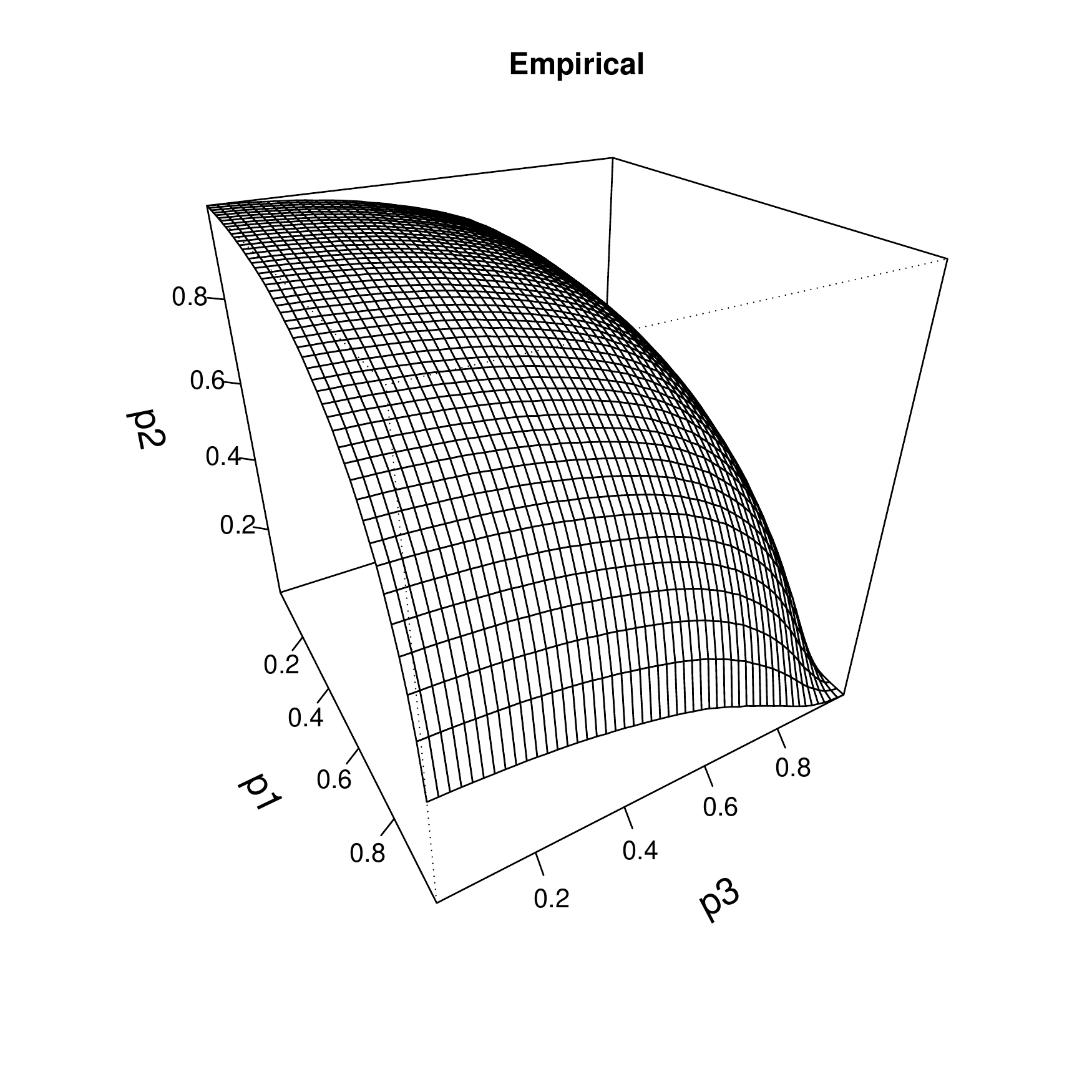}}\\
\subfigure{\includegraphics[page = 1, width=4.5cm]{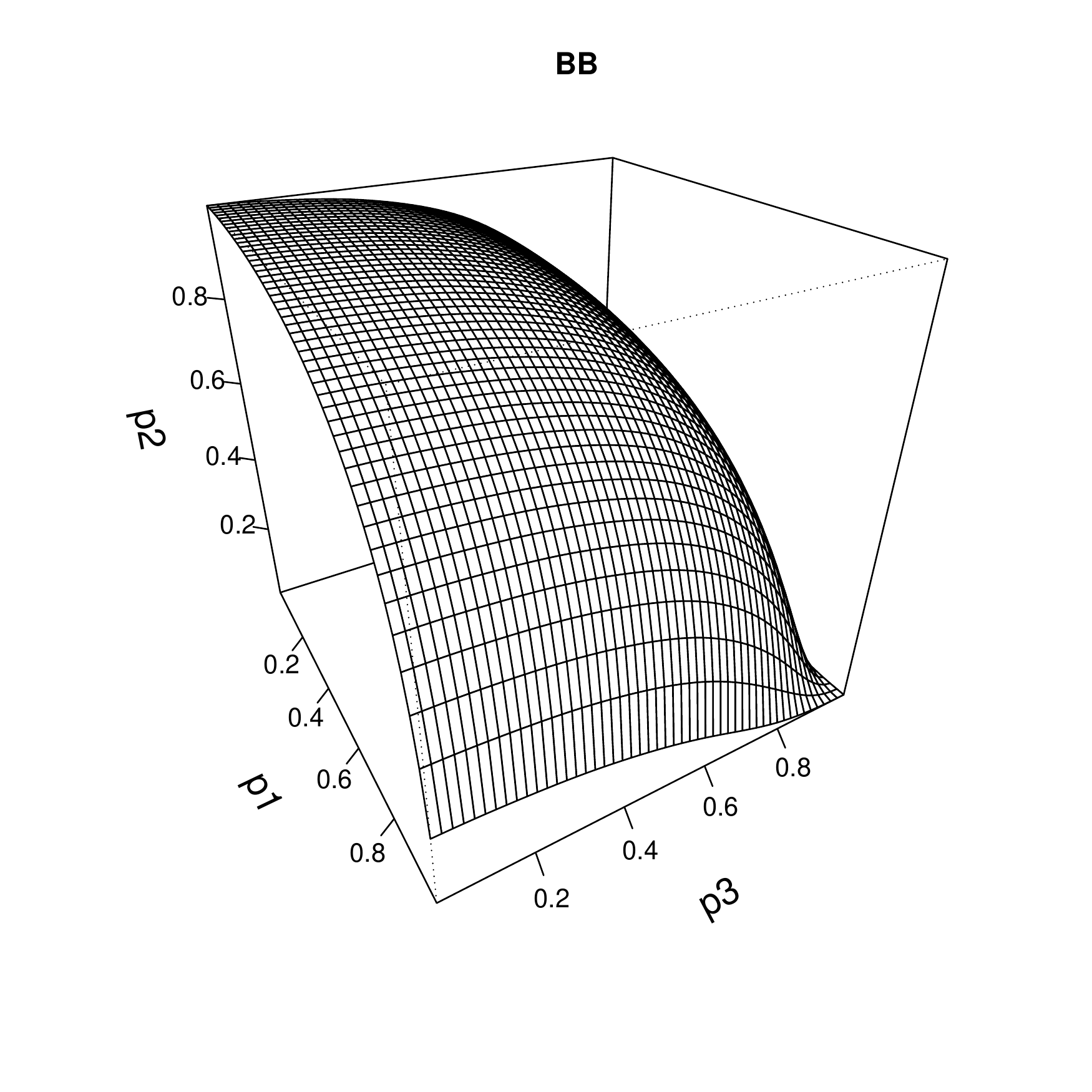}}
\subfigure{\includegraphics[page = 1, width=4.5cm]{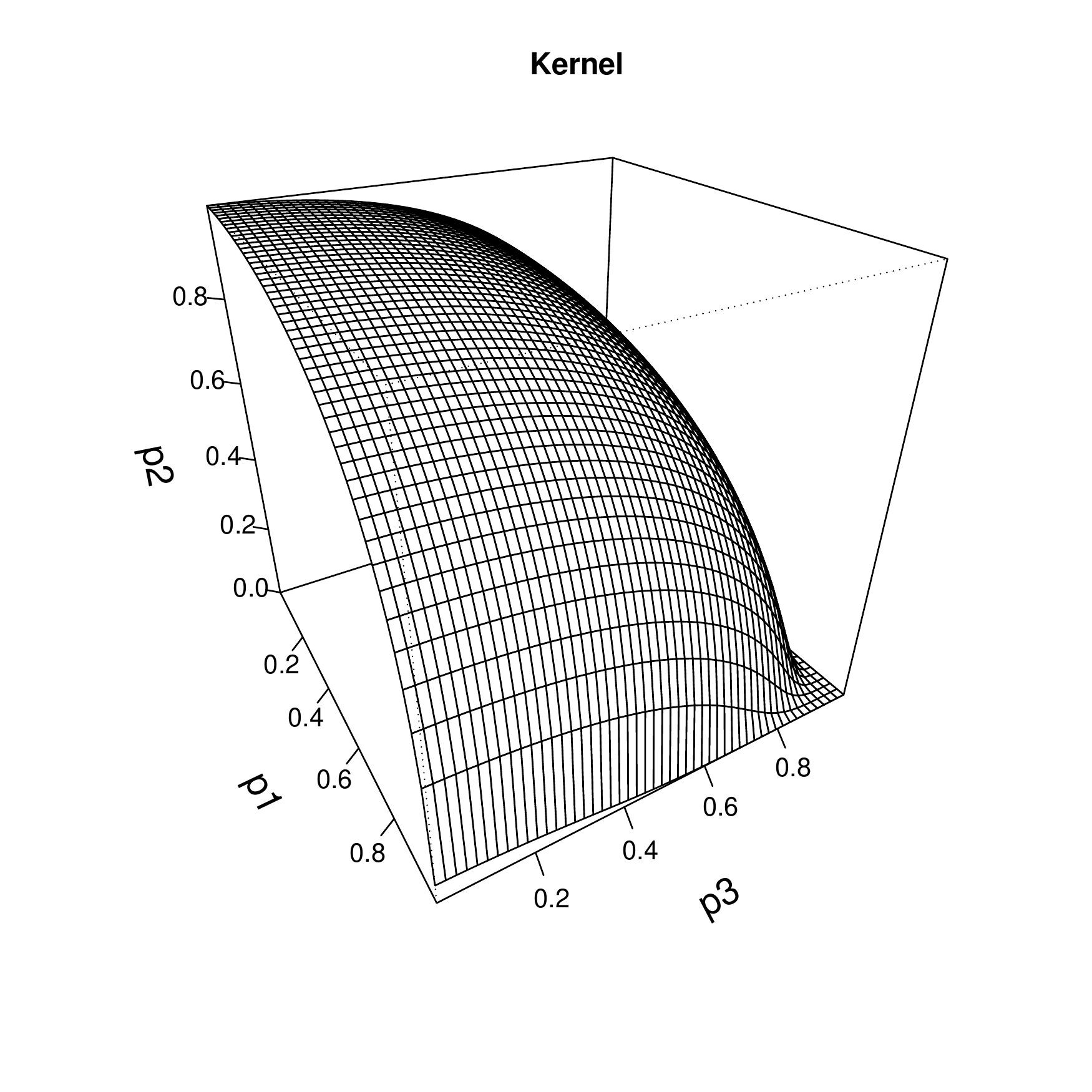}}
\subfigure{\includegraphics[page = 1, width=4.5cm]{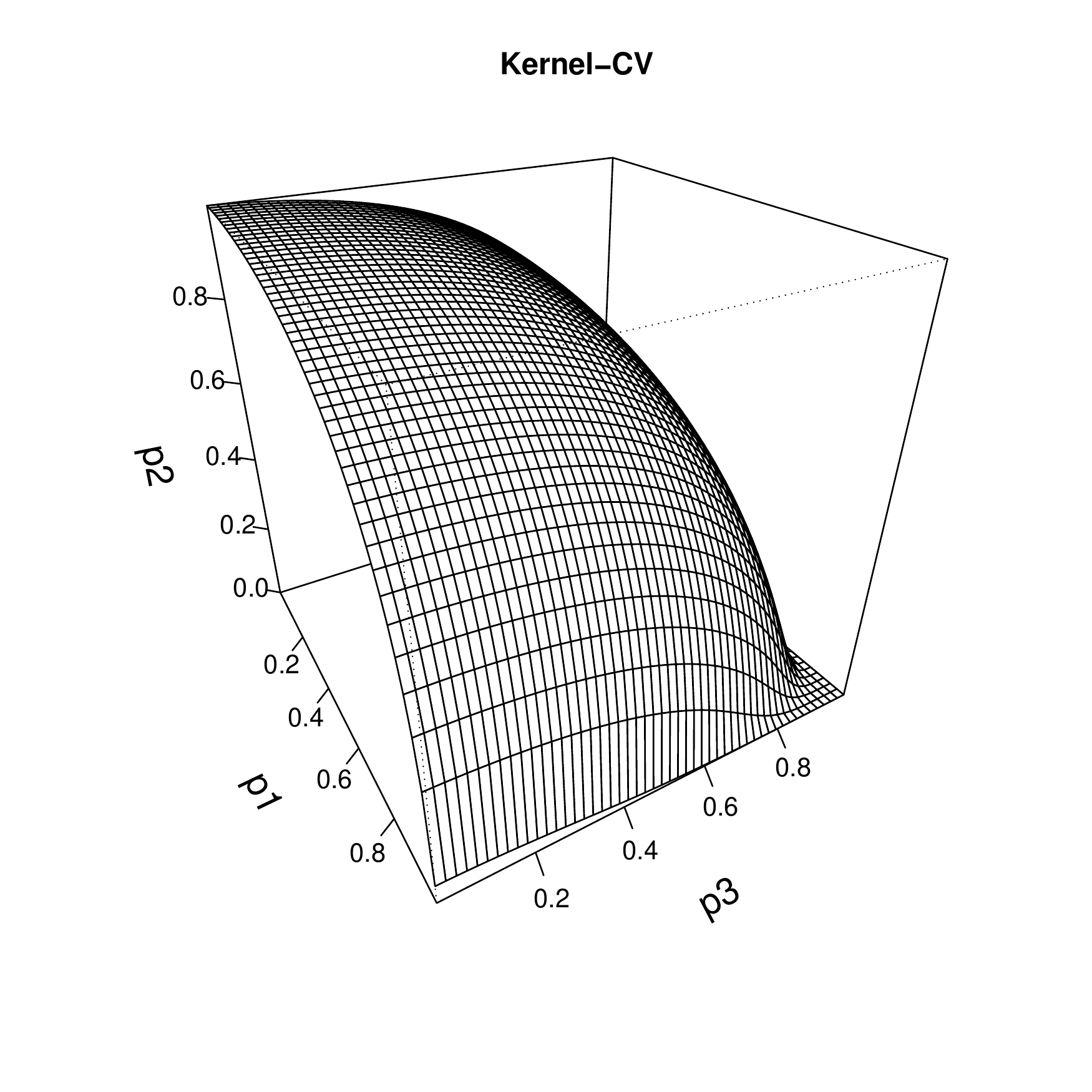}}
\subfigure{\includegraphics[page = 1, width=4.5cm]{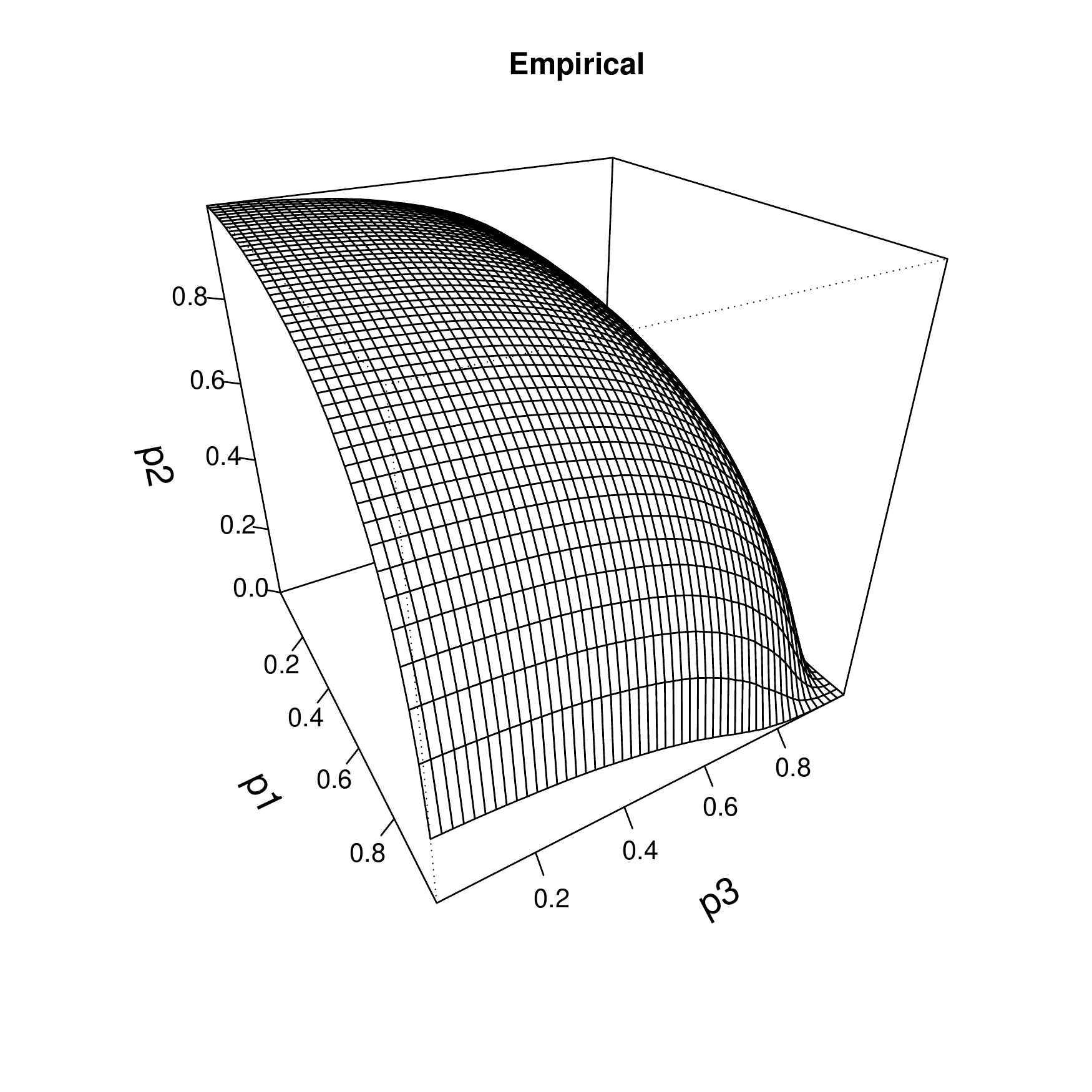}}\\
\subfigure{\includegraphics[page = 1, width=4.5cm]{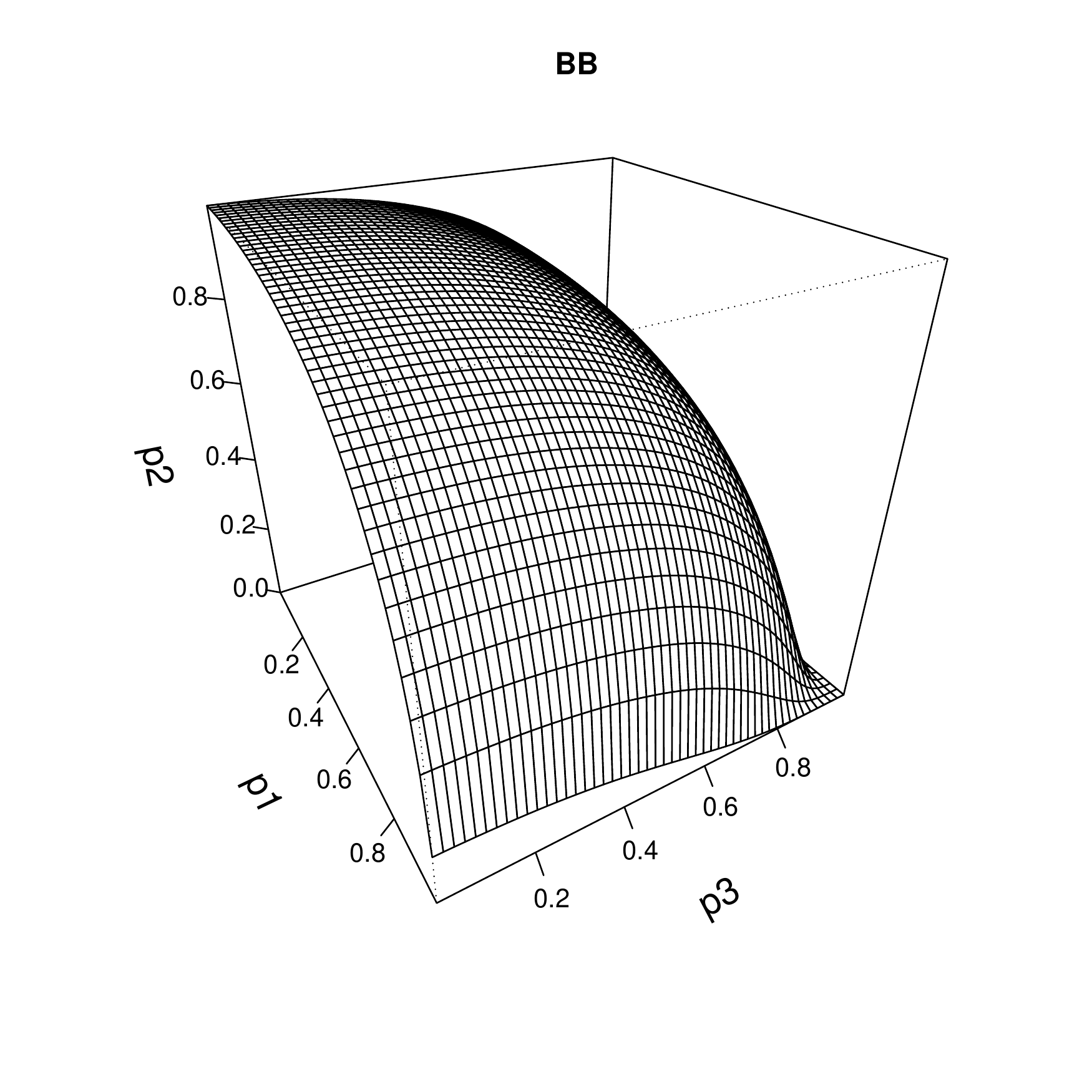}}
\subfigure{\includegraphics[page = 1, width=4.5cm]{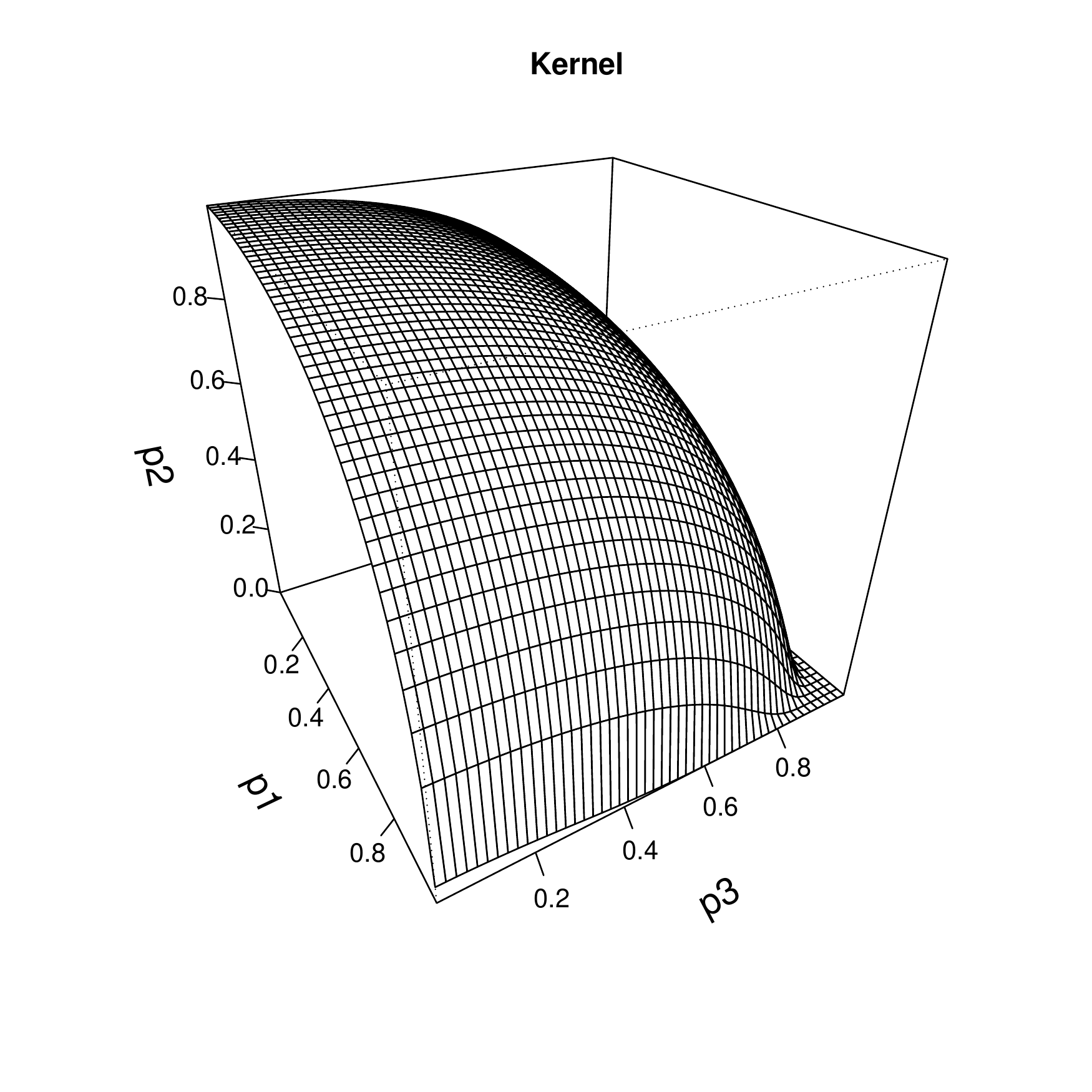}}
\subfigure{\includegraphics[page = 1, width=4.5cm]{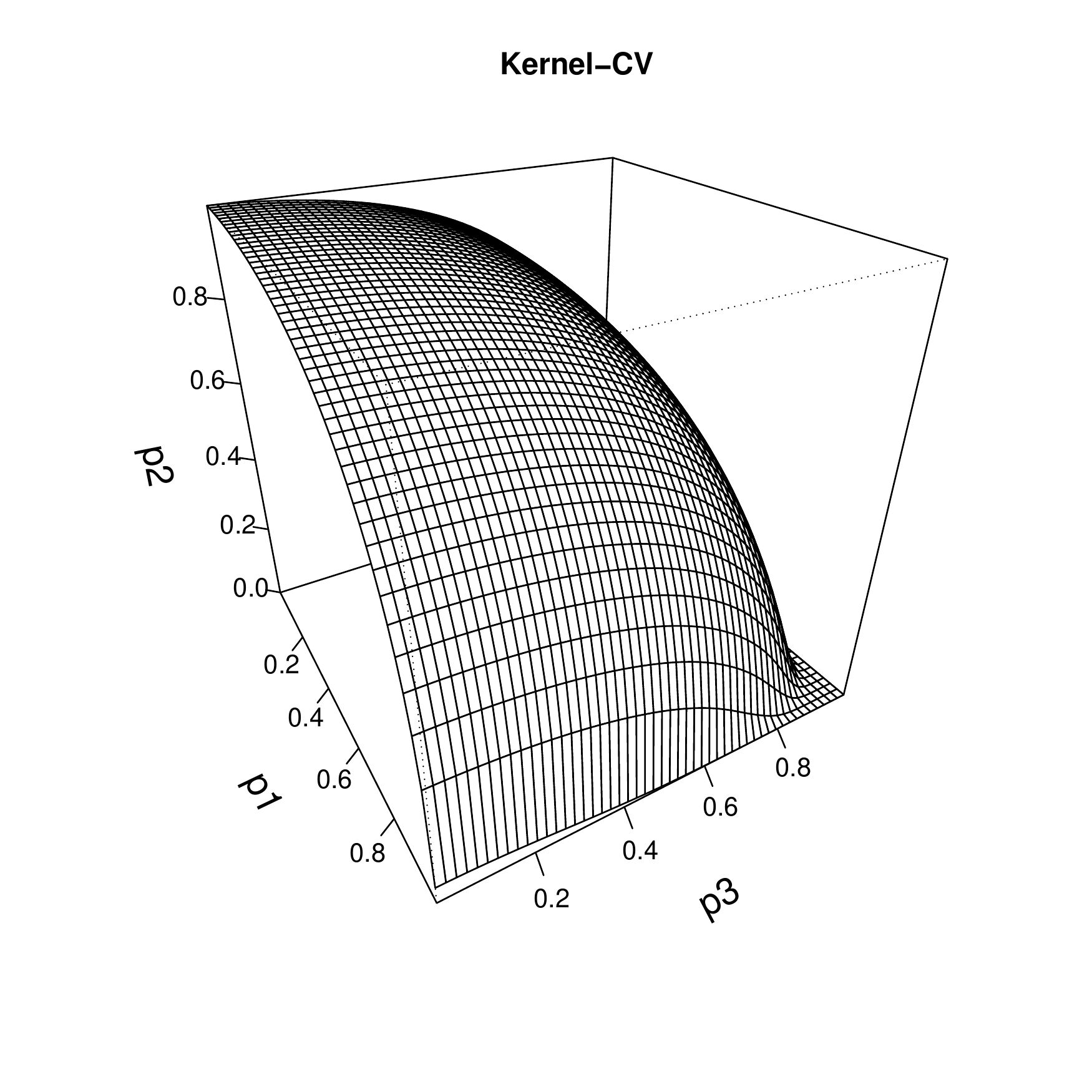}}
\subfigure{\includegraphics[page = 1, width=4.5cm]{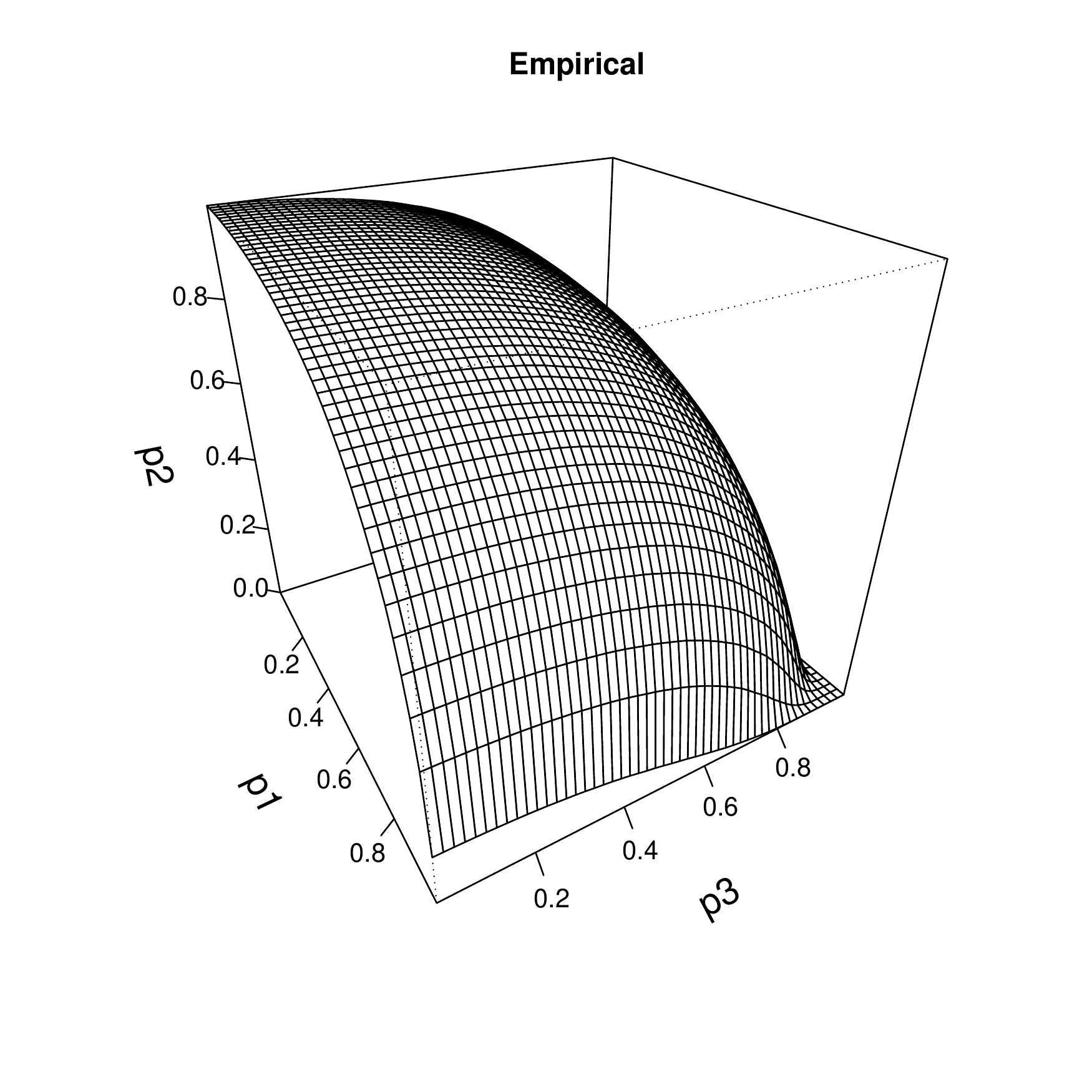}}
\caption{\footnotesize{Scenario 1. True ROC surface and mean across the 300 estimated ROC surfaces. First row: $n_1=n_2=n_3$=50. Second row: $n_1=n_2=n_3$=100. Third row: $n_1=n_2=n_3$=200. Here Kernel denotes the kernel estimate with bandwidth calculated using equation (2) of the main manuscript and Kernel-CV stands for the kernel estimate with the bandwidth selected by least squares cross-validation.}}
\end{center}
\end{figure}

\begin{figure}[H]
\begin{center}
\subfigure{\includegraphics[page = 1, width=4.75cm]{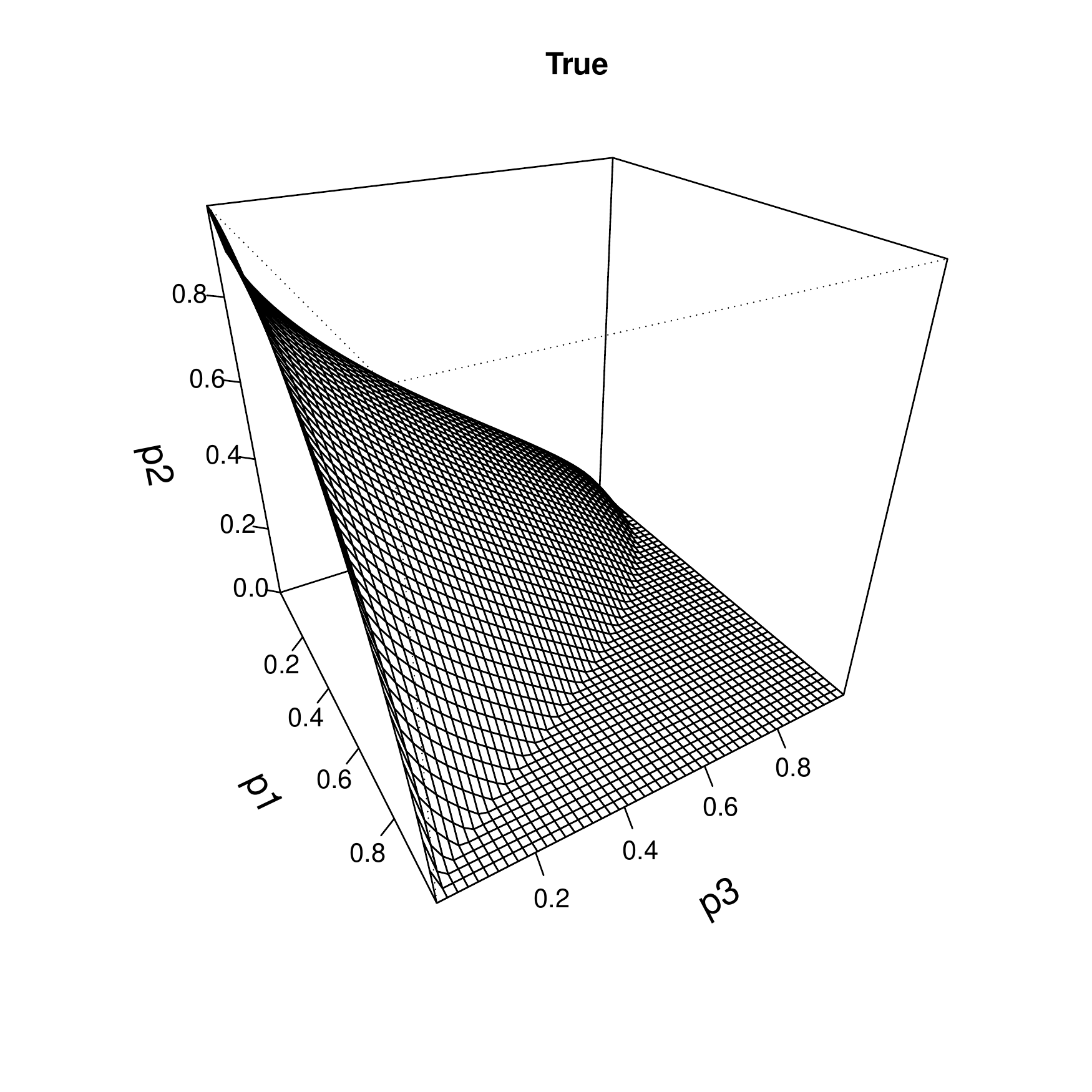}}\\
\subfigure{\includegraphics[page = 1, width=4.5cm]{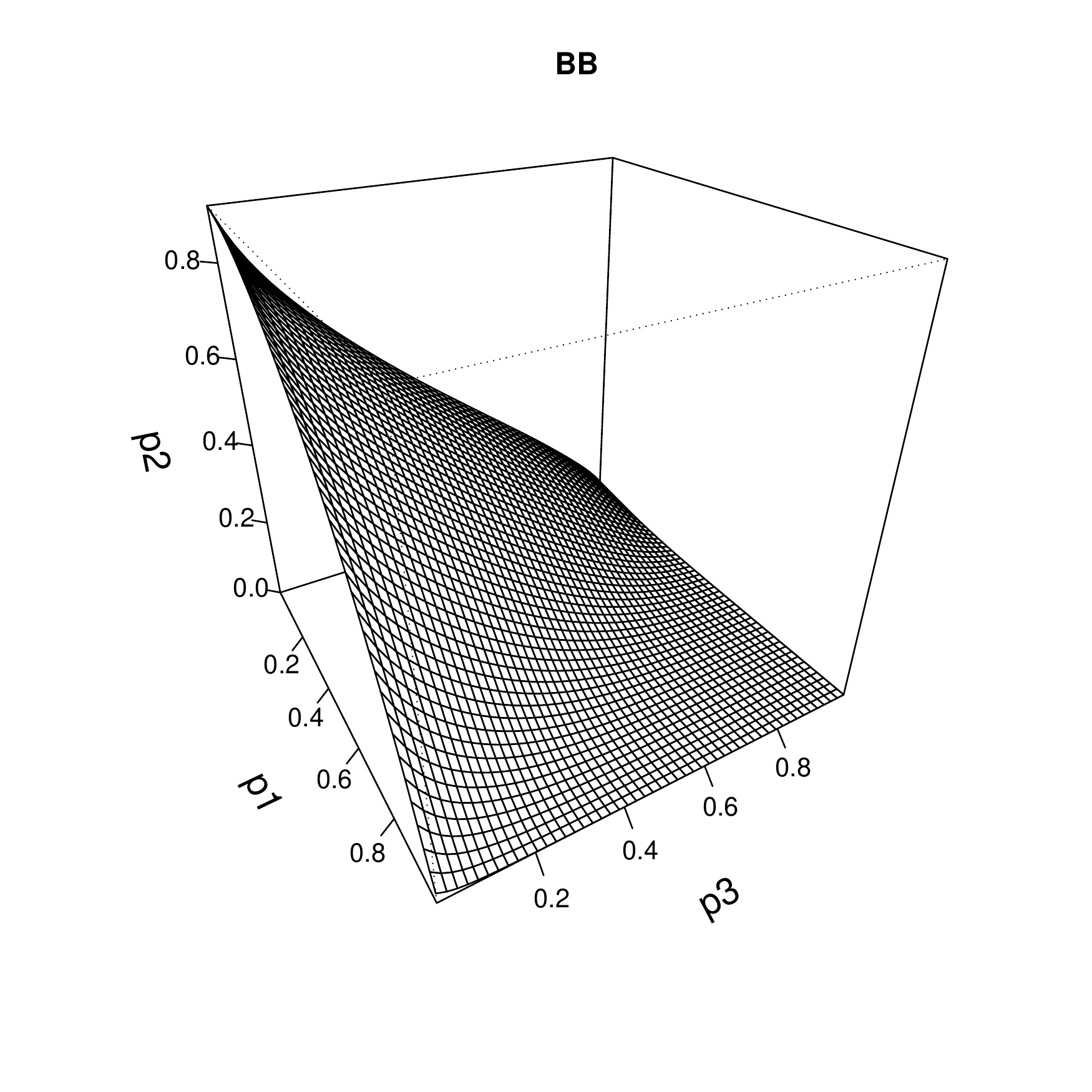}}
\subfigure{\includegraphics[page = 1, width=4.5cm]{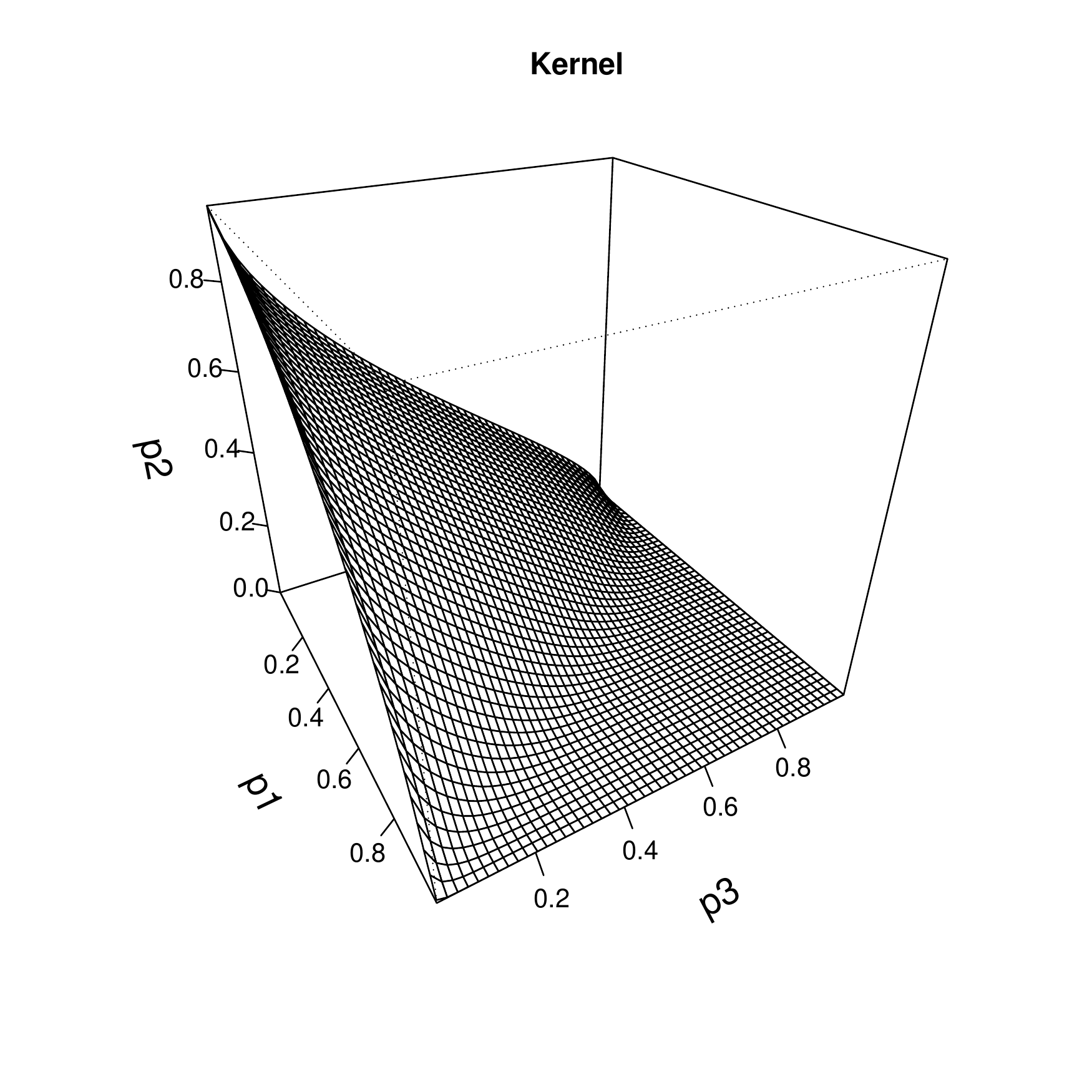}}
\subfigure{\includegraphics[page = 1, width=4.5cm]{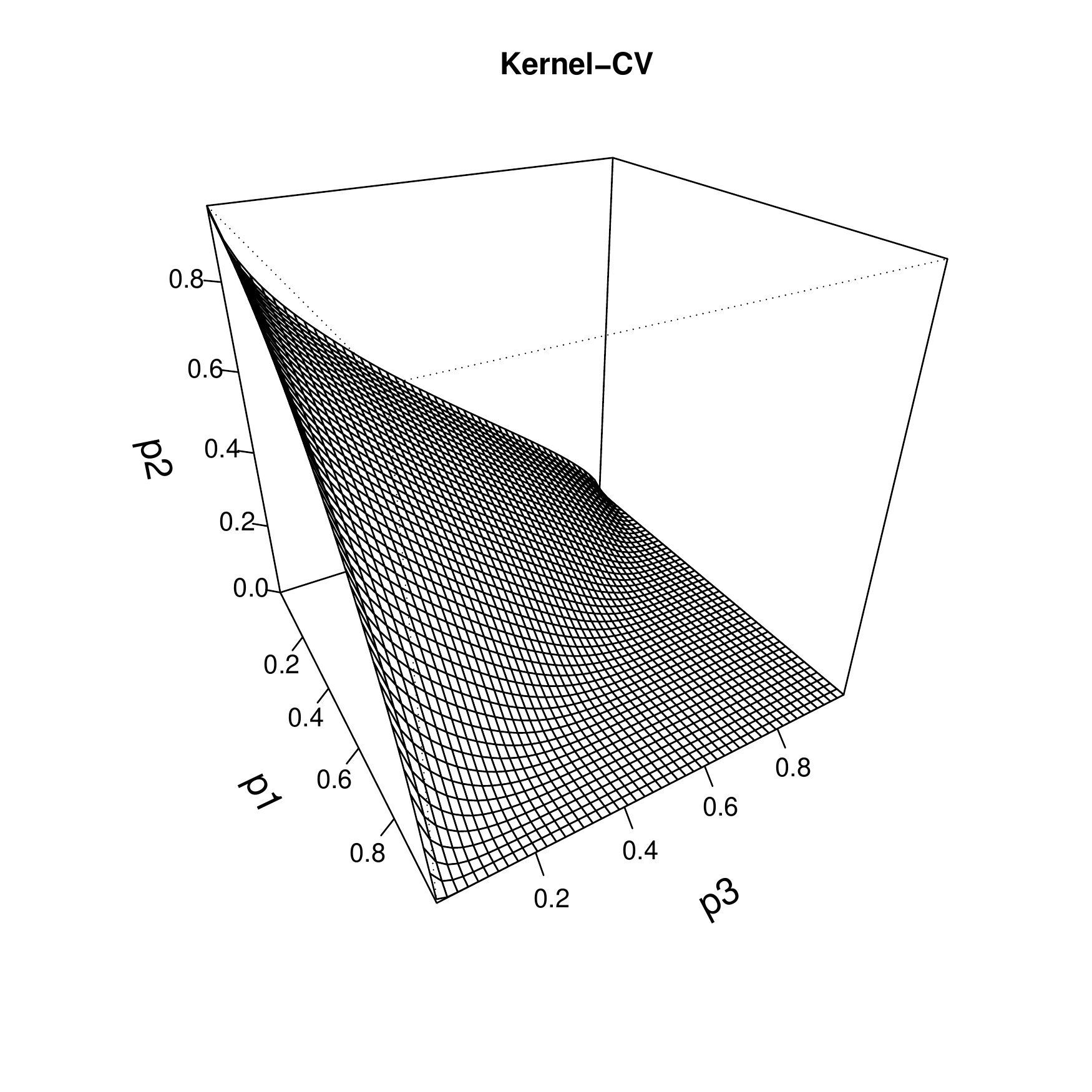}}
\subfigure{\includegraphics[page = 1, width=4.5cm]{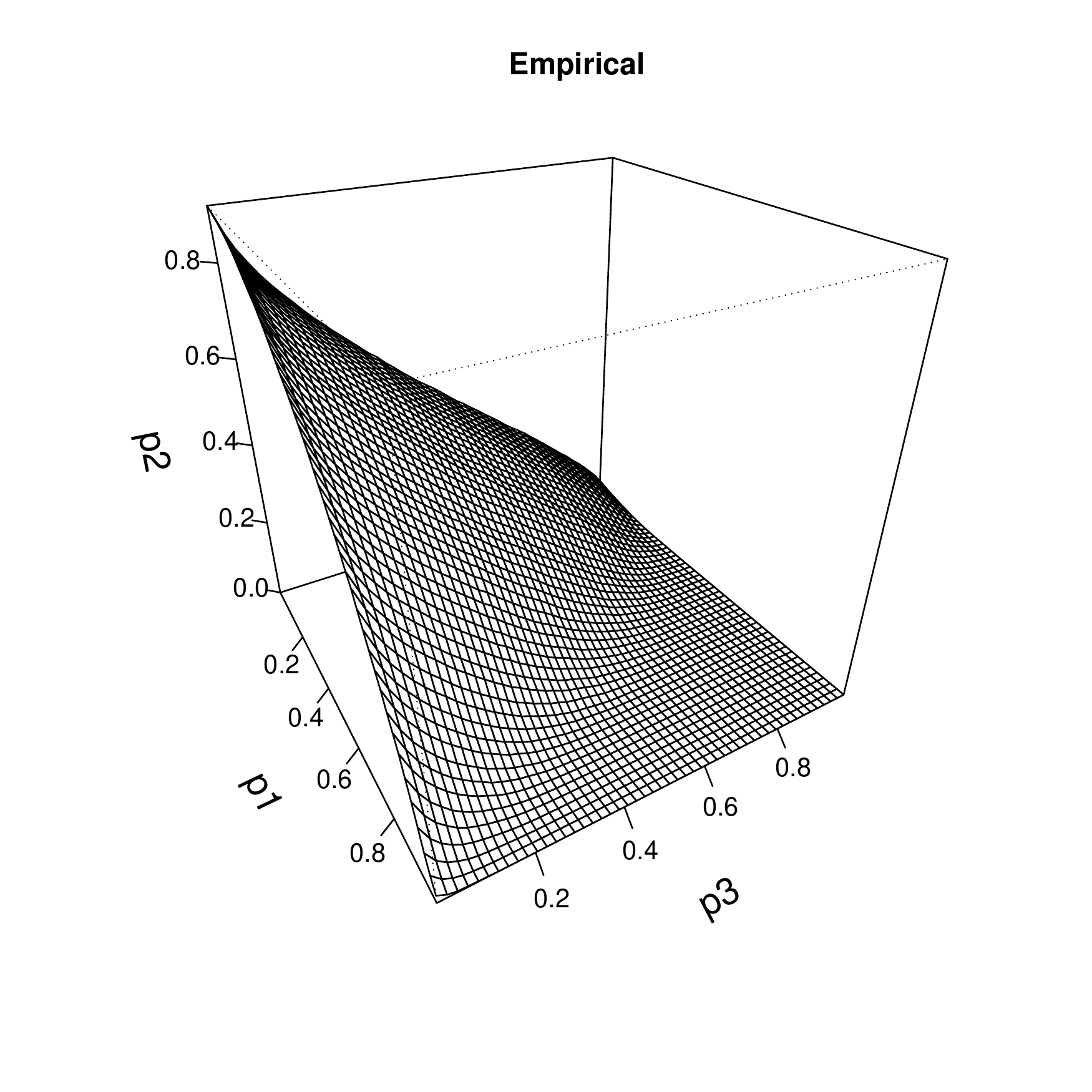}}\\
\subfigure{\includegraphics[page = 1, width=4.5cm]{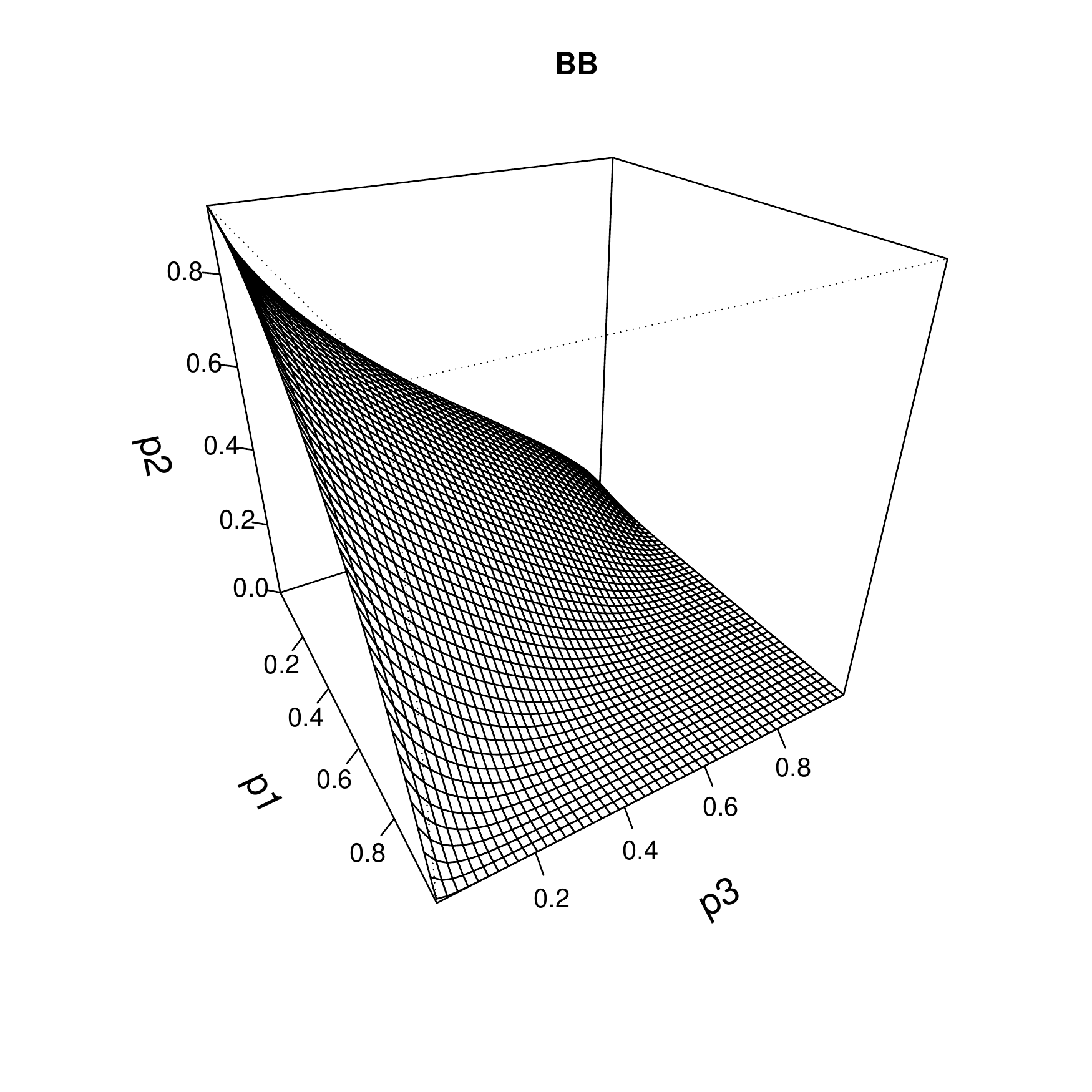}}
\subfigure{\includegraphics[page = 1, width=4.5cm]{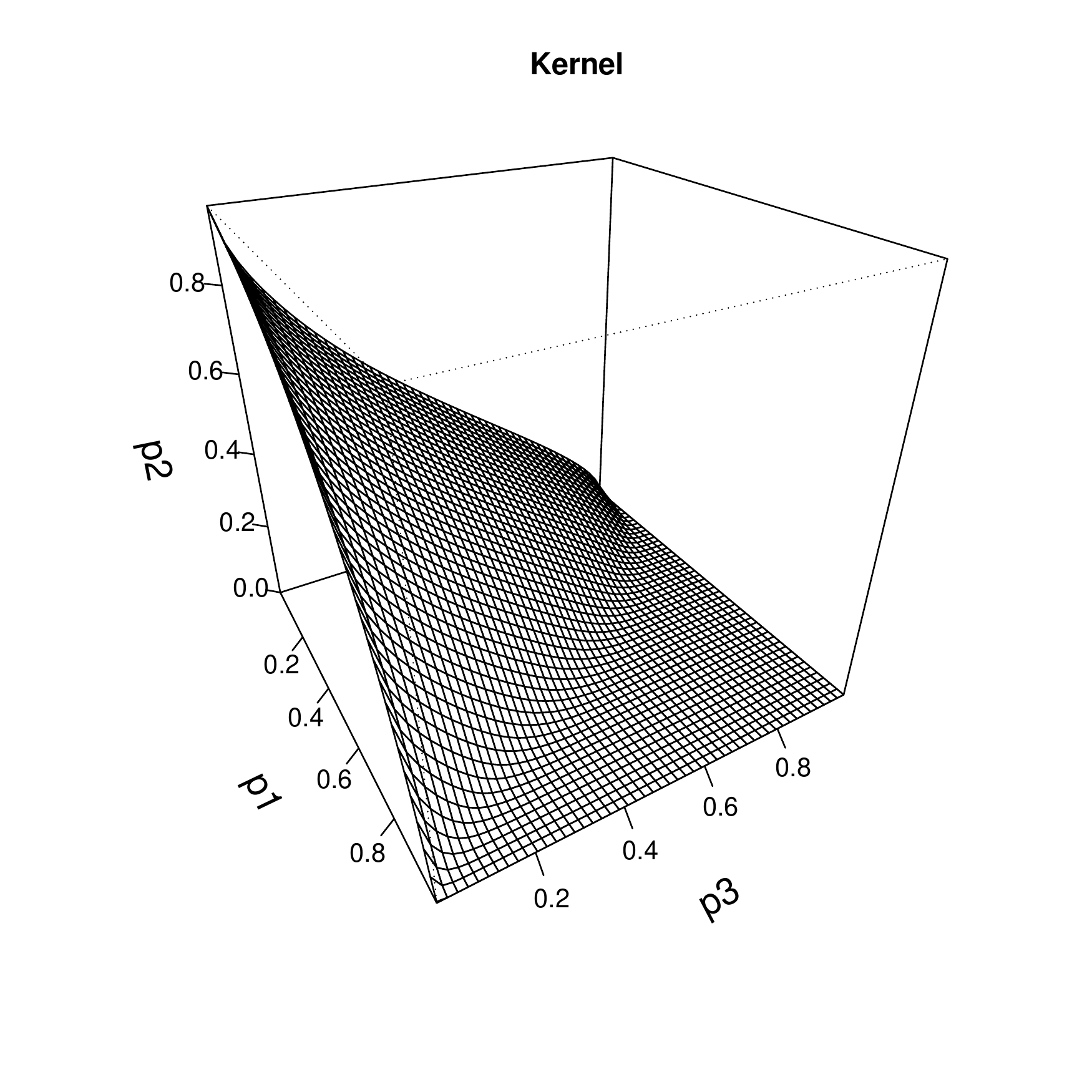}}
\subfigure{\includegraphics[page = 1, width=4.5cm]{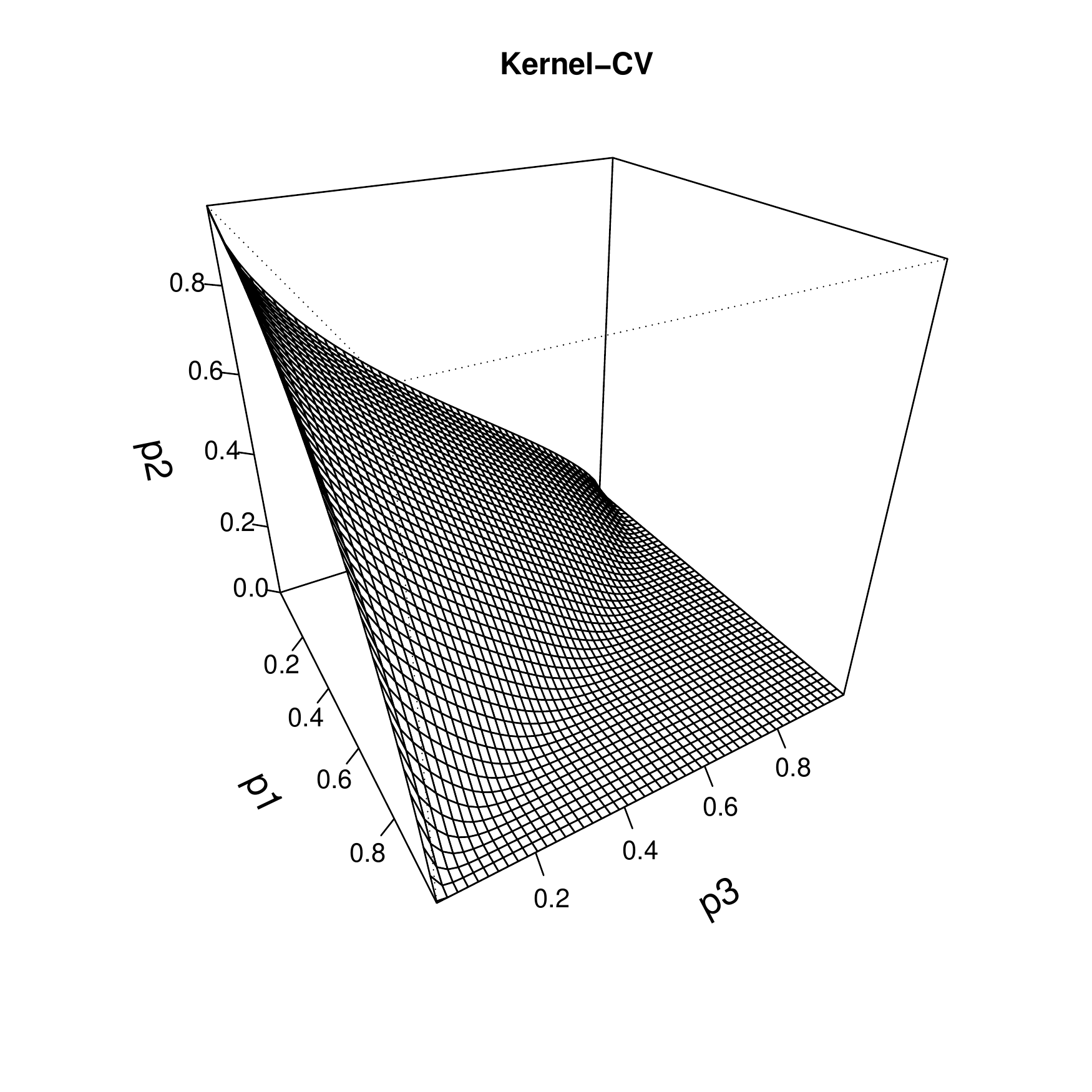}}
\subfigure{\includegraphics[page = 1, width=4.5cm]{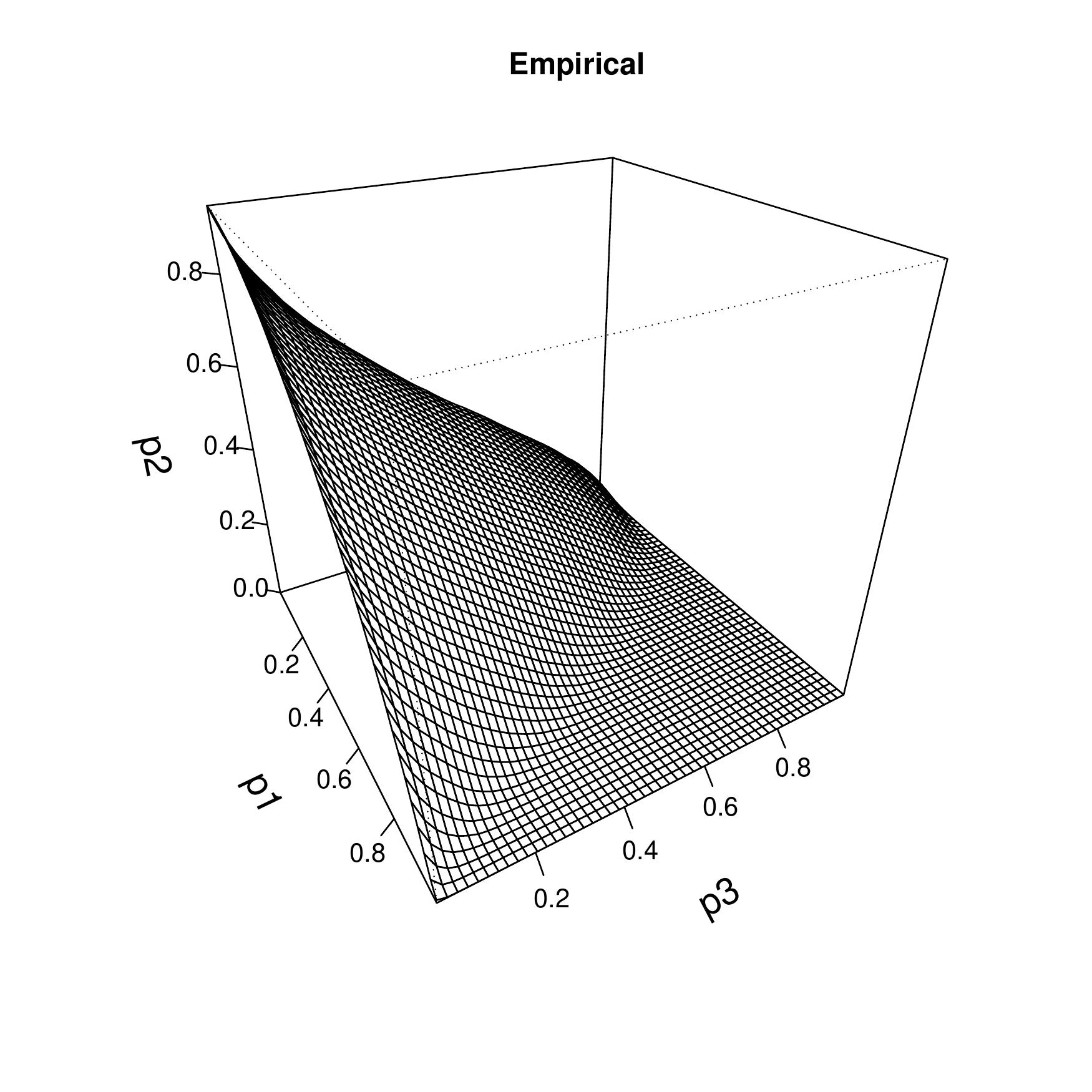}}\\
\subfigure{\includegraphics[page = 1, width=4.5cm]{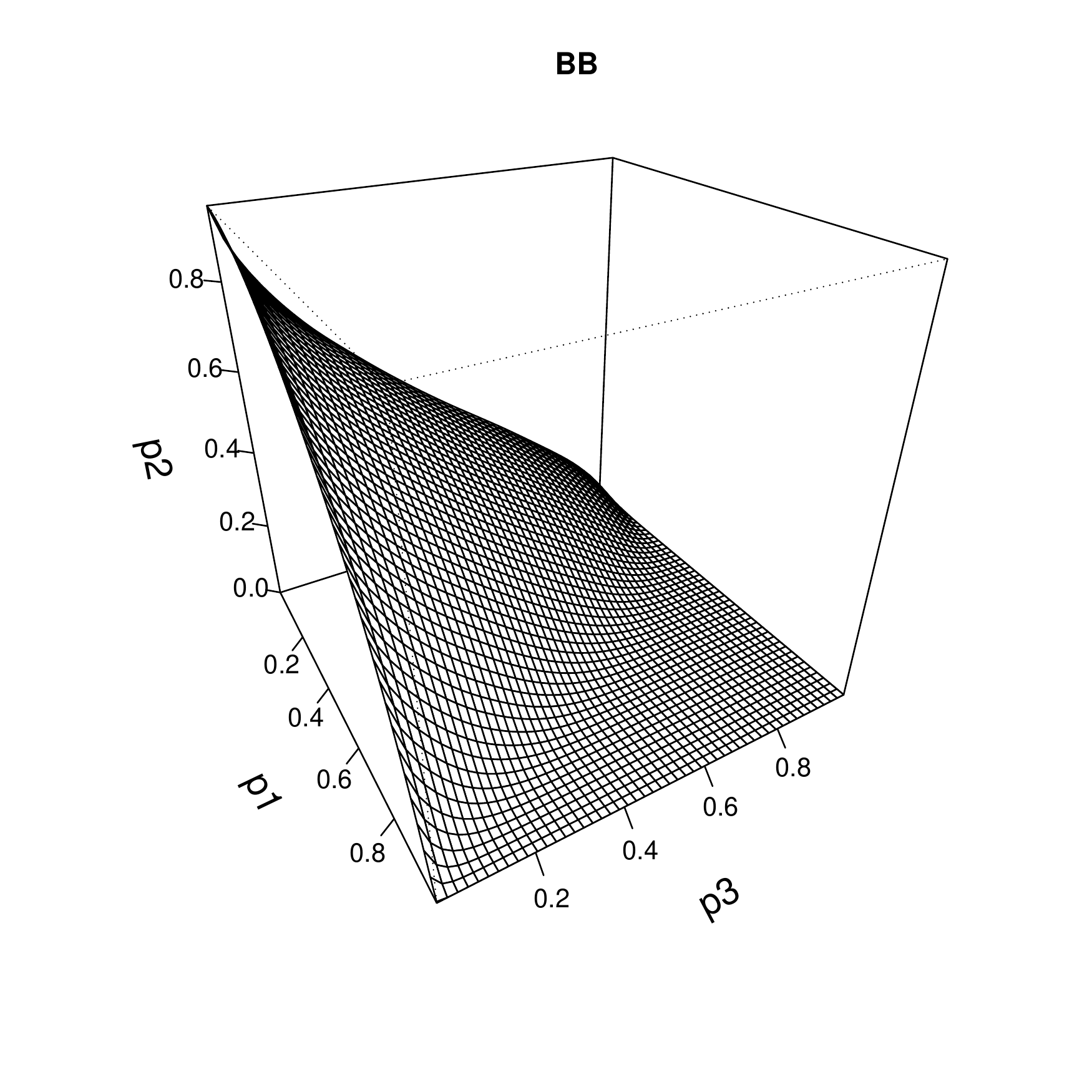}}
\subfigure{\includegraphics[page = 1, width=4.5cm]{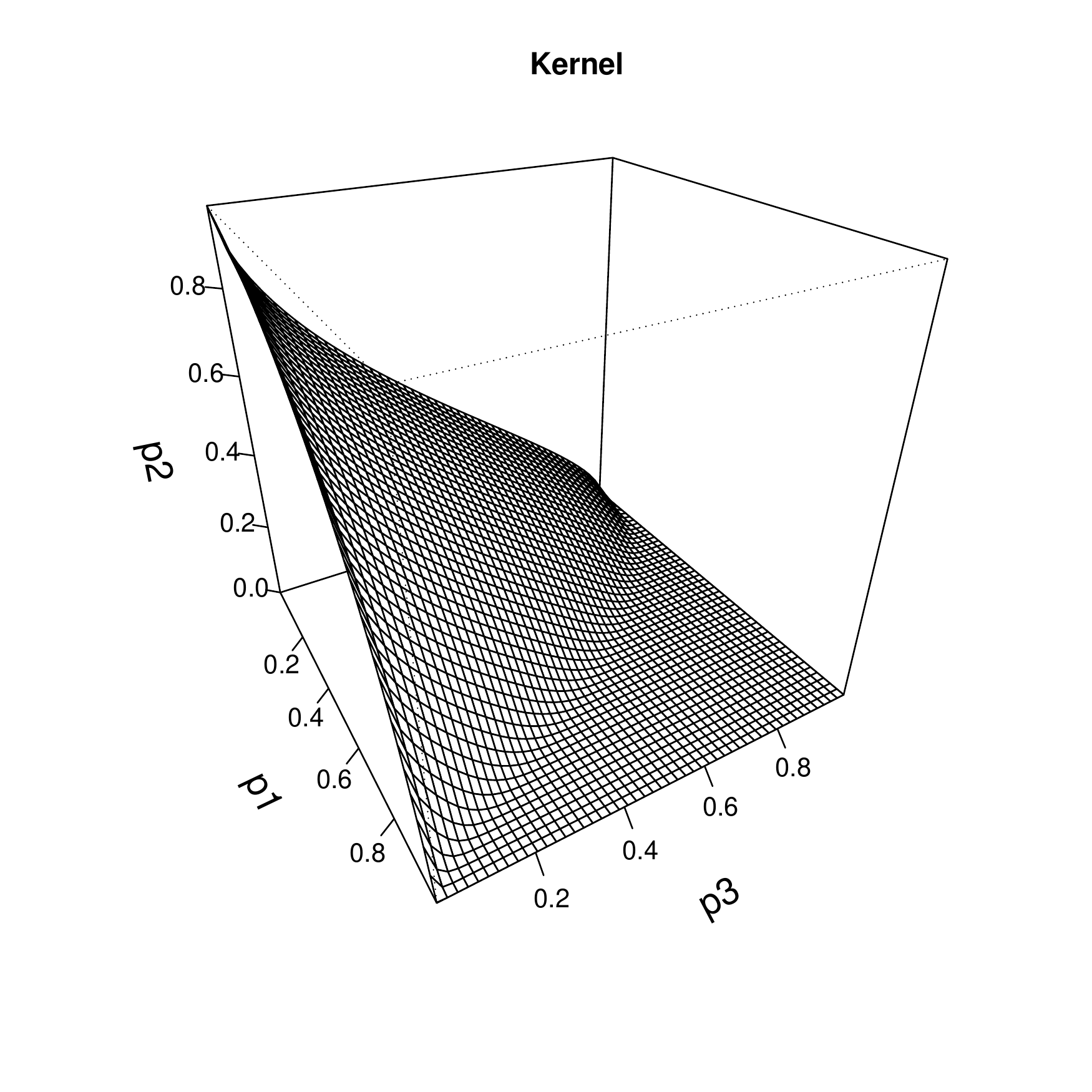}}
\subfigure{\includegraphics[page = 1, width=4.5cm]{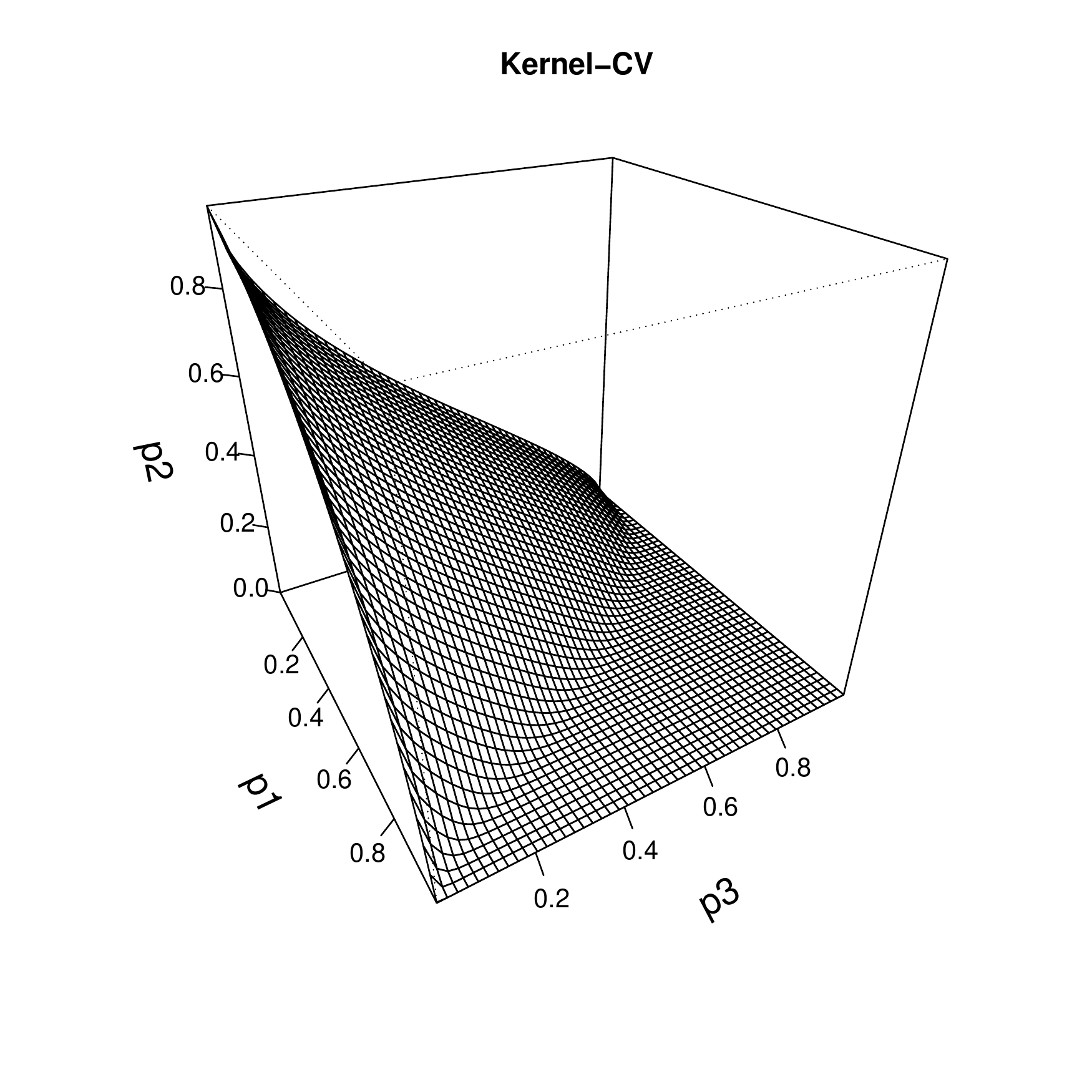}}
\subfigure{\includegraphics[page = 1, width=4.5cm]{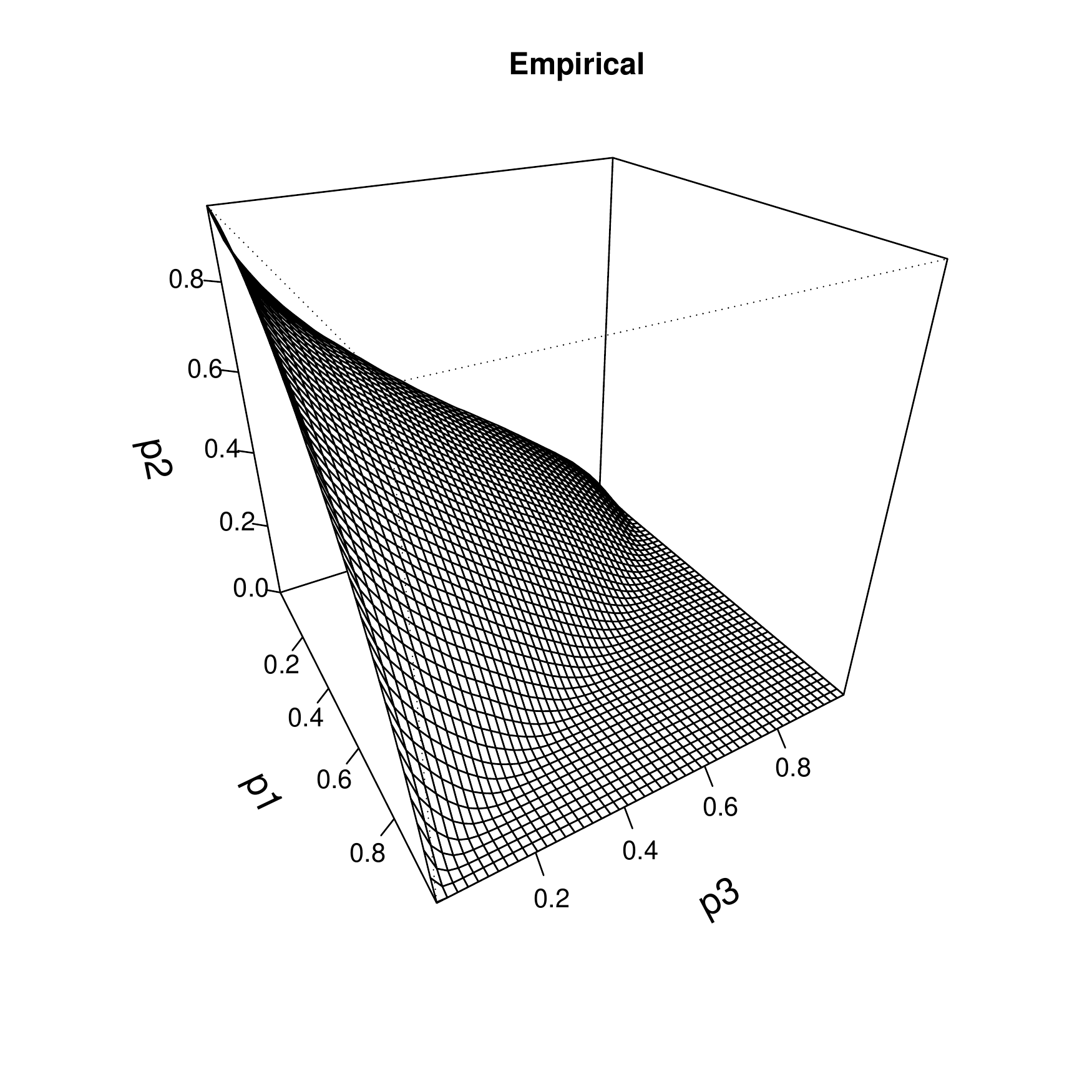}}
\caption{\footnotesize{Scenario 2. True ROC surface and mean across the 300 estimated ROC surfaces. First row: $n_1=n_2=n_3$=50. Second row: $n_1=n_2=n_3$=100. Third row: $n_1=n_2=n_3$=200. Here Kernel denotes the kernel estimate with bandwidth calculated using equation (2) of the main manuscript and Kernel-CV stands for the kernel estimate with the bandwidth selected by least squares cross-validation.}}
\end{center}
\end{figure}

\begin{figure}[H]
\begin{center}
\subfigure{\includegraphics[page = 1, width=4.75cm]{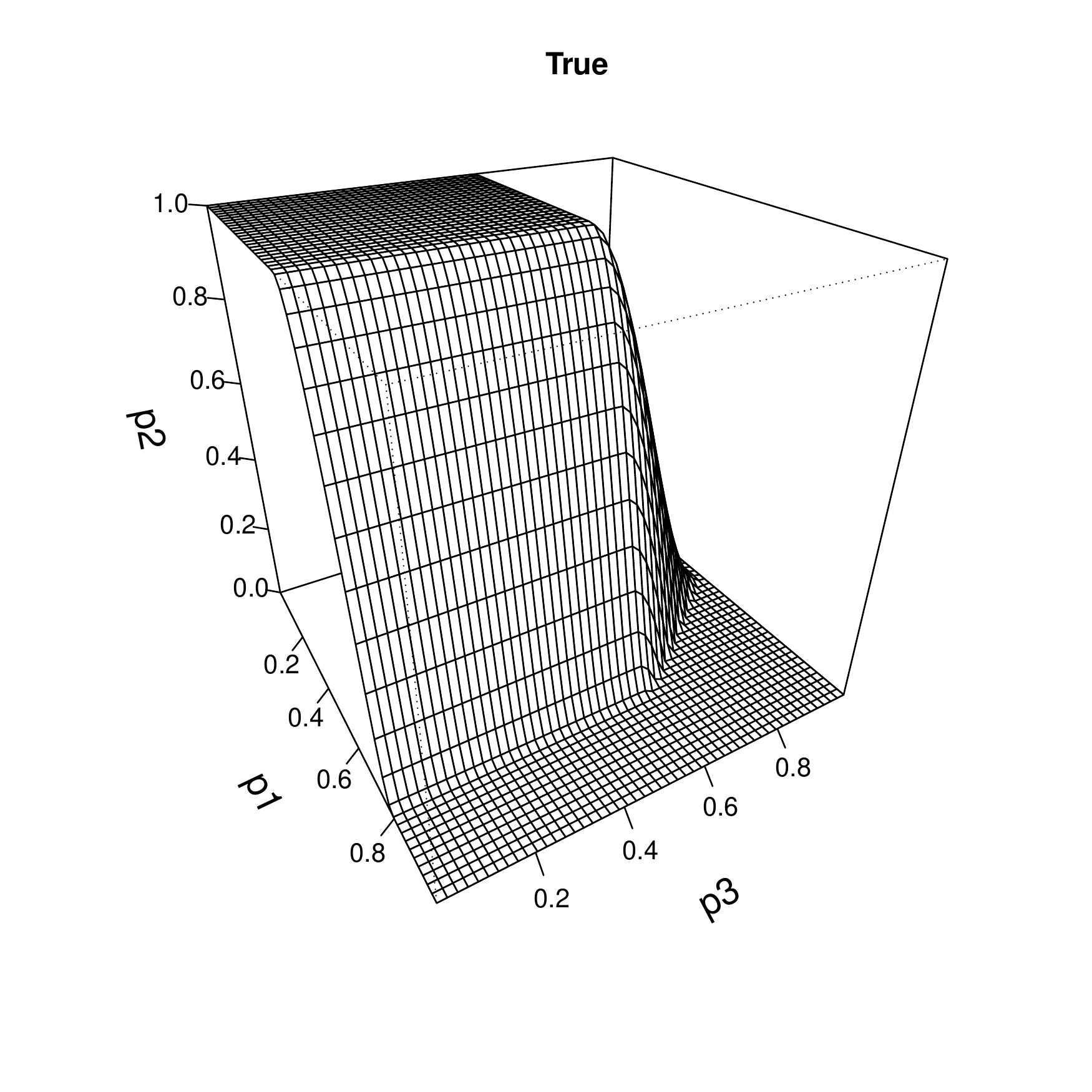}}\\
\subfigure{\includegraphics[page = 1, width=4.5cm]{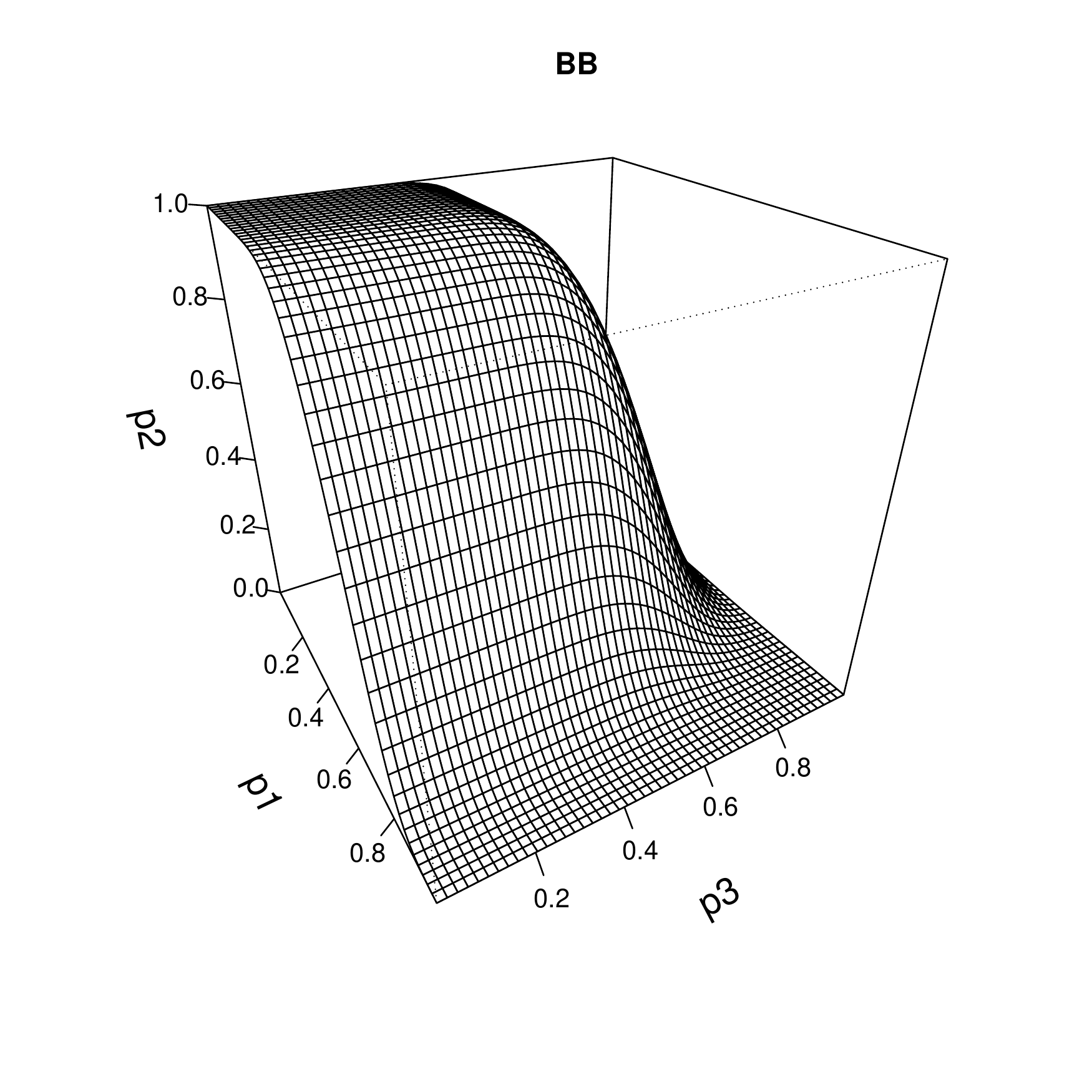}}
\subfigure{\includegraphics[page = 1, width=4.5cm]{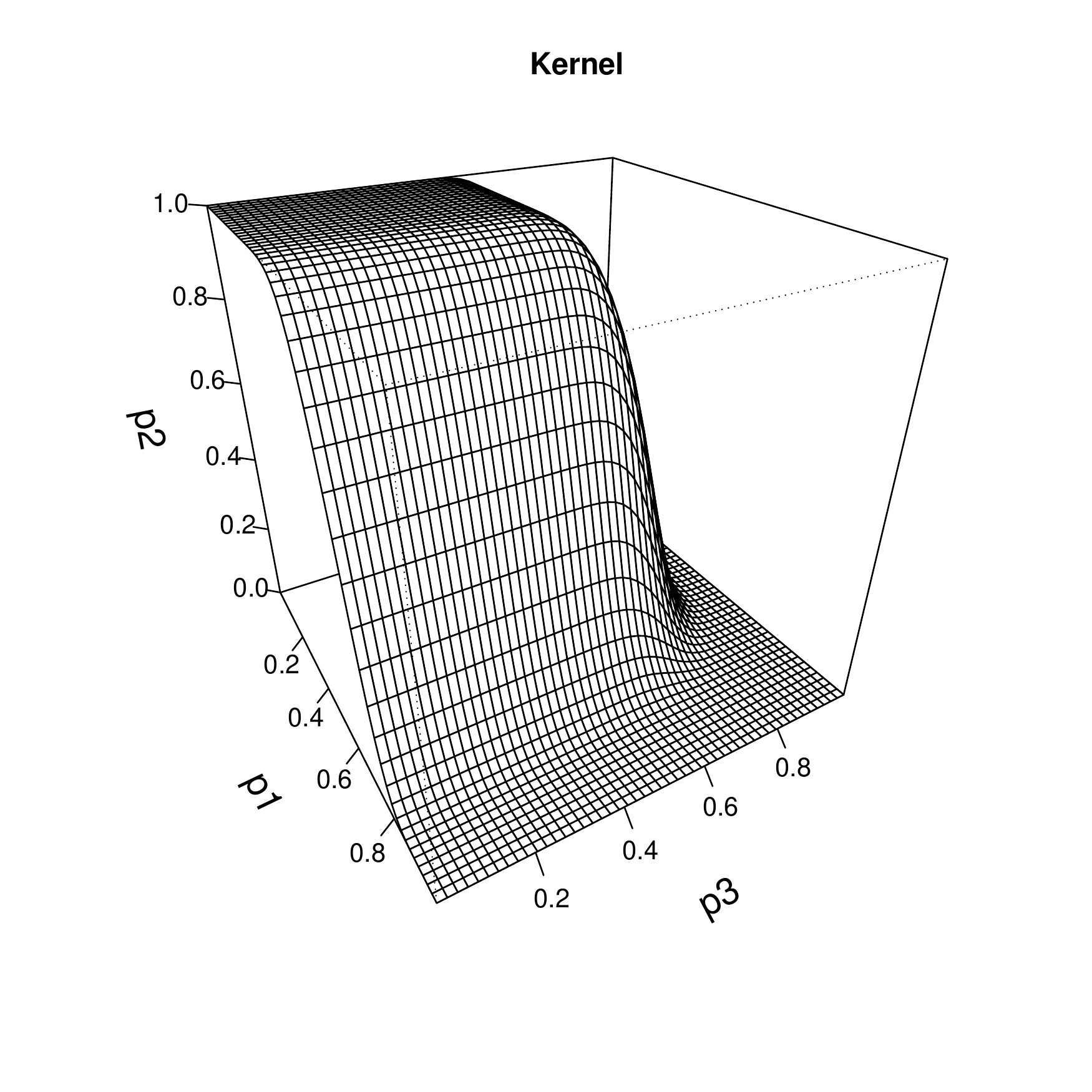}}
\subfigure{\includegraphics[page = 1, width=4.5cm]{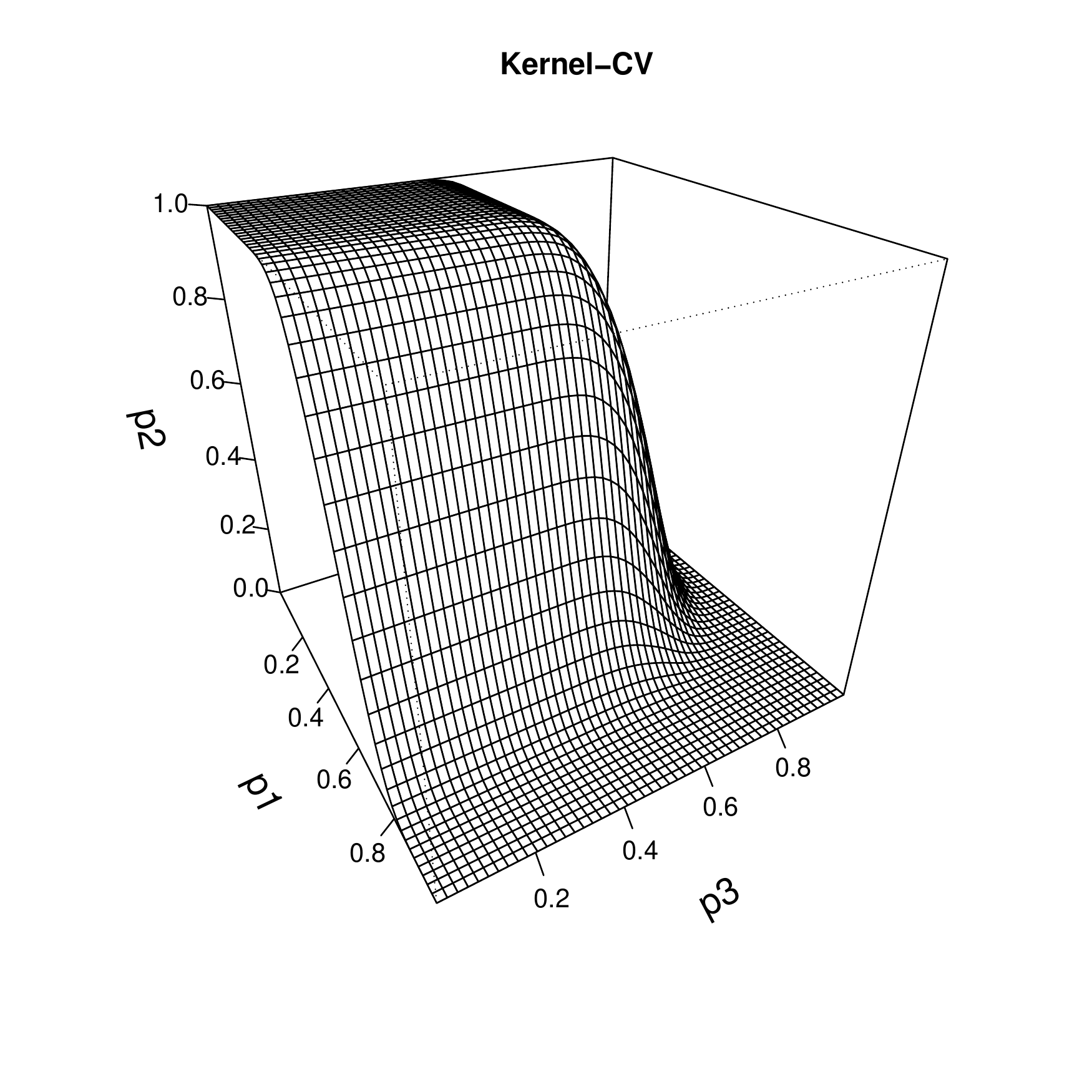}}
\subfigure{\includegraphics[page = 1, width=4.5cm]{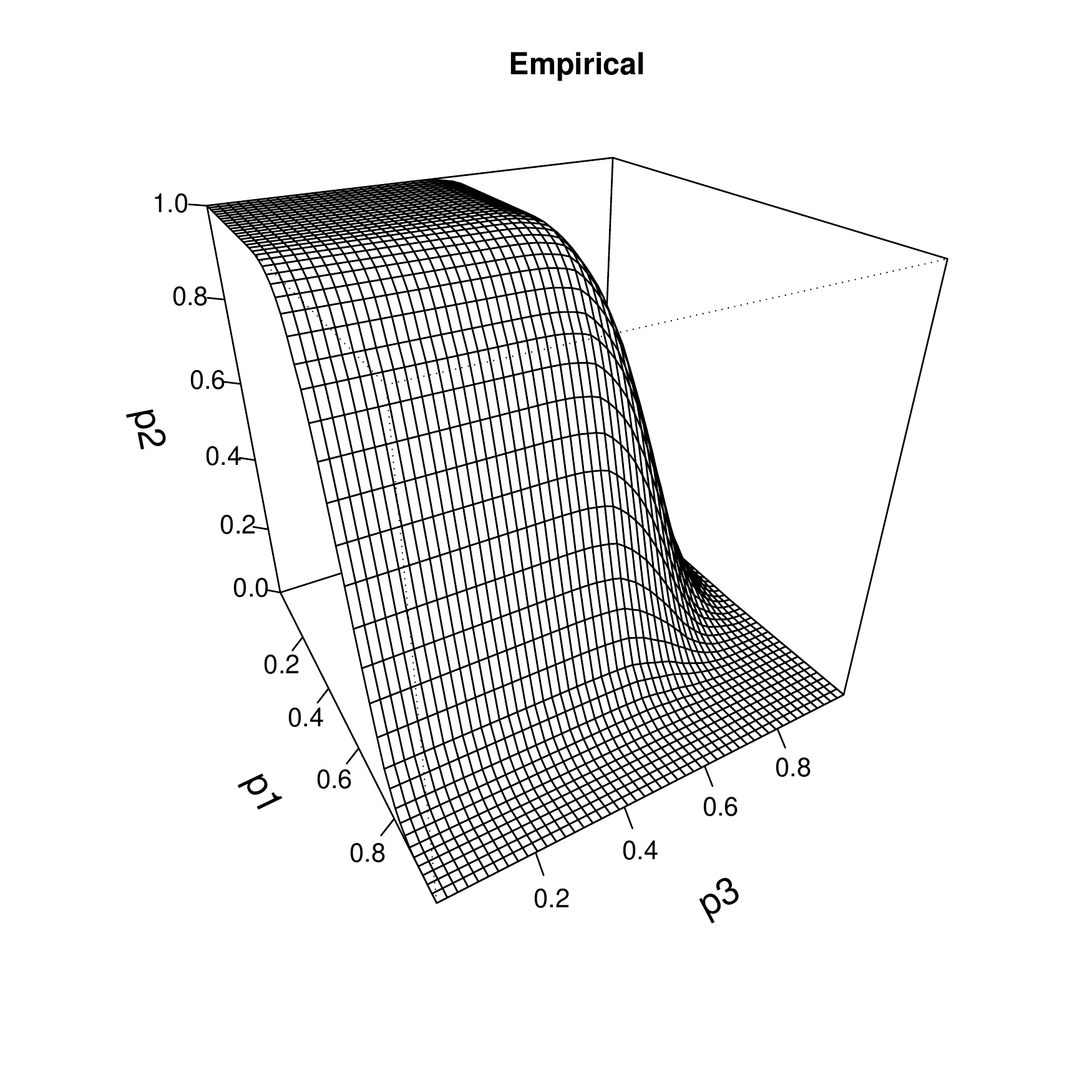}}\\
\subfigure{\includegraphics[page = 1, width=4.5cm]{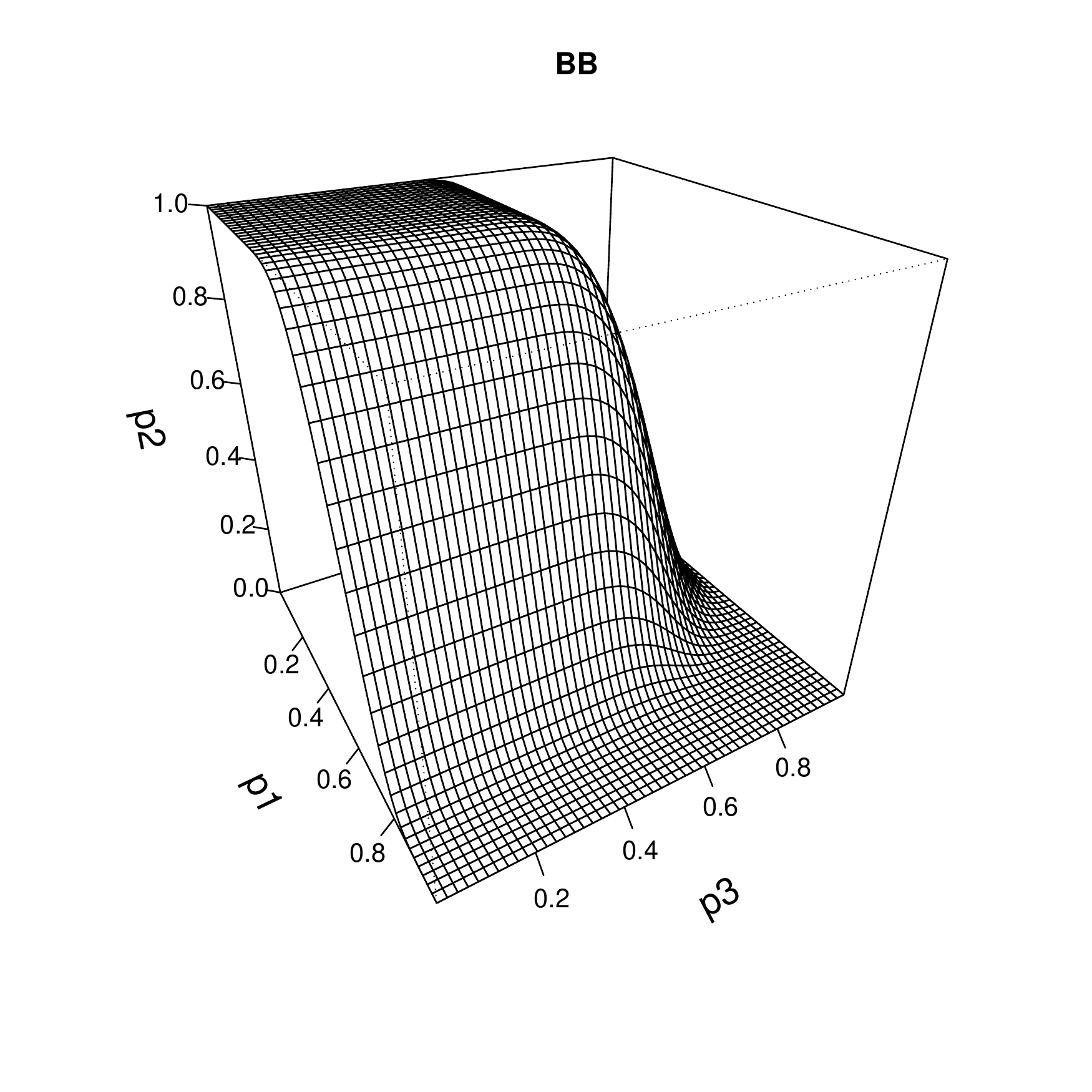}}
\subfigure{\includegraphics[page = 1, width=4.5cm]{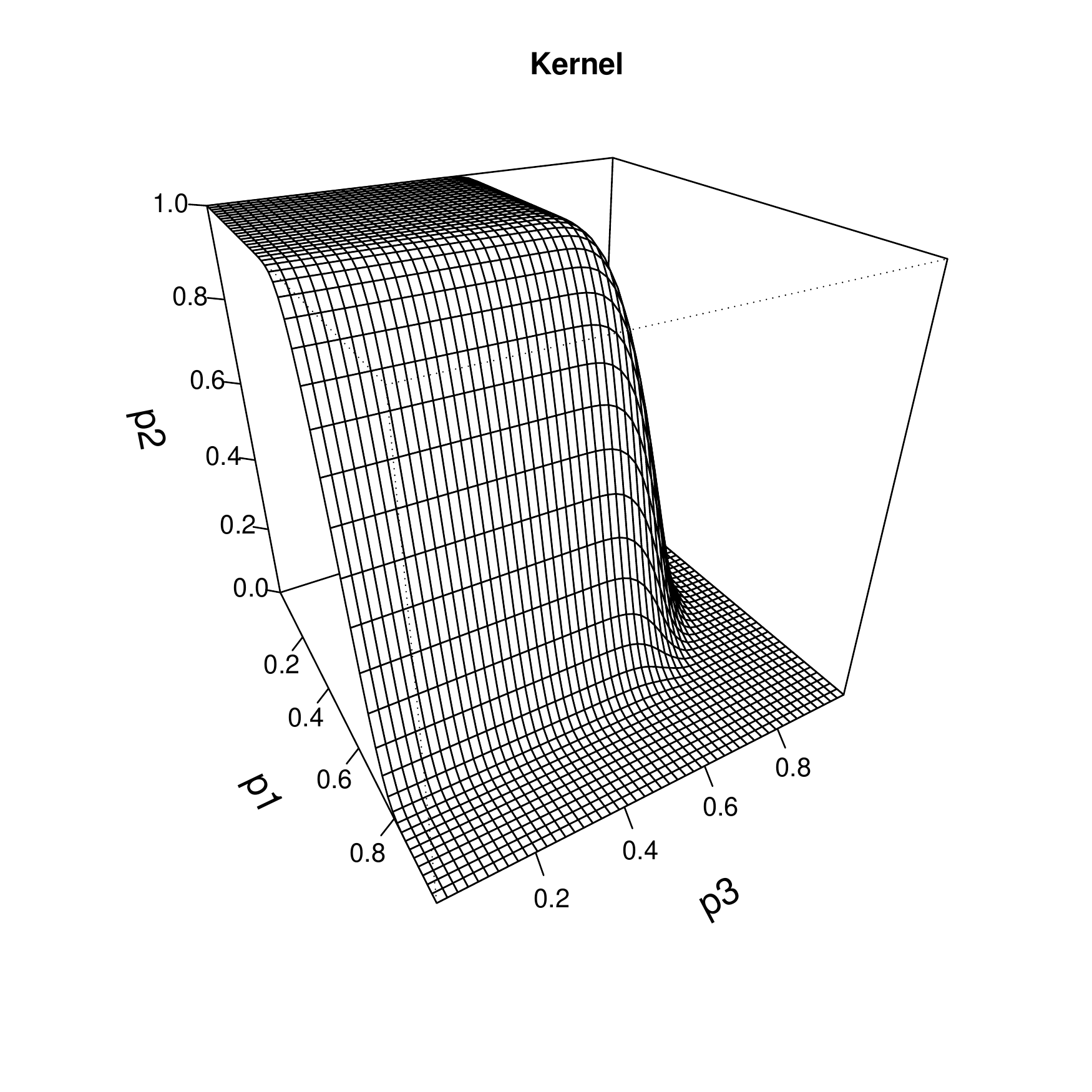}}
\subfigure{\includegraphics[page = 1, width=4.5cm]{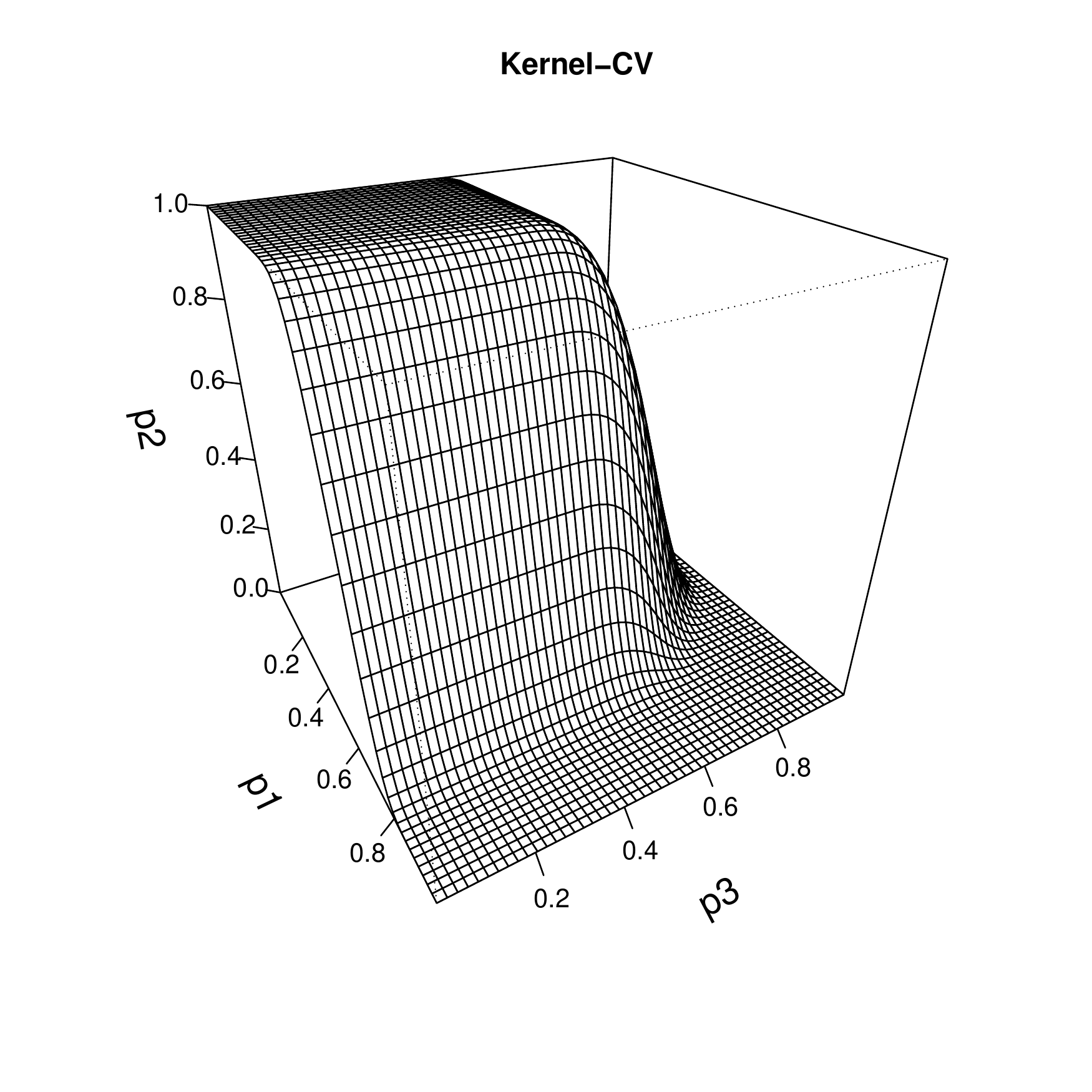}}
\subfigure{\includegraphics[page = 1, width=4.5cm]{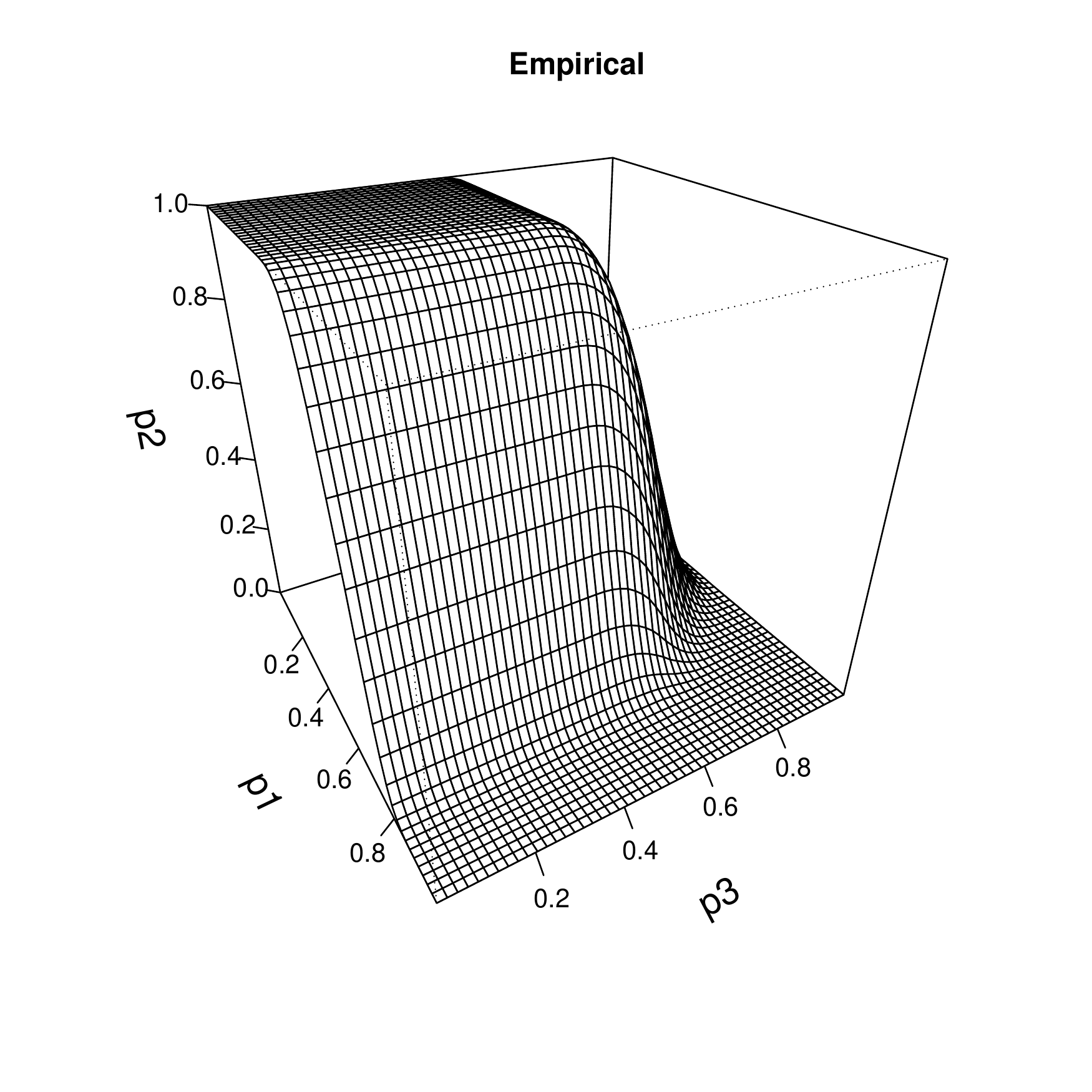}}\\
\subfigure{\includegraphics[page = 1, width=4.5cm]{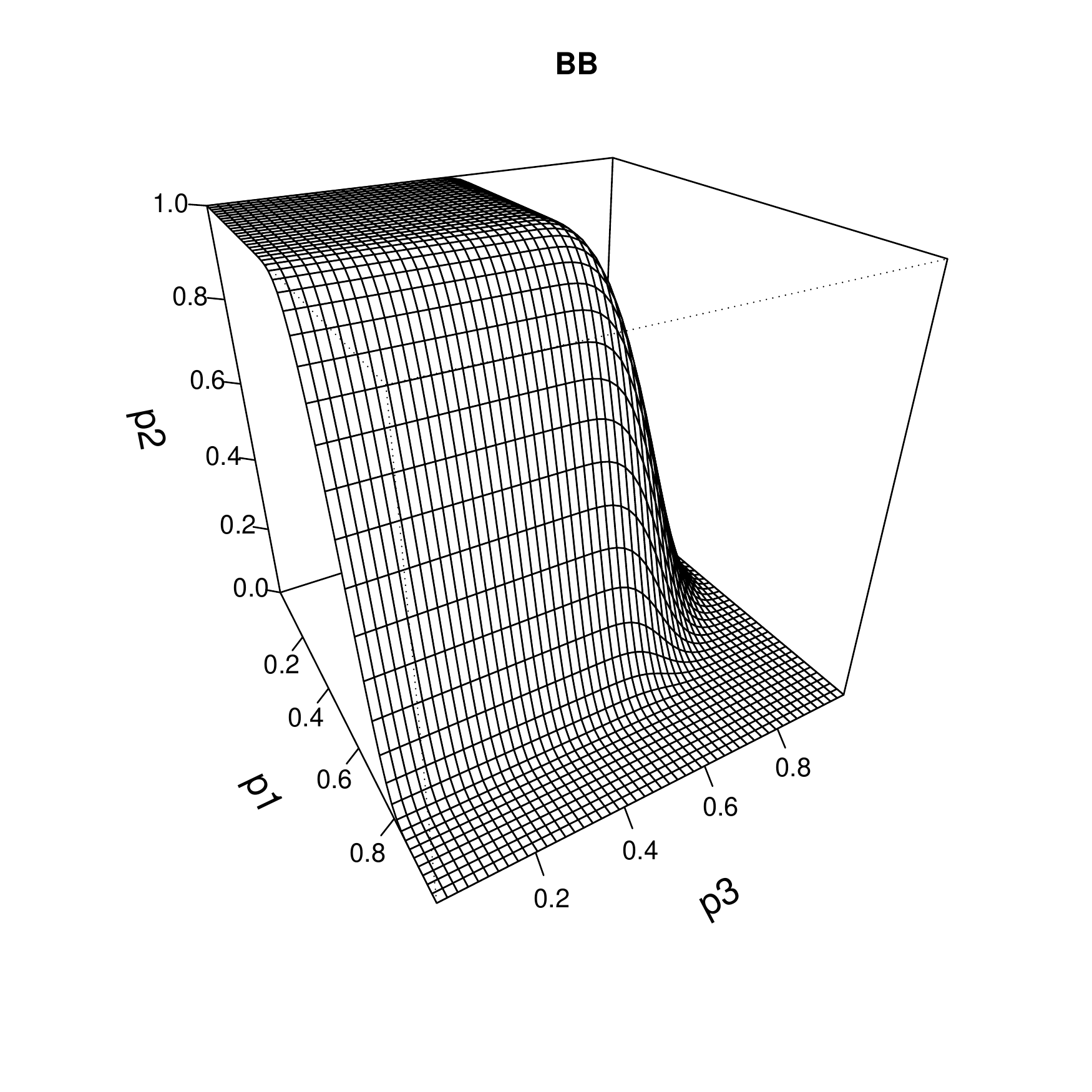}}
\subfigure{\includegraphics[page = 1, width=4.5cm]{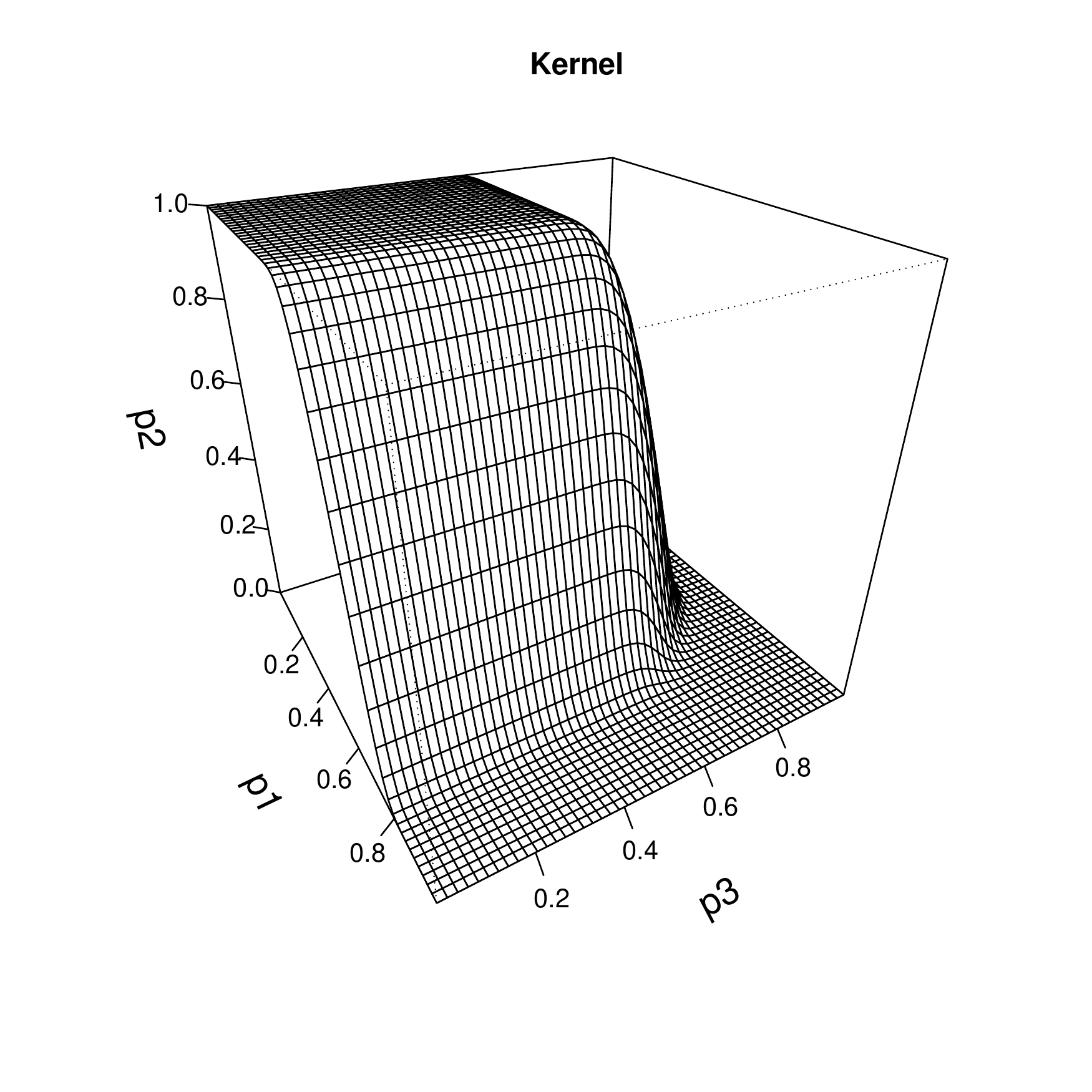}}
\subfigure{\includegraphics[page = 1, width=4.5cm]{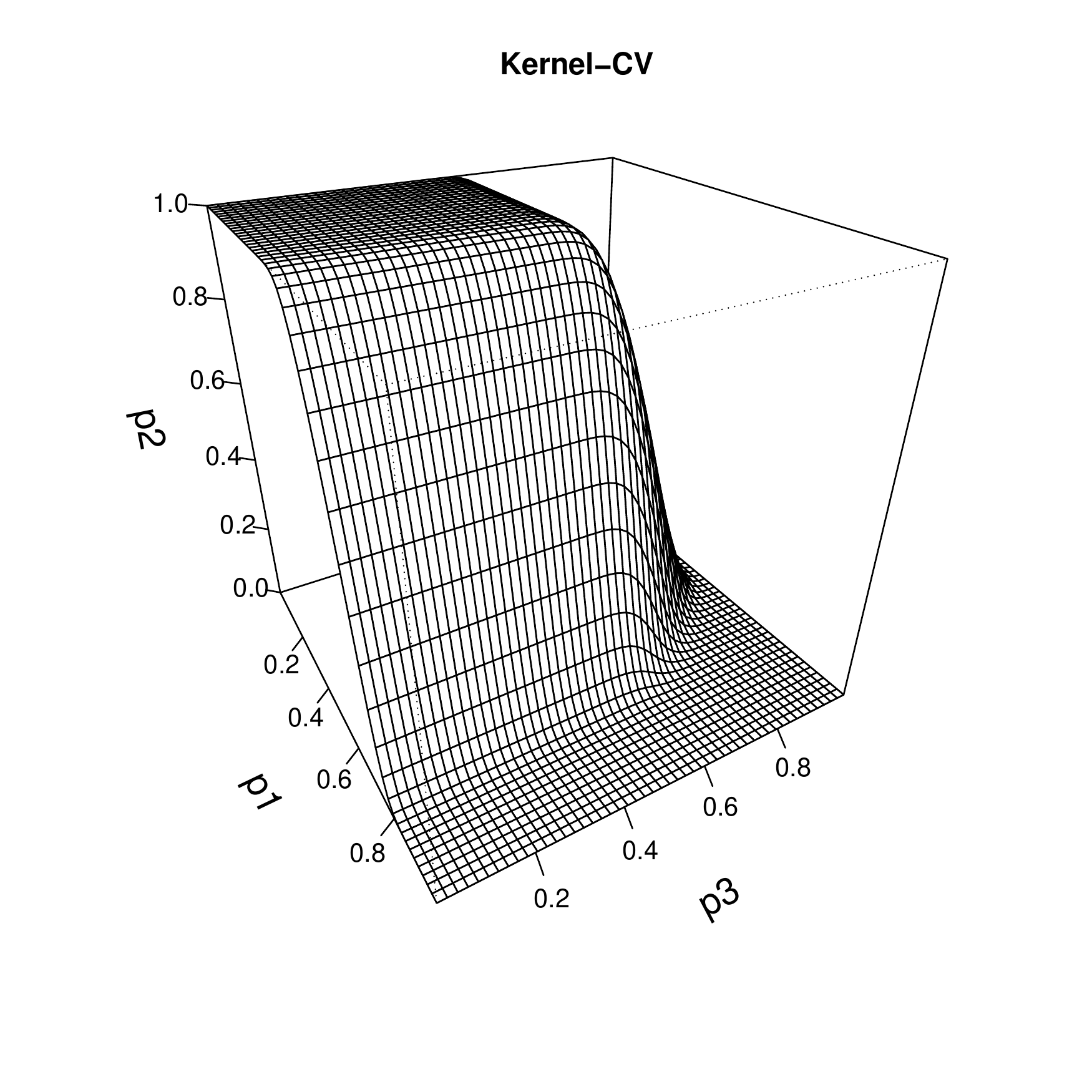}}
\subfigure{\includegraphics[page = 1, width=4.5cm]{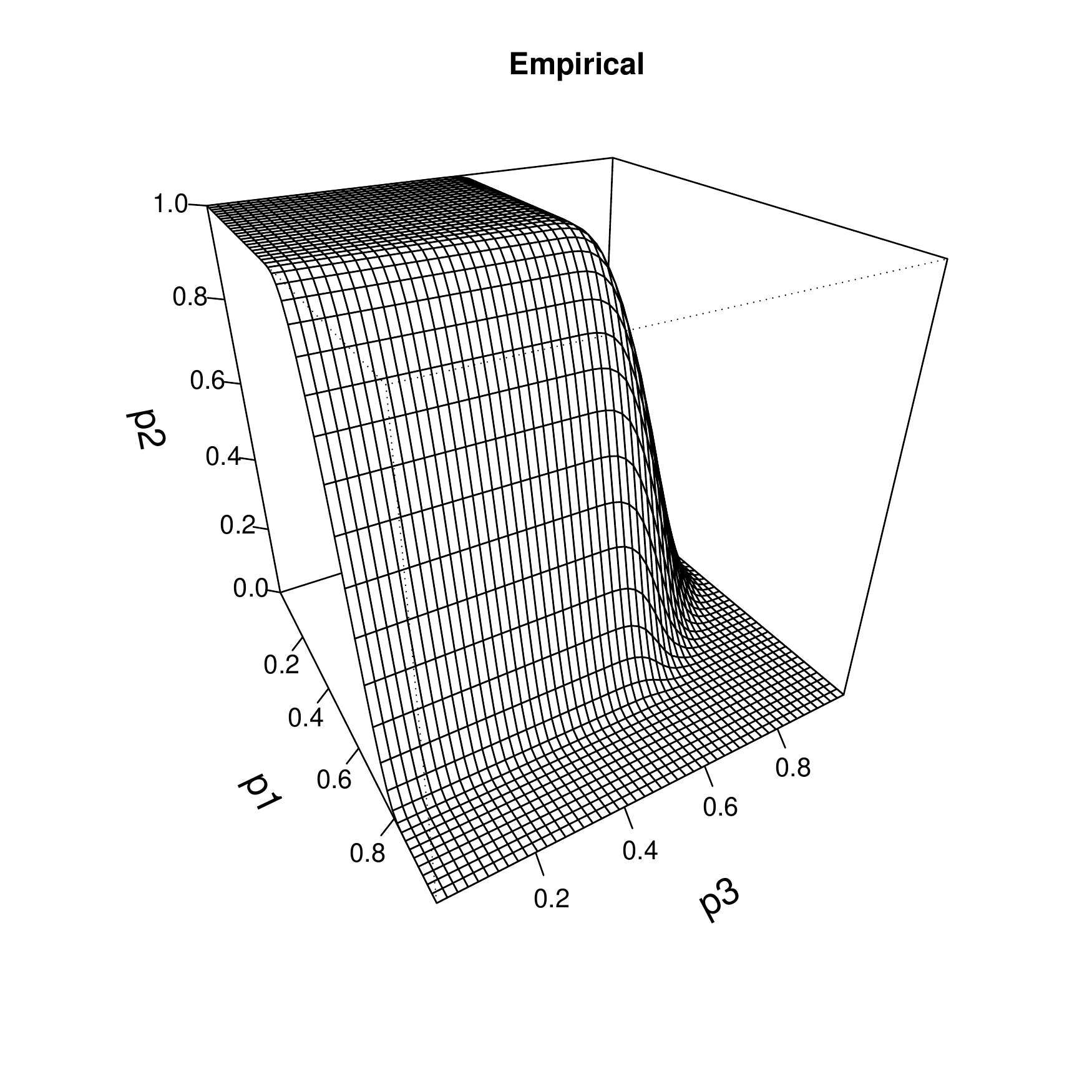}}
\caption{\footnotesize{Scenario 3. True ROC surface and mean across the 300 estimated ROC surfaces. First row: $n_1=n_2=n_3$=50. Second row: $n_1=n_2=n_3$=100. Third row: $n_1=n_2=n_3$=200. Here Kernel denotes the kernel estimate with bandwidth calculated using equation (2) of the main manuscript and Kernel-CV stands for the kernel estimate with the bandwidth selected by least squares cross-validation.}}
\end{center}
\end{figure}

\begin{figure}[H]
\begin{center}
\subfigure{\includegraphics[page = 1, width=4.75cm]{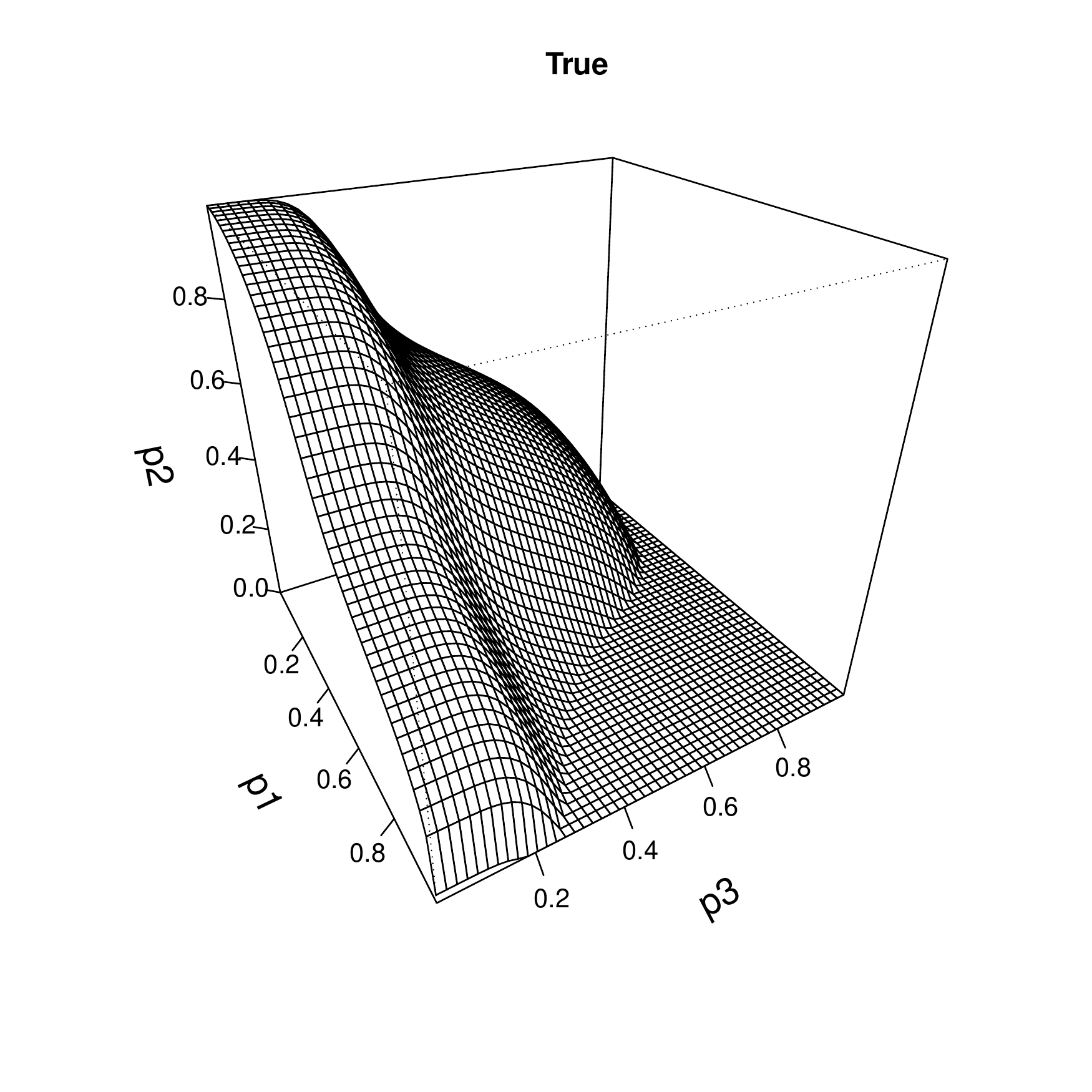}}\\
\subfigure{\includegraphics[page = 1, width=4.5cm]{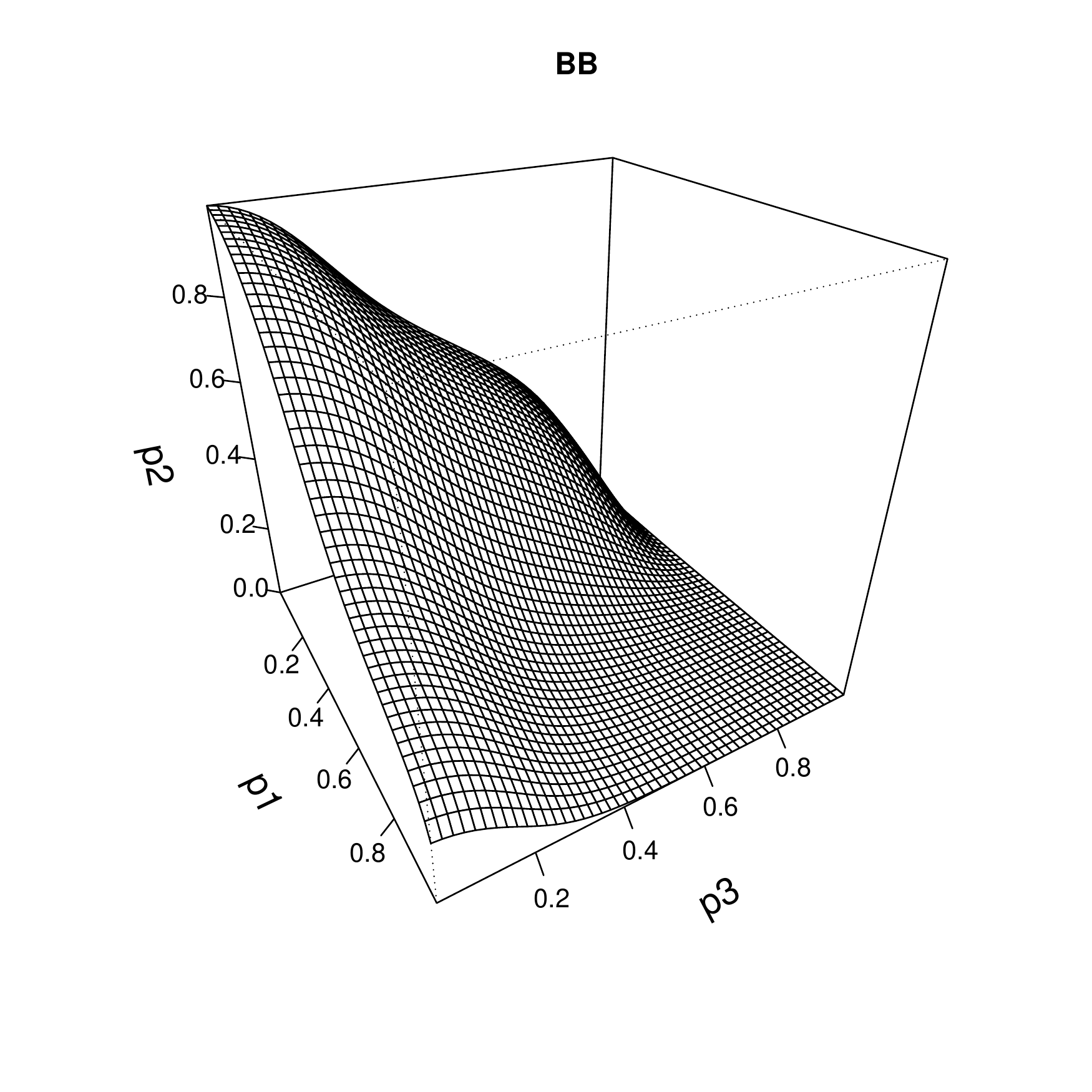}}
\subfigure{\includegraphics[page = 1, width=4.5cm]{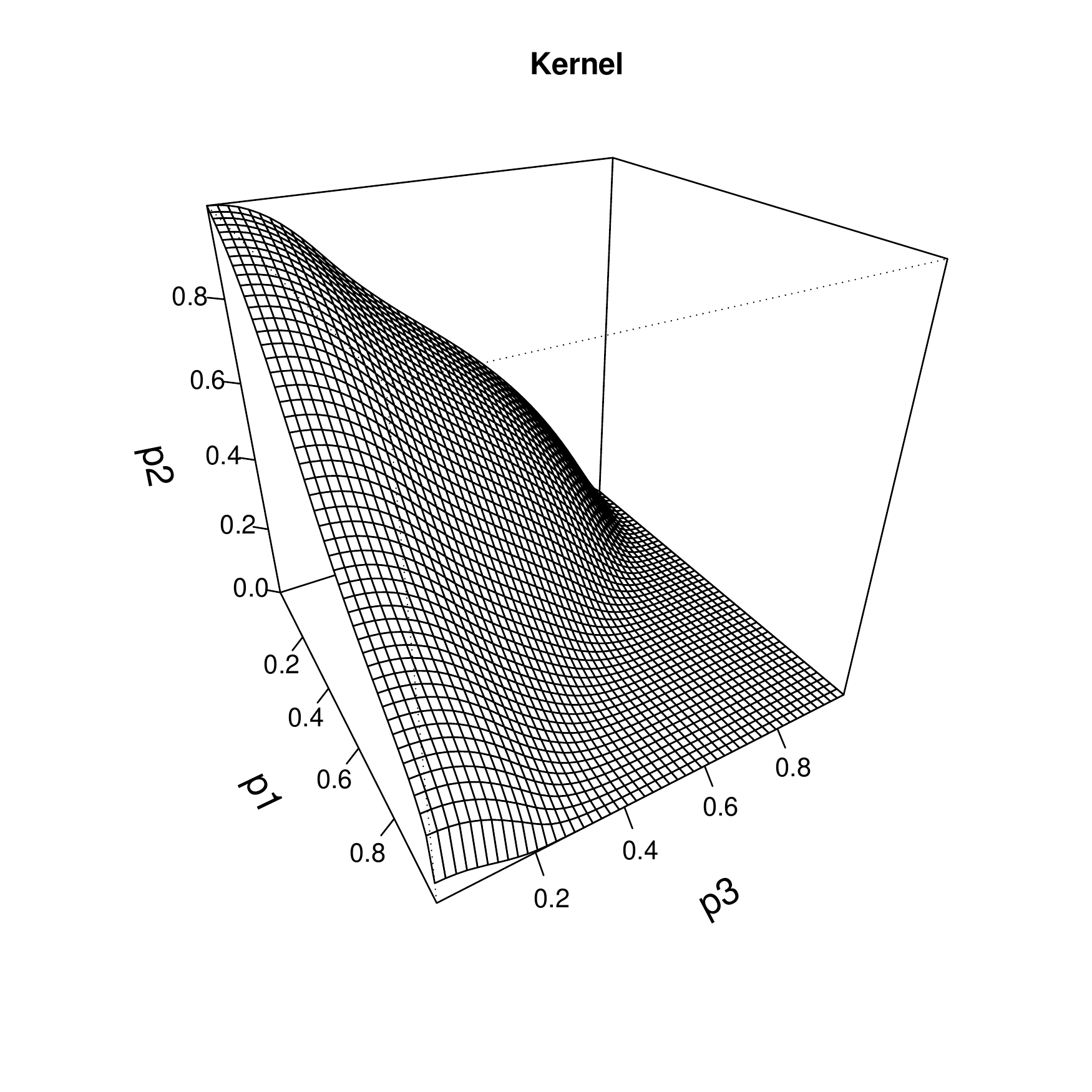}}
\subfigure{\includegraphics[page = 1, width=4.5cm]{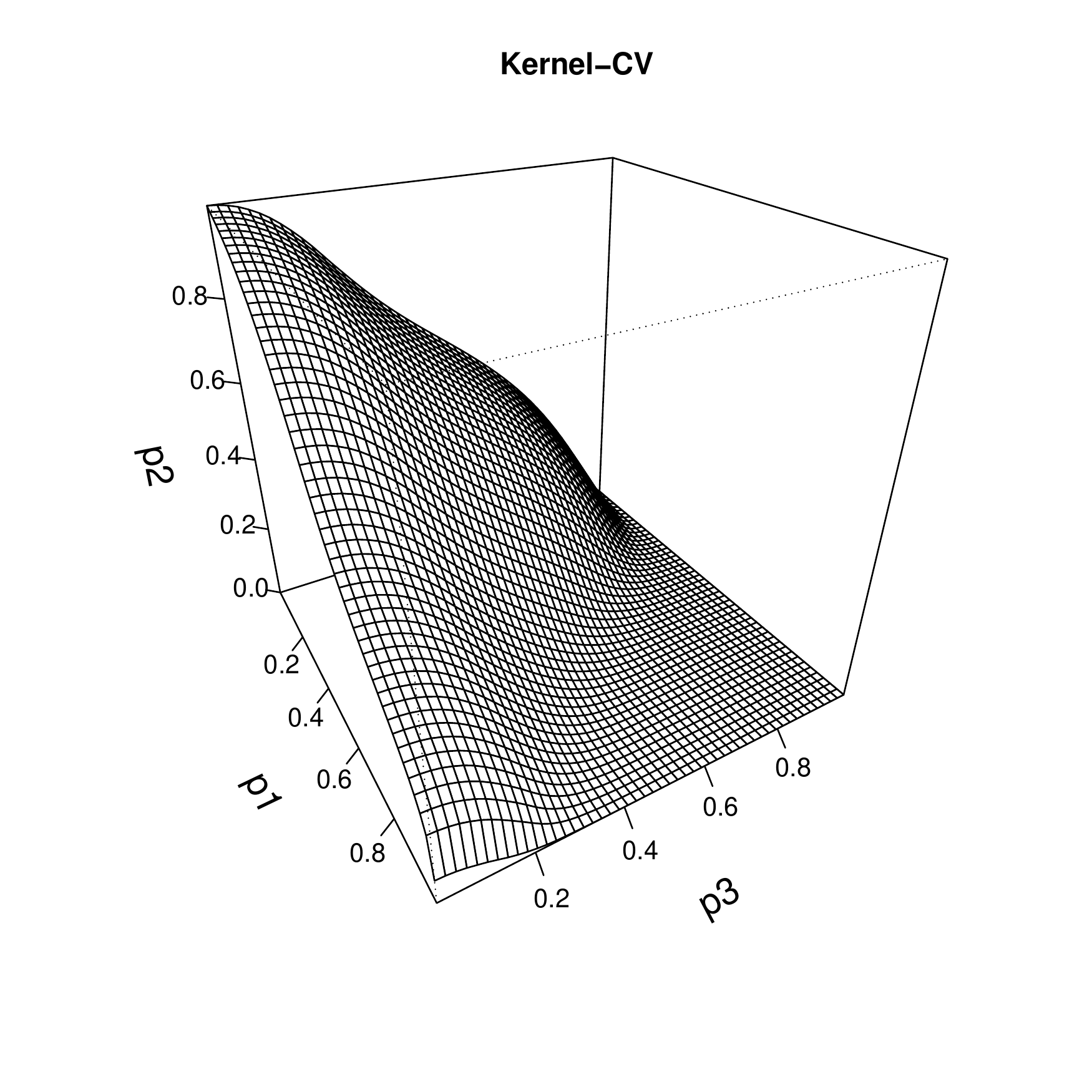}}
\subfigure{\includegraphics[page = 1, width=4.5cm]{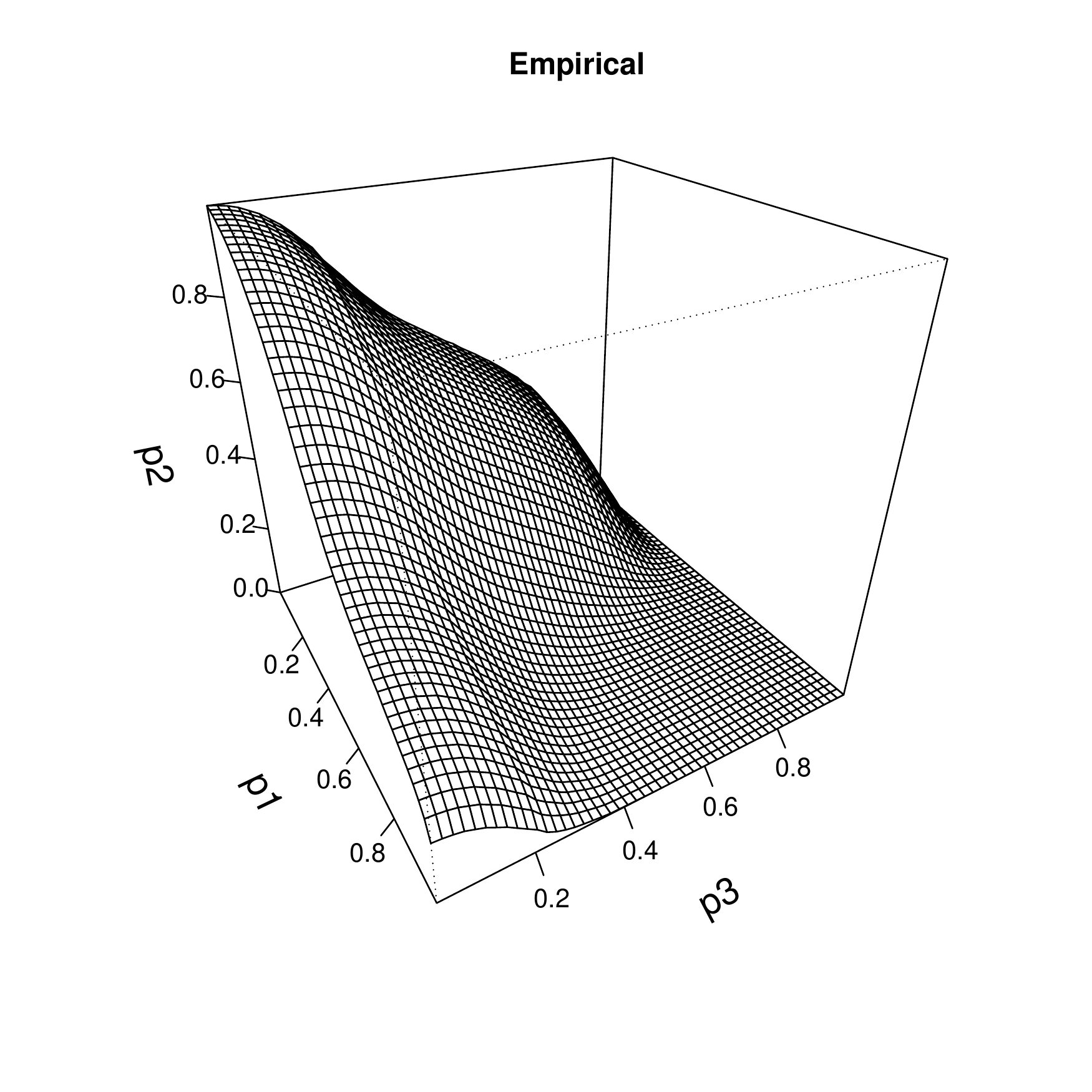}}\\
\subfigure{\includegraphics[page = 1, width=4.5cm]{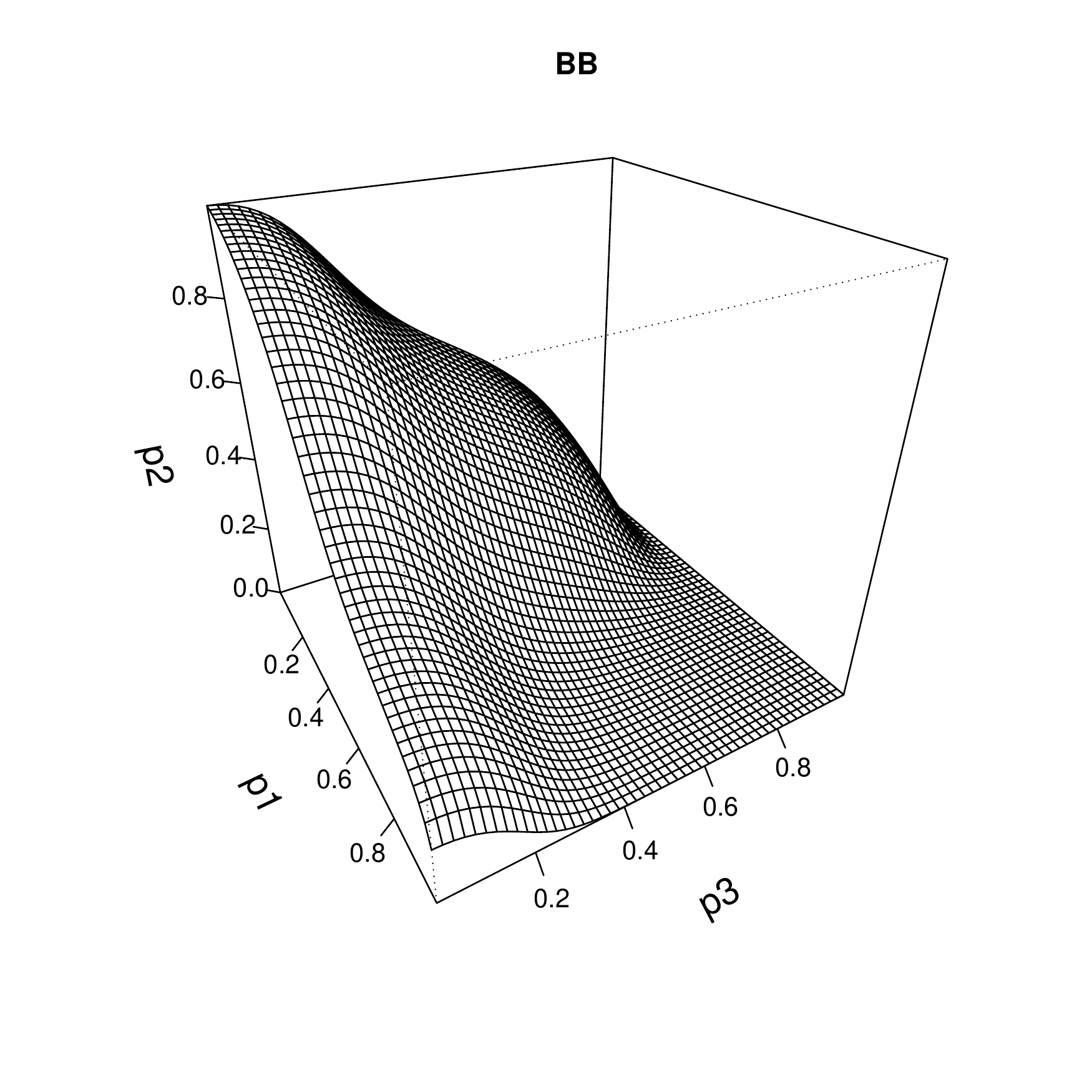}}
\subfigure{\includegraphics[page = 1, width=4.5cm]{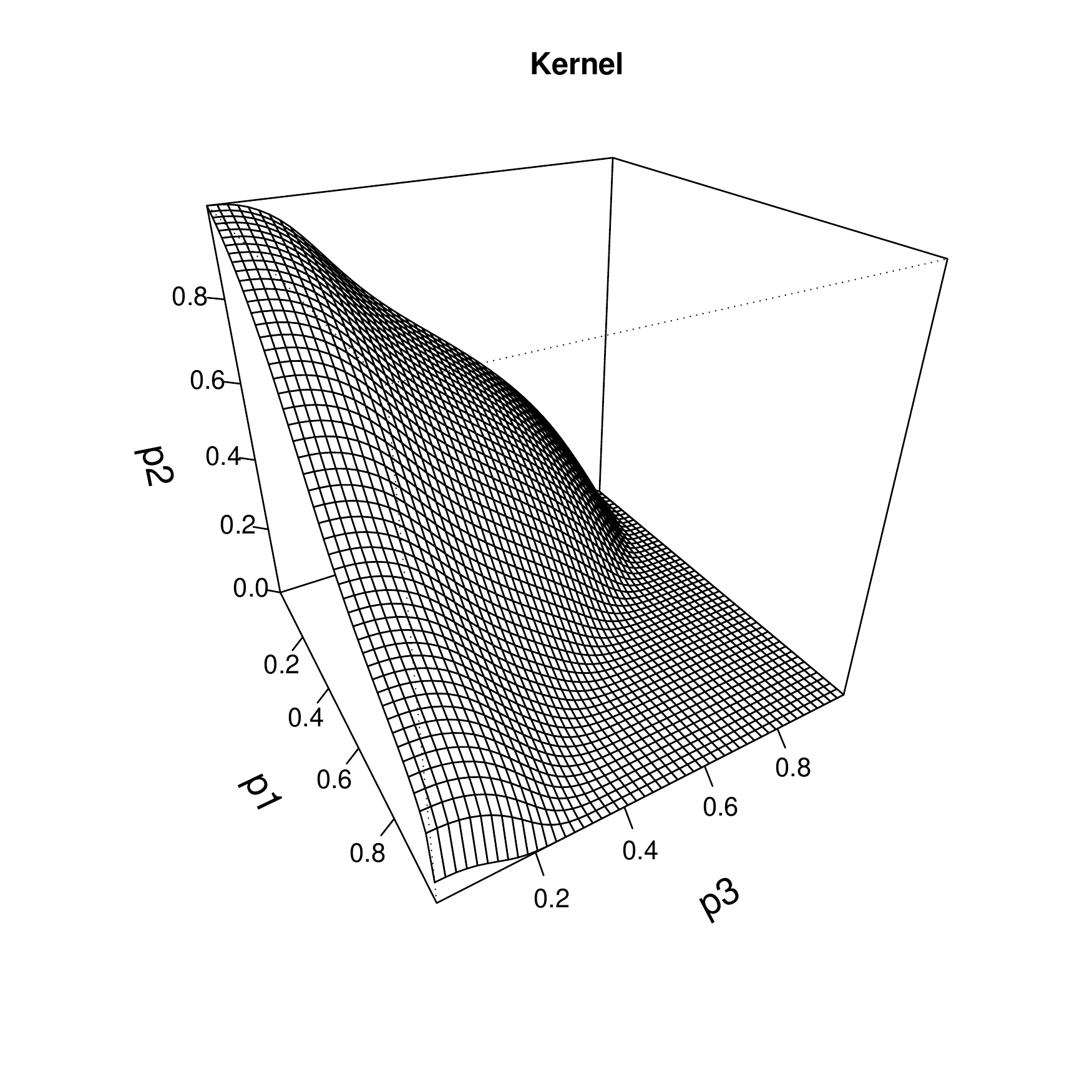}}
\subfigure{\includegraphics[page = 1, width=4.5cm]{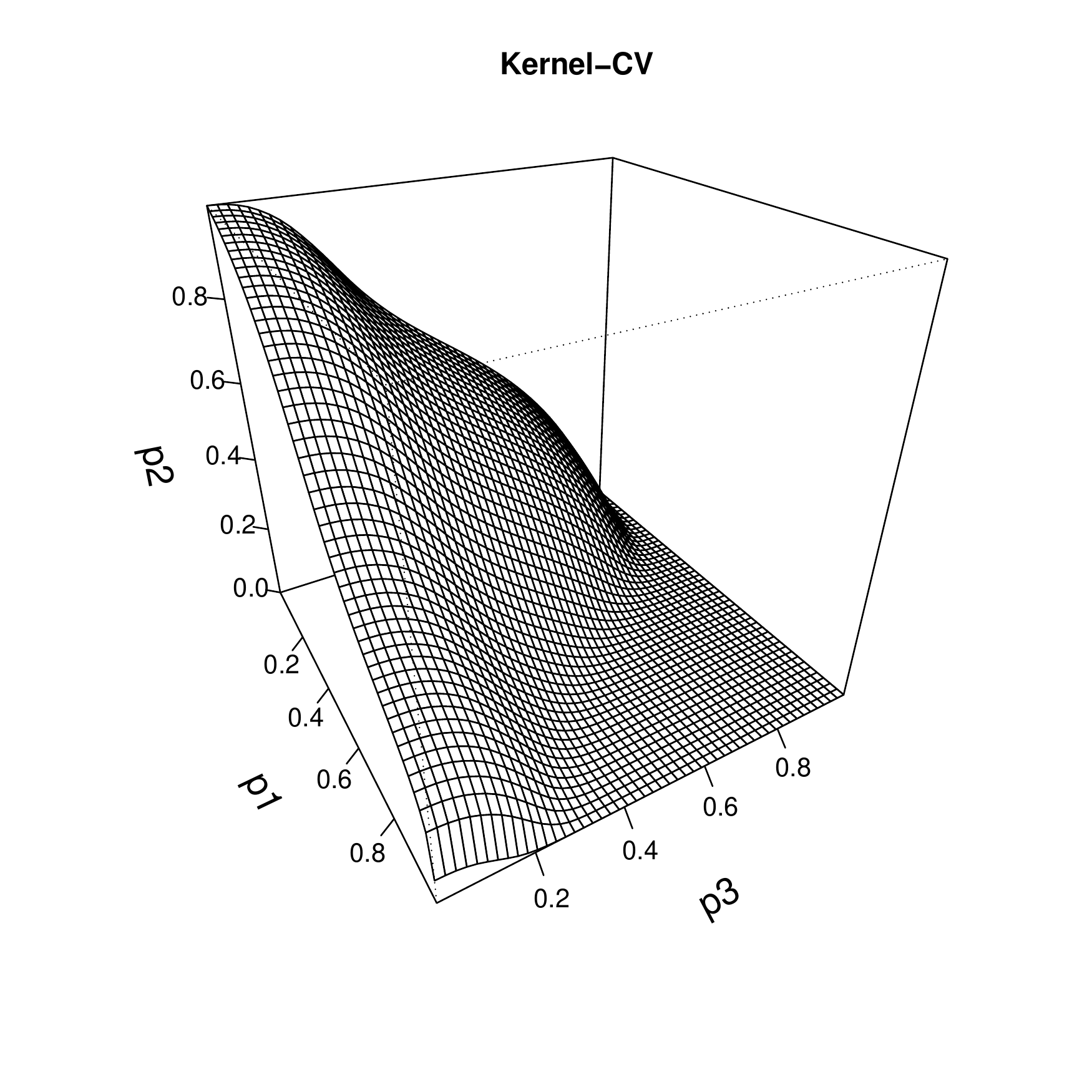}}
\subfigure{\includegraphics[page = 1, width=4.5cm]{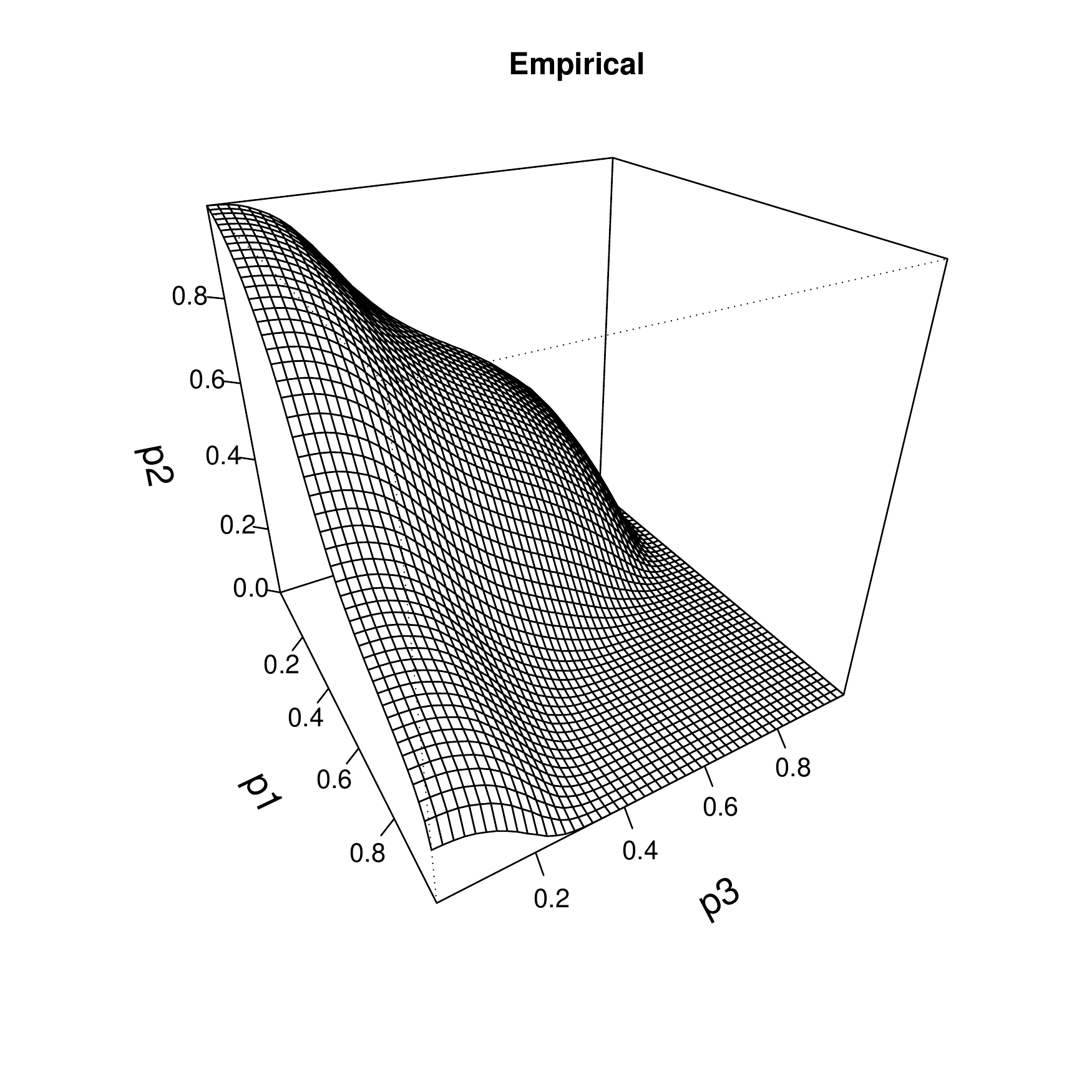}}\\
\subfigure{\includegraphics[page = 1, width=4.5cm]{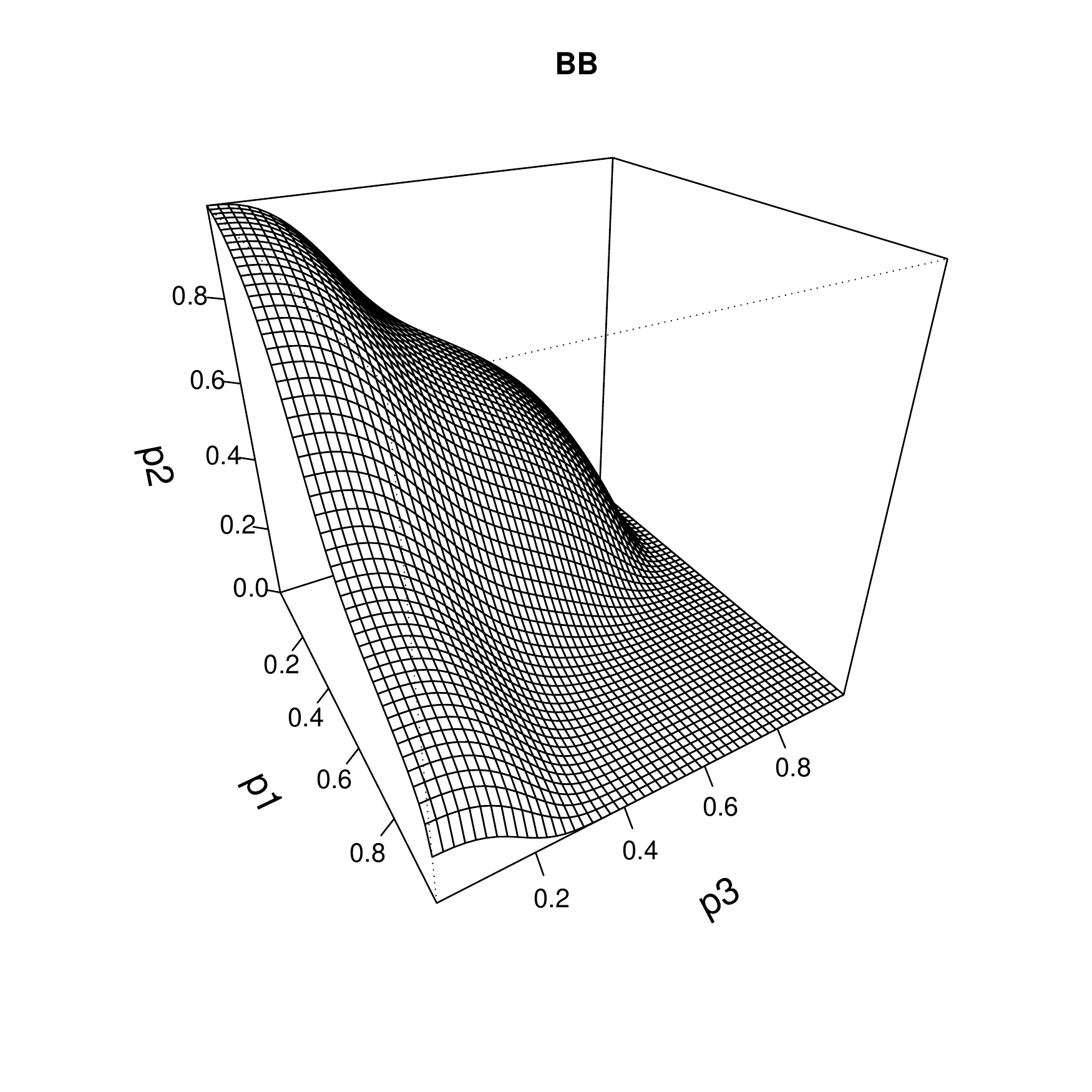}}
\subfigure{\includegraphics[page = 1, width=4.5cm]{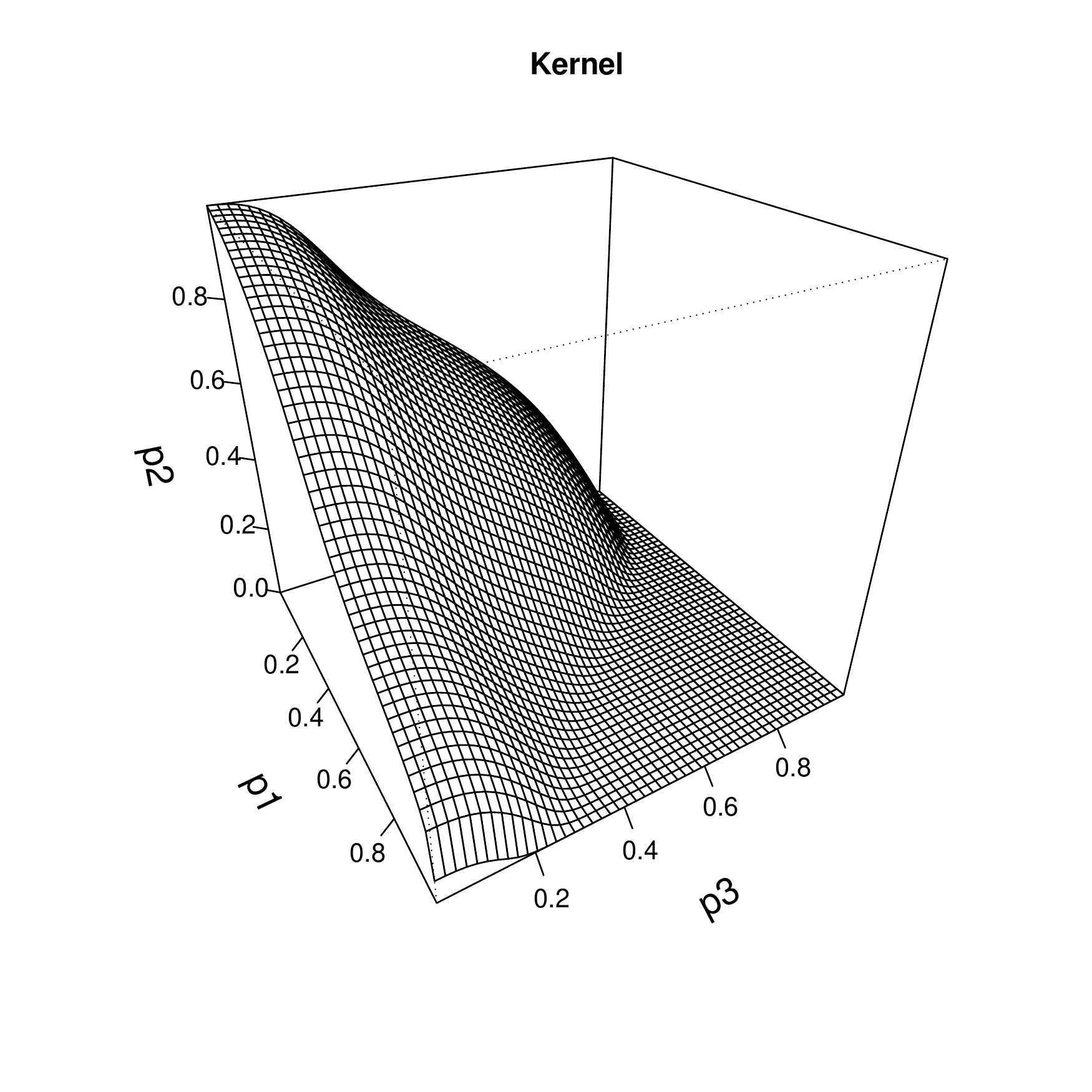}}
\subfigure{\includegraphics[page = 1, width=4.5cm]{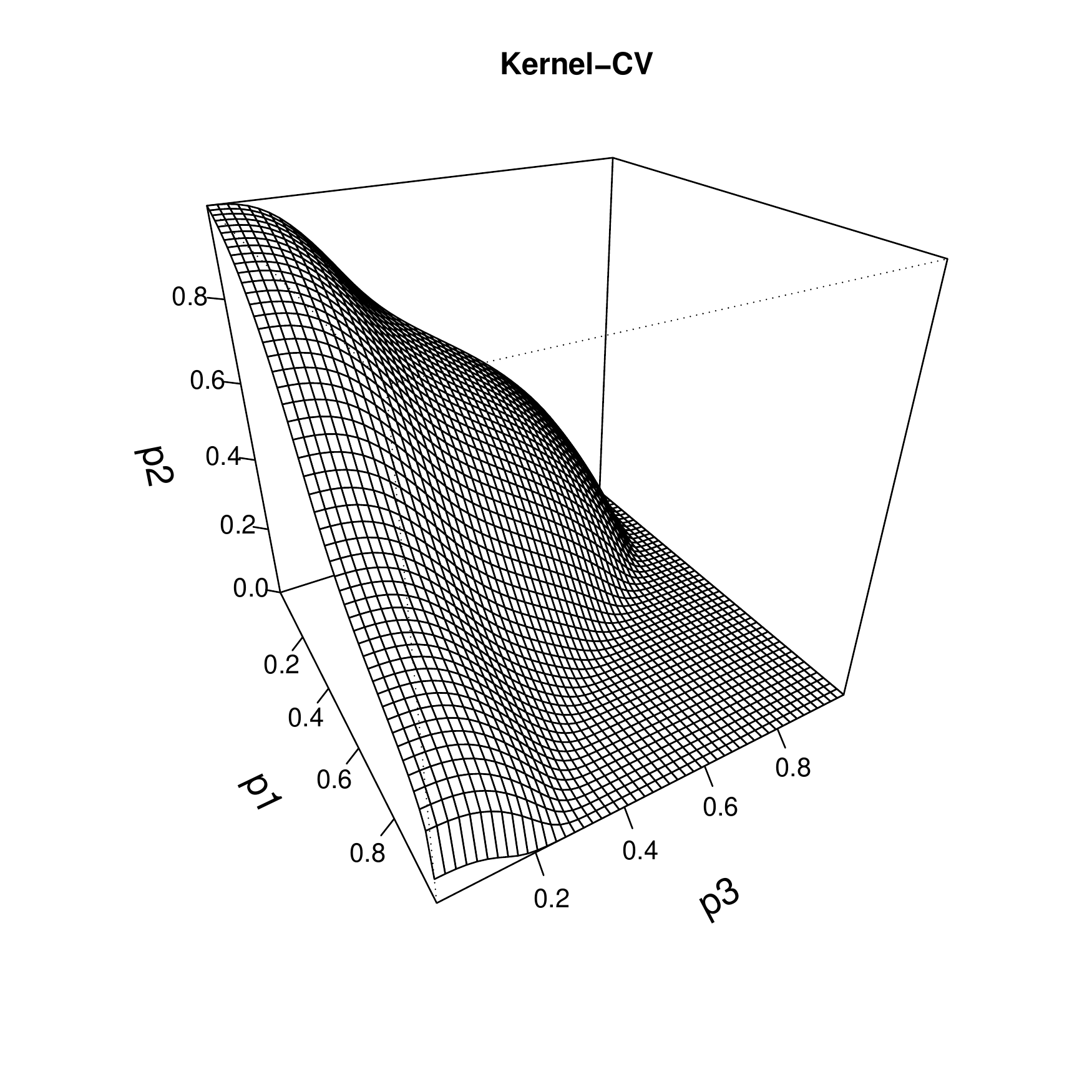}}
\subfigure{\includegraphics[page = 1, width=4.5cm]{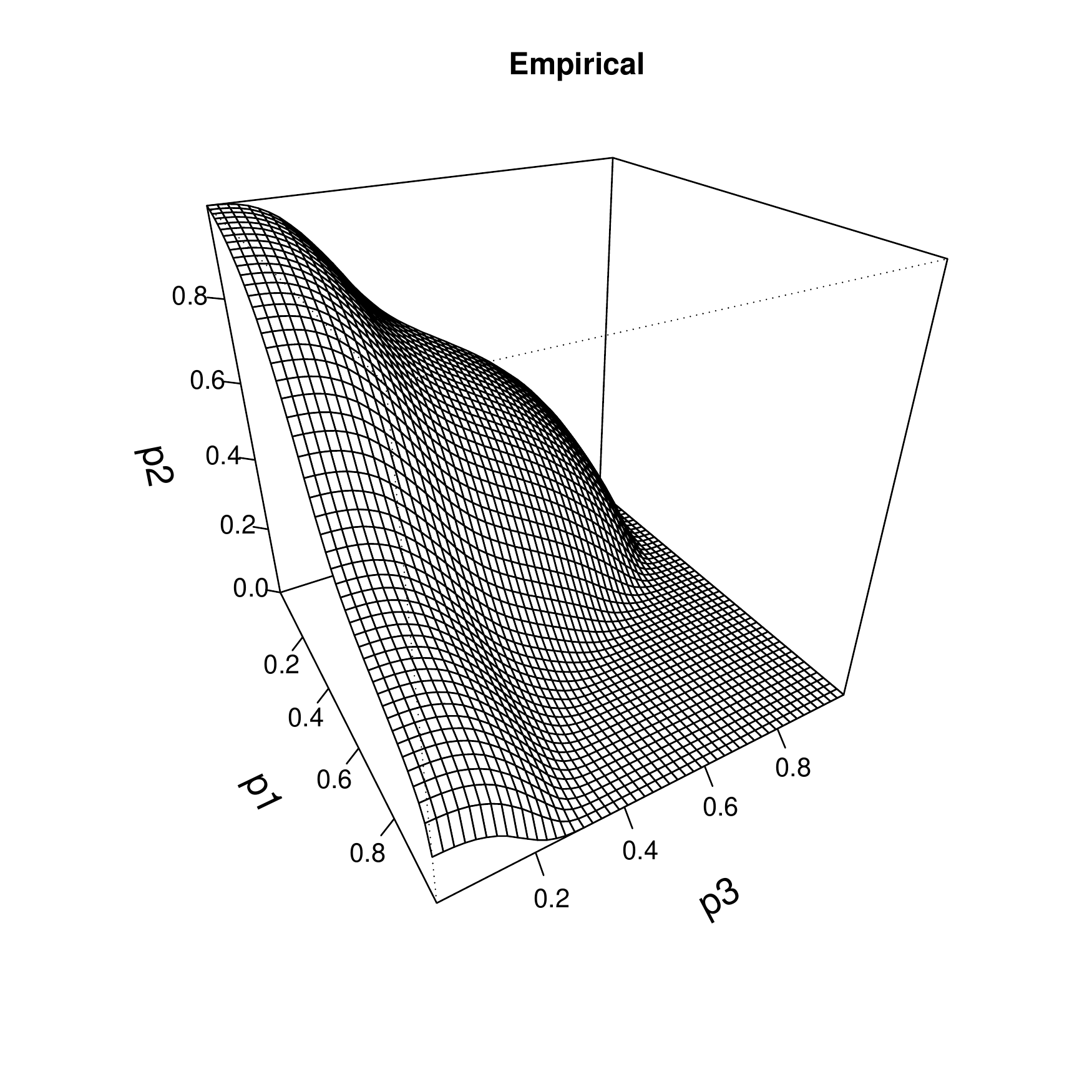}}
\caption{\footnotesize{Scenario 4. True ROC surface and mean across the 300 estimated ROC surfaces. First row: $n_1=n_2=n_3$=50. Second row: $n_1=n_2=n_3$=100. Third row: $n_1=n_2=n_3$=200. Here Kernel denotes the kernel estimate with bandwidth calculated using equation (2) of the main manuscript and Kernel-CV stands for the kernel estimate with the bandwidth selected by least squares cross-validation.}}
\end{center}
\end{figure}

\begin{figure}[H]
\begin{center}
\subfigure{\includegraphics[page = 1, width=5.85cm]{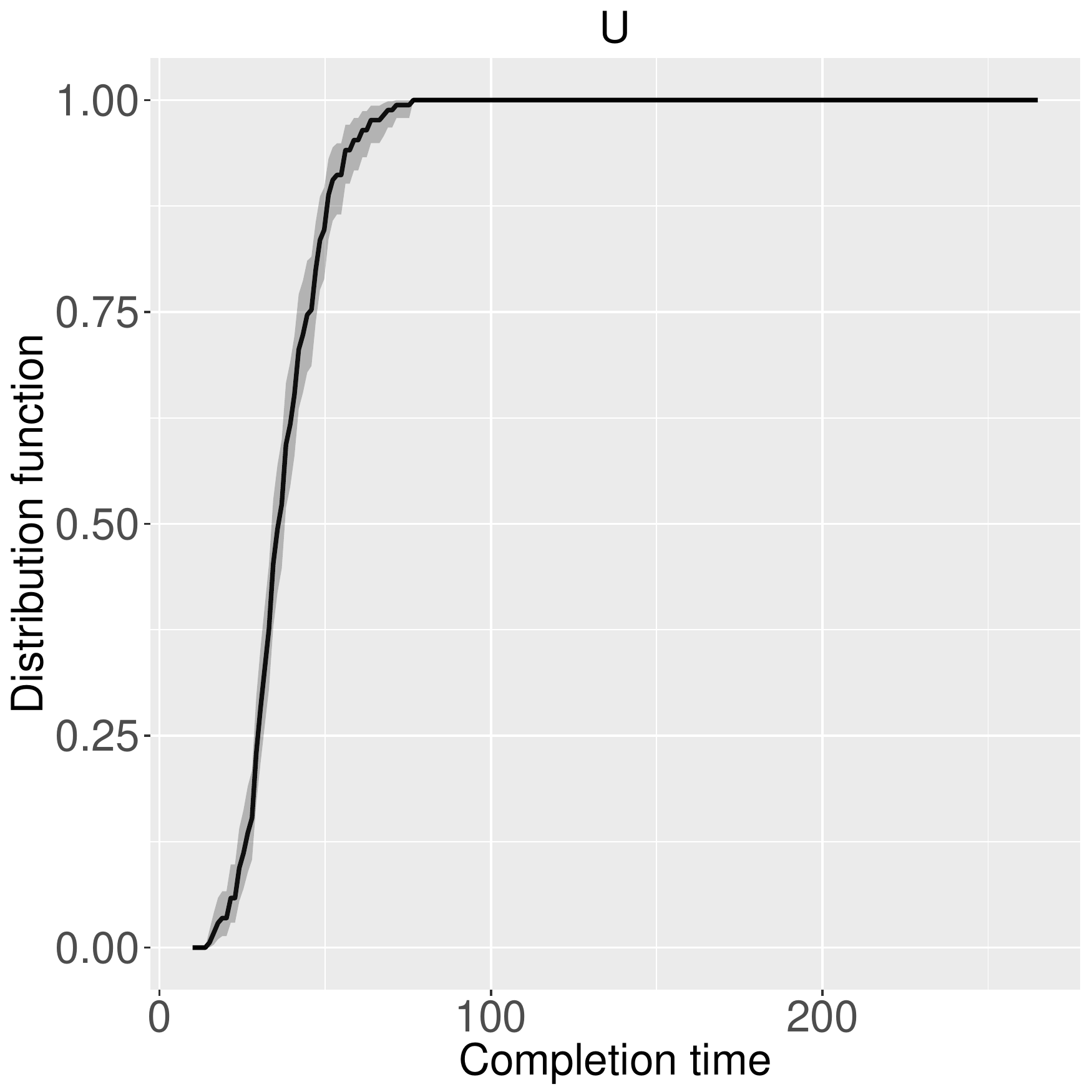}}\hspace{0.4cm}
\subfigure{\includegraphics[page = 1, width=5.85cm]{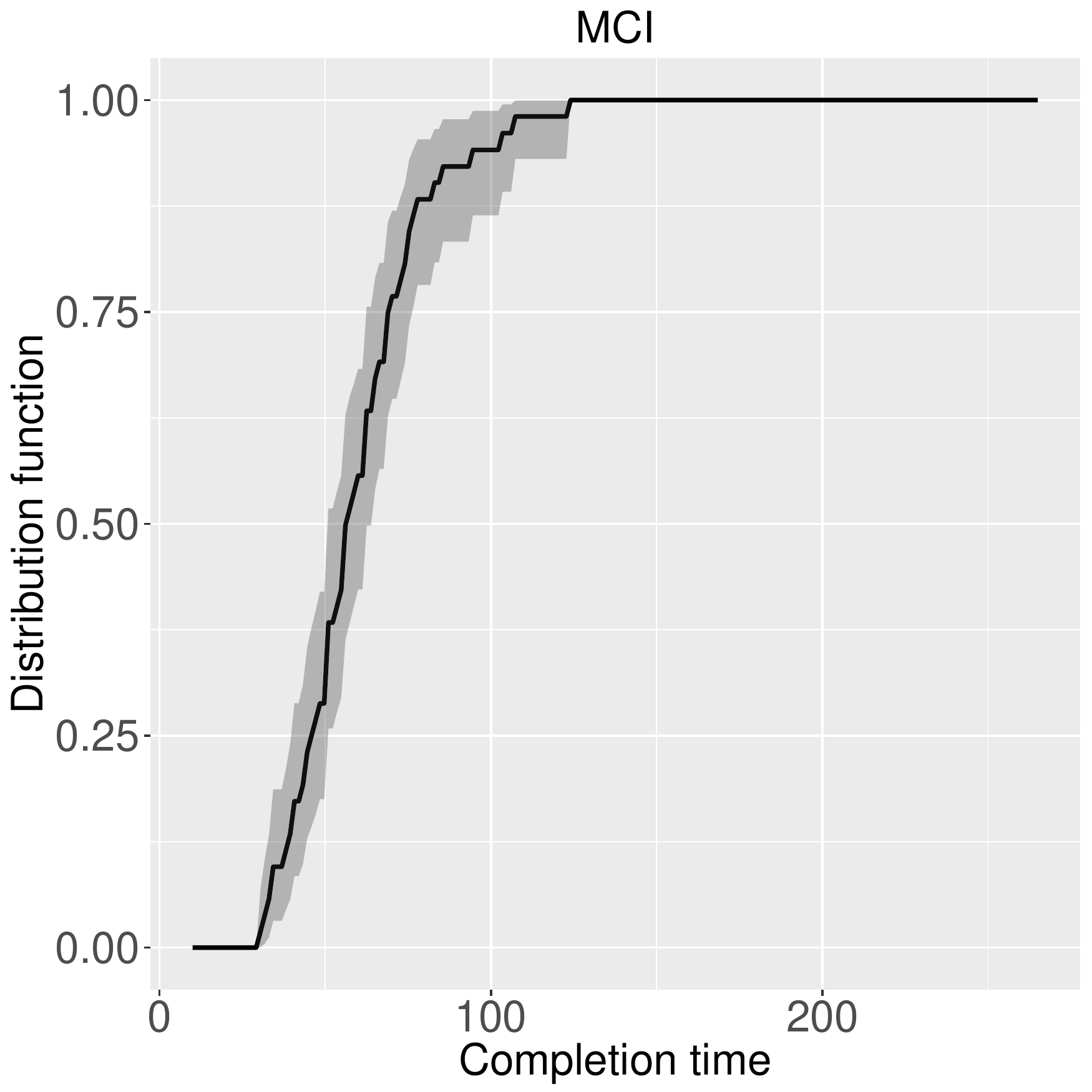}}\\
\subfigure{\includegraphics[page = 1, width=5.85cm]{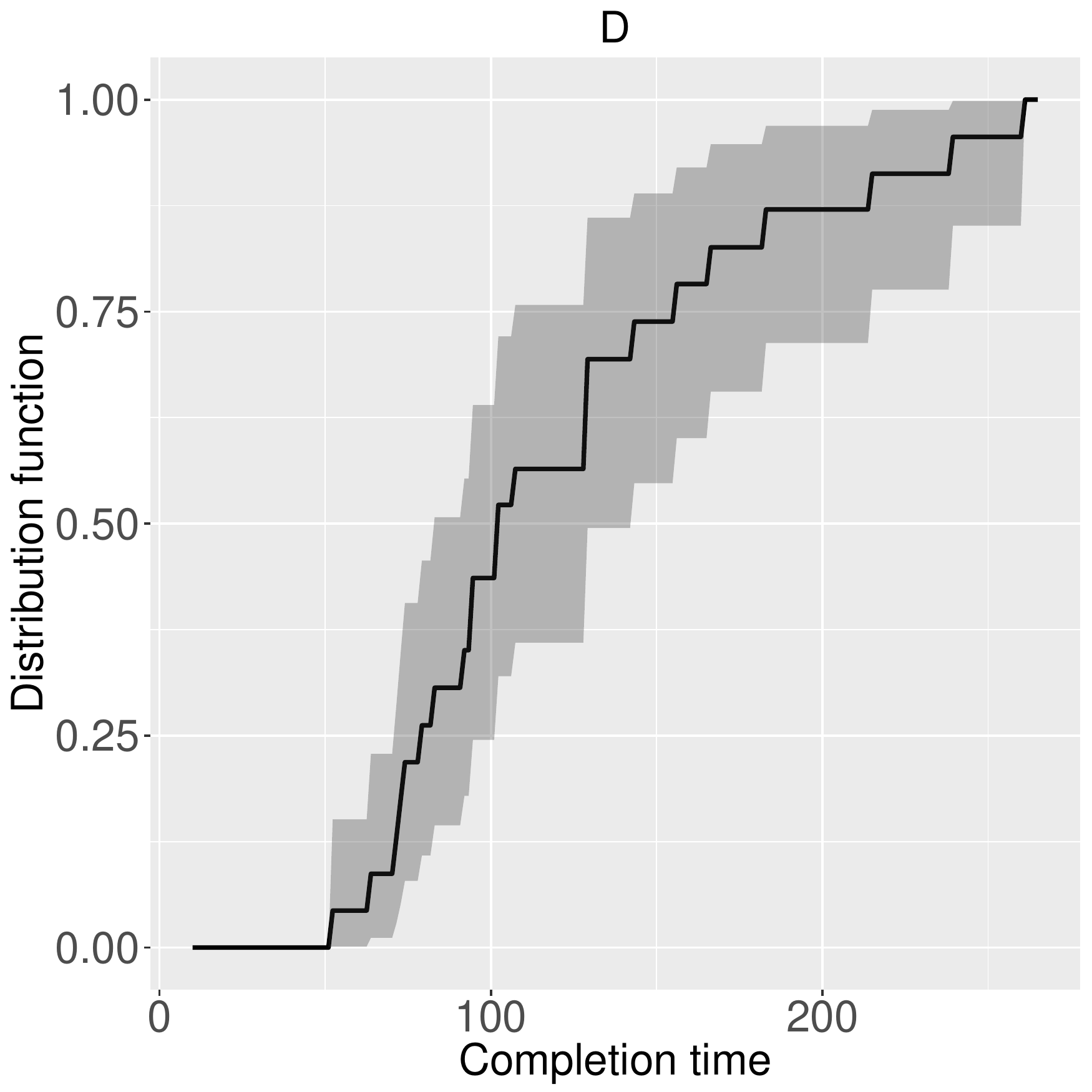}}\hspace{0.4cm}
\subfigure{\includegraphics[page = 1, width=5.85cm]{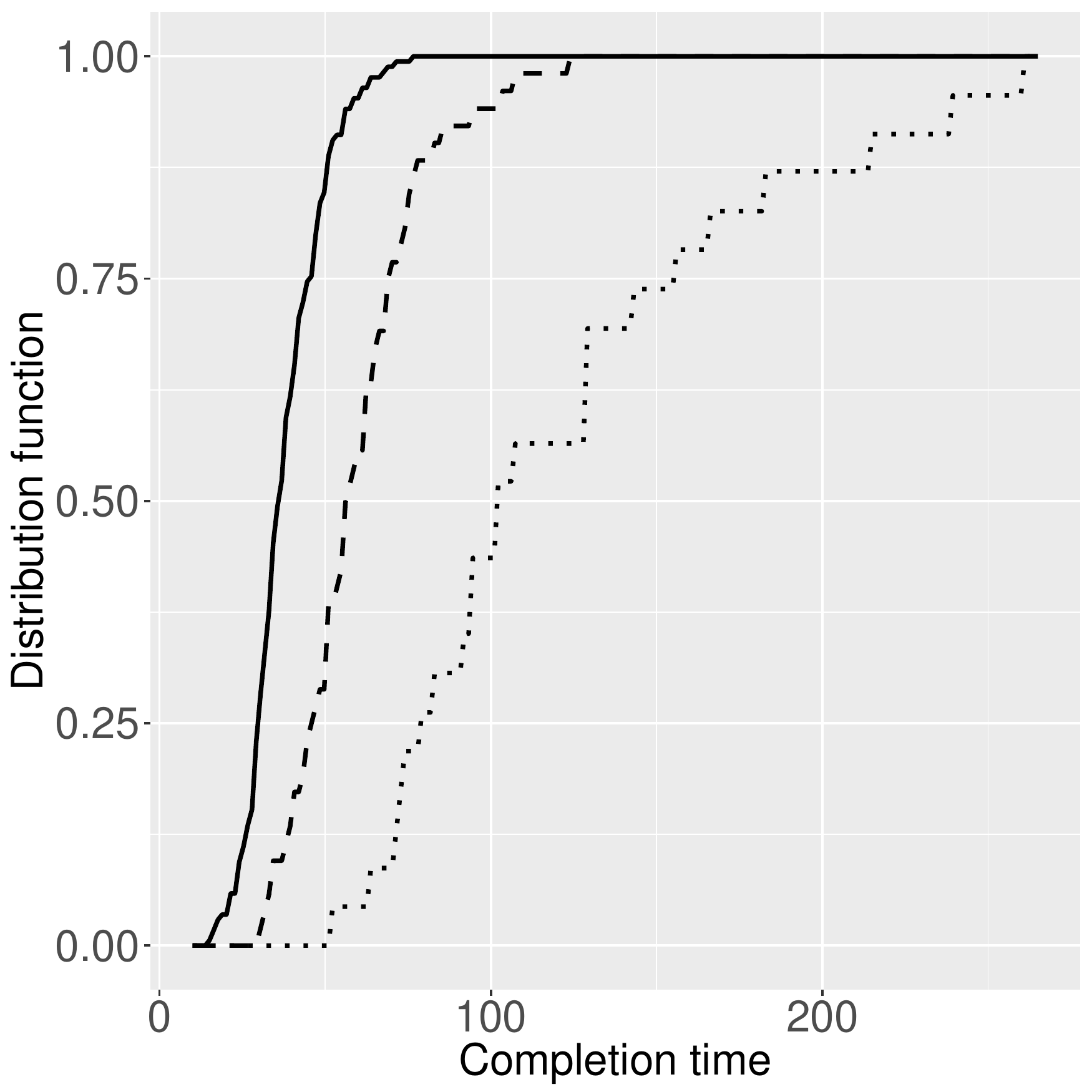}}
\caption{\footnotesize{TMT Part A data: BB estimate and 95\% pointwise probability band of (a) $F_1$ (unimpaired group), (b) $F_2$ (mild cognitive impairment group), and (c) $F_3$ (dementia group).}}
\label{cdfs}
\end{center}
\end{figure}

\begin{knitrout}\small
\definecolor{shadecolor}{rgb}{0.969, 0.969, 0.969}\color{fgcolor}\begin{kframe}
\begin{alltt}
\hlcom{#Data}
\hlstd{U}\hlkwb{=}\hlkwd{c}\hlstd{(}\hlnum{34}\hlstd{,} \hlnum{58}\hlstd{,} \hlnum{18}\hlstd{,} \hlnum{29}\hlstd{,} \hlnum{30}\hlstd{,} \hlnum{37}\hlstd{,} \hlnum{41}\hlstd{,} \hlnum{36}\hlstd{,} \hlnum{15}\hlstd{,} \hlnum{36}\hlstd{,} \hlnum{40}\hlstd{,} \hlnum{36}\hlstd{,} \hlnum{32}\hlstd{,} \hlnum{26}\hlstd{,} \hlnum{28}\hlstd{,} \hlnum{25}\hlstd{,} \hlnum{40}\hlstd{,}\hlnum{34}\hlstd{,} \hlnum{27}\hlstd{,} \hlnum{27}\hlstd{,} \hlnum{35}\hlstd{,}
    \hlnum{17}\hlstd{,} \hlnum{56}\hlstd{,} \hlnum{31}\hlstd{,} \hlnum{29}\hlstd{,} \hlnum{34}\hlstd{,} \hlnum{46}\hlstd{,} \hlnum{29}\hlstd{,} \hlnum{44}\hlstd{,} \hlnum{38}\hlstd{,} \hlnum{31}\hlstd{,} \hlnum{29}\hlstd{,} \hlnum{50}\hlstd{,} \hlnum{50}\hlstd{,} \hlnum{41}\hlstd{,} \hlnum{28}\hlstd{,} \hlnum{34}\hlstd{,} \hlnum{44}\hlstd{,} \hlnum{43}\hlstd{,} \hlnum{34}\hlstd{,} \hlnum{67}\hlstd{,} \hlnum{76}\hlstd{,}
    \hlnum{33}\hlstd{,} \hlnum{28}\hlstd{,}\hlnum{51}\hlstd{,} \hlnum{45}\hlstd{,} \hlnum{61}\hlstd{,} \hlnum{36}\hlstd{,} \hlnum{47}\hlstd{,} \hlnum{30}\hlstd{,} \hlnum{35}\hlstd{,} \hlnum{39}\hlstd{,} \hlnum{42}\hlstd{,} \hlnum{40}\hlstd{,} \hlnum{42}\hlstd{,} \hlnum{41}\hlstd{,}\hlnum{17}\hlstd{,} \hlnum{25}\hlstd{,} \hlnum{48}\hlstd{,} \hlnum{61}\hlstd{,} \hlnum{48}\hlstd{,} \hlnum{34}\hlstd{,} \hlnum{31}\hlstd{,} \hlnum{35}\hlstd{,}
    \hlnum{48}\hlstd{,} \hlnum{30}\hlstd{,} \hlnum{33}\hlstd{,} \hlnum{34}\hlstd{,} \hlnum{34}\hlstd{,} \hlnum{58}\hlstd{,} \hlnum{28}\hlstd{,} \hlnum{28}\hlstd{,} \hlnum{24}\hlstd{,}\hlnum{55}\hlstd{,} \hlnum{21}\hlstd{,} \hlnum{21}\hlstd{,} \hlnum{37}\hlstd{,} \hlnum{25}\hlstd{,} \hlnum{38}\hlstd{,} \hlnum{40}\hlstd{,} \hlnum{55}\hlstd{,} \hlnum{35}\hlstd{,} \hlnum{39}\hlstd{,} \hlnum{34}\hlstd{,}\hlnum{28}\hlstd{,} \hlnum{37}\hlstd{,}
    \hlnum{37}\hlstd{,} \hlnum{46}\hlstd{,} \hlnum{37}\hlstd{,} \hlnum{51}\hlstd{,} \hlnum{37}\hlstd{,} \hlnum{30}\hlstd{,} \hlnum{46}\hlstd{,} \hlnum{37}\hlstd{,} \hlnum{24}\hlstd{,} \hlnum{38}\hlstd{,} \hlnum{23}\hlstd{,} \hlnum{52}\hlstd{,} \hlnum{40}\hlstd{,} \hlnum{34}\hlstd{,} \hlnum{29}\hlstd{,}\hlnum{44}\hlstd{,} \hlnum{30}\hlstd{,} \hlnum{24}\hlstd{,} \hlnum{35}\hlstd{,} \hlnum{21}\hlstd{,} \hlnum{48}\hlstd{,} \hlnum{47}\hlstd{,}
    \hlnum{16}\hlstd{,} \hlnum{34}\hlstd{,} \hlnum{30}\hlstd{,} \hlnum{28}\hlstd{,}\hlnum{35}\hlstd{,} \hlnum{36}\hlstd{,} \hlnum{34}\hlstd{,} \hlnum{27}\hlstd{,} \hlnum{31}\hlstd{,} \hlnum{37}\hlstd{,} \hlnum{26}\hlstd{,} \hlnum{50}\hlstd{,} \hlnum{44}\hlstd{,} \hlnum{42}\hlstd{,} \hlnum{32}\hlstd{,} \hlnum{42}\hlstd{,} \hlnum{48}\hlstd{,} \hlnum{43}\hlstd{,} \hlnum{49}\hlstd{,} \hlnum{23}\hlstd{,} \hlnum{49}\hlstd{,} \hlnum{16}\hlstd{,}
    \hlnum{26}\hlstd{,} \hlnum{52}\hlstd{,} \hlnum{34}\hlstd{,} \hlnum{55}\hlstd{,} \hlnum{51}\hlstd{,} \hlnum{46}\hlstd{,} \hlnum{63}\hlstd{,} \hlnum{42}\hlstd{,} \hlnum{41}\hlstd{,} \hlnum{53}\hlstd{,}\hlnum{38}\hlstd{,} \hlnum{21}\hlstd{,} \hlnum{68}\hlstd{,} \hlnum{56}\hlstd{,} \hlnum{46}\hlstd{,} \hlnum{31}\hlstd{,} \hlnum{33}\hlstd{,} \hlnum{52}\hlstd{,} \hlnum{33}\hlstd{,} \hlnum{30}\hlstd{,} \hlnum{50}\hlstd{,} \hlnum{71}\hlstd{,}
    \hlnum{29}\hlstd{,} \hlnum{48}\hlstd{,} \hlnum{63}\hlstd{,} \hlnum{39}\hlstd{,} \hlnum{31}\hlstd{,} \hlnum{32}\hlstd{,} \hlnum{32}\hlstd{,} \hlnum{43}\hlstd{,} \hlnum{26}\hlstd{,} \hlnum{35}\hlstd{,} \hlnum{40}\hlstd{,} \hlnum{39}\hlstd{,} \hlnum{31}\hlstd{,} \hlnum{31}\hlstd{,} \hlnum{30}\hlstd{,} \hlnum{24}\hlstd{,} \hlnum{47}\hlstd{,} \hlnum{30}\hlstd{)}

\hlstd{MCI}\hlkwb{=}\hlkwd{c}\hlstd{(}\hlnum{66}\hlstd{,} \hlnum{34}\hlstd{,} \hlnum{44}\hlstd{,} \hlnum{56}\hlstd{,} \hlnum{75}\hlstd{,} \hlnum{45}\hlstd{,} \hlnum{48}\hlstd{,} \hlnum{43}\hlstd{,} \hlnum{62}\hlstd{,} \hlnum{68}\hlstd{,} \hlnum{85}\hlstd{,} \hlnum{107}\hlstd{,} \hlnum{34}\hlstd{,} \hlnum{82}\hlstd{,} \hlnum{68}\hlstd{,} \hlnum{103}\hlstd{,} \hlnum{51}\hlstd{,} \hlnum{57}\hlstd{,} \hlnum{50}\hlstd{,} \hlnum{30}\hlstd{,}
      \hlnum{38}\hlstd{,} \hlnum{59}\hlstd{,} \hlnum{31}\hlstd{,} \hlnum{68}\hlstd{,} \hlnum{65}\hlstd{,} \hlnum{62}\hlstd{,} \hlnum{51}\hlstd{,} \hlnum{74}\hlstd{,} \hlnum{46}\hlstd{,} \hlnum{70}\hlstd{,} \hlnum{40}\hlstd{,} \hlnum{54}\hlstd{,} \hlnum{51}\hlstd{,} \hlnum{56}\hlstd{,} \hlnum{40}\hlstd{,} \hlnum{72}\hlstd{,} \hlnum{123}\hlstd{,} \hlnum{62}\hlstd{,} \hlnum{64}\hlstd{,} \hlnum{76}\hlstd{,}
      \hlnum{77}\hlstd{,} \hlnum{75}\hlstd{,} \hlnum{55}\hlstd{,} \hlnum{94}\hlstd{,} \hlnum{44}\hlstd{,} \hlnum{51}\hlstd{,} \hlnum{62}\hlstd{,} \hlnum{33}\hlstd{,} \hlnum{58}\hlstd{,} \hlnum{53}\hlstd{,} \hlnum{39}\hlstd{,} \hlnum{55}\hlstd{)}

\hlstd{D}\hlkwb{=}\hlkwd{c}\hlstd{(}\hlnum{182}\hlstd{,} \hlnum{63}\hlstd{,} \hlnum{166}\hlstd{,} \hlnum{143}\hlstd{,} \hlnum{94} \hlstd{,}\hlnum{155}\hlstd{,} \hlnum{78}\hlstd{,} \hlnum{91}\hlstd{,} \hlnum{239}\hlstd{,} \hlnum{261}\hlstd{,} \hlnum{101}\hlstd{,} \hlnum{129}\hlstd{,} \hlnum{73}\hlstd{,} \hlnum{214}\hlstd{,} \hlnum{82}\hlstd{,} \hlnum{72}\hlstd{,} \hlnum{107}\hlstd{,}
    \hlnum{129}\hlstd{,} \hlnum{128}\hlstd{,} \hlnum{52}\hlstd{,} \hlnum{94}\hlstd{,} \hlnum{71}\hlstd{,} \hlnum{101}\hlstd{)}

\hlstd{y1}\hlkwb{=}\hlstd{U; y2}\hlkwb{=}\hlstd{MCI; y3}\hlkwb{=}\hlstd{D}
\hlstd{p}\hlkwb{=}\hlkwd{seq}\hlstd{(}\hlnum{0.0001}\hlstd{,}\hlnum{0.9999}\hlstd{,}\hlkwc{len}\hlstd{=}\hlnum{50}\hlstd{)}

\hlcom{#Method}
\hlstd{rocsbb}\hlkwb{=}\hlkwa{function}\hlstd{(}\hlkwc{y1}\hlstd{,}\hlkwc{y2}\hlstd{,}\hlkwc{y3}\hlstd{,}\hlkwc{p1}\hlstd{,}\hlkwc{p3}\hlstd{,}\hlkwc{B}\hlstd{)\{}
\hlstd{np}\hlkwb{=}\hlkwd{length}\hlstd{(p1); n1}\hlkwb{=}\hlkwd{length}\hlstd{(y1); n2}\hlkwb{=}\hlkwd{length}\hlstd{(y2); n3}\hlkwb{=}\hlkwd{length}\hlstd{(y3)}
\hlstd{rocbb}\hlkwb{=}\hlkwd{array}\hlstd{(}\hlnum{0}\hlstd{,}\hlkwd{c}\hlstd{(np,np,B)); vusb}\hlkwb{=}\hlkwd{numeric}\hlstd{(B)}
\hlkwa{for}\hlstd{(b} \hlkwa{in} \hlnum{1}\hlopt{:}\hlstd{B)\{}
\hlstd{aux3}\hlkwb{=}\hlkwd{rexp}\hlstd{(n3,}\hlnum{1}\hlstd{); v3}\hlkwb{=}\hlstd{aux3}\hlopt{/}\hlkwd{sum}\hlstd{(aux3)}
\hlstd{aux1}\hlkwb{=}\hlkwd{rexp}\hlstd{(n1,}\hlnum{1}\hlstd{); v1}\hlkwb{=}\hlstd{aux1}\hlopt{/}\hlkwd{sum}\hlstd{(aux1)}
\hlstd{u3}\hlkwb{=}\hlkwd{numeric}\hlstd{(n2); u1}\hlkwb{=}\hlkwd{numeric}\hlstd{(n2)}
\hlkwa{for}\hlstd{(j} \hlkwa{in} \hlnum{1}\hlopt{:}\hlstd{n2)\{}
\hlstd{u3[j]}\hlkwb{=}\hlkwd{sum}\hlstd{(v3}\hlopt{*}\hlstd{(y3}\hlopt{>}\hlstd{y2[j]))}
\hlstd{u1[j]}\hlkwb{=}\hlkwd{sum}\hlstd{(v1}\hlopt{*}\hlstd{(y1}\hlopt{<=}\hlstd{y2[j]))}
\hlstd{\}}
\hlstd{aux3a}\hlkwb{=}\hlkwd{rexp}\hlstd{(n2,}\hlnum{1}\hlstd{); omega3}\hlkwb{=}\hlstd{aux3a}\hlopt{/}\hlkwd{sum}\hlstd{(aux3a)}
\hlstd{aux1a}\hlkwb{=}\hlkwd{rexp}\hlstd{(n2,}\hlnum{1}\hlstd{); omega1}\hlkwb{=}\hlstd{aux1a}\hlopt{/}\hlkwd{sum}\hlstd{(aux1a)}
\hlkwa{for}\hlstd{(i} \hlkwa{in} \hlnum{1}\hlopt{:}\hlstd{np)\{}
\hlkwa{for}\hlstd{(j} \hlkwa{in} \hlnum{1}\hlopt{:}\hlstd{np)\{}
\hlstd{rocbb[i,j,b]}\hlkwb{=}\hlkwd{sum}\hlstd{(omega3}\hlopt{*}\hlstd{(u3}\hlopt{>}\hlstd{p3[j]))}\hlopt{-}\hlkwd{sum}\hlstd{(omega1}\hlopt{*}\hlstd{(u1}\hlopt{<=}\hlstd{p1[i]))}
\hlkwa{if}\hlstd{(rocbb[i,j,b]}\hlopt{<}\hlnum{0}\hlstd{) rocbb[i,j,b]}\hlkwb{=}\hlnum{0}
\hlstd{\}}
\hlstd{\}}
\hlstd{vusb[b]}\hlkwb{=}\hlkwd{sum}\hlstd{(rocbb[,,b])}\hlopt{/}\hlstd{(np}\hlopt{*}\hlstd{np)}
\hlstd{\}}
\hlstd{rocbbf}\hlkwb{=}\hlkwd{apply}\hlstd{(rocbb,}\hlnum{1}\hlopt{:}\hlnum{2}\hlstd{,mean); vusbf}\hlkwb{=}\hlkwd{mean}\hlstd{(vusb); int}\hlkwb{=}\hlkwd{quantile}\hlstd{(vusb,}\hlkwd{c}\hlstd{(}\hlnum{0.025}\hlstd{,}\hlnum{0.975}\hlstd{))}

\hlkwd{return}\hlstd{(}\hlkwd{list}\hlstd{(rocbbf,vusbf,int))}
\hlstd{\}}

\hlstd{res}\hlkwb{=}\hlkwd{rocsbb}\hlstd{(}\hlkwc{y1}\hlstd{=y1,}\hlkwc{y2}\hlstd{=y2,}\hlkwc{y3}\hlstd{=y3,}\hlkwc{p1}\hlstd{=p,}\hlkwc{p3}\hlstd{=p,}\hlkwc{B}\hlstd{=}\hlnum{5000}\hlstd{)}

\hlcom{#ROC surface}
\hlkwd{persp}\hlstd{(p,p,res[[}\hlnum{1}\hlstd{]],}\hlkwc{phi}\hlstd{=}\hlnum{30}\hlstd{,}\hlkwc{theta}\hlstd{=}\hlnum{60}\hlstd{,} \hlkwc{xlab} \hlstd{=} \hlstr{"\textbackslash{}n\textbackslash{}n p1"}\hlstd{,} \hlkwc{ylab} \hlstd{=} \hlstr{"\textbackslash{}n\textbackslash{}n p3"}\hlstd{,} \hlkwc{zlab} \hlstd{=} \hlstr{"\textbackslash{}n\textbackslash{}n p2"}\hlstd{,}
      \hlkwc{ticktype}\hlstd{=}\hlstr{"detailed"}\hlstd{,}\hlkwc{cex.lab}\hlstd{=}\hlnum{1.4}\hlstd{)}

\hlcom{#mean VUS and 95% credible interval}
\hlstd{vusm}\hlkwb{=}\hlstd{res[[}\hlnum{2}\hlstd{]]; vusci}\hlkwb{=}\hlstd{res[[}\hlnum{3}\hlstd{]]}
\hlstd{vusm; vusci}
\end{alltt}
\end{kframe}
\end{knitrout}
\end{document}